\documentclass[%
 reprint,
 superscriptaddress,
 amsmath,amssymb,
 aps,
prb,
]{revtex4-2}

\usepackage{dcolumn}
\usepackage{bm}
\usepackage{amsmath}
\usepackage{txfonts}
\usepackage[T1]{fontenc}
\usepackage{xspace}
\usepackage{comment}
\usepackage{bbold}
\usepackage{braket}

\ifx\pdfoutput\undefined
\usepackage[dvipdfmx]{graphicx}
\usepackage[dvipdfmx]{hyperref}
\usepackage[dvipdfmx]{color}
\usepackage[dvipdfmx]{xcolor}
\else
\usepackage{graphicx}
\usepackage{hyperref}
\usepackage{color}
\usepackage{xcolor}
\fi

\newcommand{\QQ}{\mathcal{Q}}
\newcommand{\vp}{\varphi}
\newcommand{\tvp}{\tilde{\varphi}}
\newcommand{\Sr}{{\bf S}({\bf r})}

\graphicspath{
{../figure/}
{./figure/}}

\begin{document}

\title{
Phase degree of freedom and topology in multiple-$Q$ spin textures 
}

\author{Kotaro Shimizu}
\affiliation{Department of Applied Physics, The University of Tokyo, Tokyo 113-8656, Japan} 

\author{Shun Okumura} 
\affiliation{The Institute for Solid State Physics, The University of Tokyo, Kashiwa 277-8581, Japan}

\author{Yasuyuki Kato}
\author{Yukitoshi Motome}
\affiliation{Department of Applied Physics, The University of Tokyo, Tokyo 113-8656, Japan}

\date{\today}

\begin{abstract}
A periodic array of topological spin textures, such as skyrmions and hedgehogs, is called the multiple-$Q$ spin texture, as it is represented by a superposition of multiple spin density waves. 
Depending on the way of superposition, not only the magnetic but also the topological properties are modified, leading to a variety of quantum transport and optical phenomena caused by the emergent electromagnetic fields through the Berry phase. 
Among others, the phase degree of freedom of the superposed waves is potentially important for such modifications, but its effect has not been fully investigated thus far. 
Here we perform systematic theoretical analyses of magnetic and topological properties of the multiple-$Q$ spin textures with the phase degree of freedom in two and three dimensions. 
By introducing a hyperspace with an additional dimension corresponding to the phase degree of freedom, 
we establish a generic framework to deal with the phase shift in the multiple-$Q$ spin textures. 
Using the hyperspace representation, we elaborate the complete topological phase diagrams for the superpositions of three proper screws or sinusoidal waves in two dimensions and those of four in three dimensions. 
In the two-dimensional case, we find that the phase shift as well as the magnetization change can yield the skyrmion lattices with the skymion number of $-2$, $-1$, $1$, and $2$, corresponding to the evolution of the Dirac strings connecting hedgehogs and antihedgehogs in the three-dimensional hyperspace. 
We show that the high skyrmion numbers $\pm 2$ appear in wider parameter regions for the sinusoidal superpositions than the screw ones. 
Meanwhile, in the three-dimensional case, we clarify that the topological phase diagrams include various types of the hedgehog lattices whose total number of hedgehogs and antihedgehogs ranges up to $48$ in a cubic unit. 
Interestingly, the phase shift can generate unusual Dirac strings running on the horizontal planes perpendicular to the magnetization direction, which gives rise to unconventional pair creation of hedgehogs and antihedgehogs while increasing the magnetization in the case of the screw superpositions. 
We also show that the amplitude of the emergent magnetic field is maximized by fusion of hedgehogs and antihedgehogs on the horizontal Dirac strings in both proper screw and sinusoidal cases. 
In addition, by analyzing the numerical data in the previous studies, we demonstrate that phase shifts are indeed caused by an external magnetic field, associated with the topological transitions in the multiple-$Q$ spin textures.
Our results illuminate the topological aspects of the skyrmion and hedgehog lattices with the phase degree of freedom, which would be extended to other multiple-$Q$ textures and useful for the exploration of topologically nontrivial magnetic phases and exotic quantum phenomena.
\end{abstract}


\maketitle

\section{Introduction \label{sec:1}}

Topology is originally a mathematical concept to discuss the properties of a geometric object, but has been extended to a variety of research fields in these decades~\cite{Mermin1979, Nakahara2003, Braun2012, Xiao2010}. 
In condensed matter physics, the electronic states of solids have been discussed by topology of the electronic band structures in momentum space, which led to the discovery of topological states of matter, e.g., the quantum Hall state~\cite{Ando1974, Klitzing1980, Laughlin1981, Thouless1982} and the topological insulator~\cite{Kane2005, Bernevig2006, Fu2007TI3D, Moore2007, Fu2007TIIS, Roy2009, Hasan2010, Ando2013}. 
Topology appears also in geometric structures of the spin textures in magnets. 
The typical examples are swirling noncoplanar spin textures, such as magnetic skyrmions~\cite{Bogdanov1989, Bogdanov1994, Bogdanov1995, Roessler2006} and Bloch points~\cite{Feldtkeller1965, Doring1968, Kotiuga1989} (or equivalently,  magnetic hedgehogs~\cite{Volovik1987, Kanazawa2016, Fujishiro2019}). 
These topological spin textures are characterized by an integer called the topological invariant: for instance, the skyrmion number for skyrmions~\cite{Rajaraman1987, Braun2012,Nagaosa2013} and the monopole charge for hedgehogs~\cite{Volovik1987, Braun2012}. 
The topological invariant is robust against perturbation, which ensures the topological protection of the spin textures. 
Moreover, the noncoplanar spin structures can generate 
the so-called emergent electromagnetic fields through the Berry phase mechanism~\cite{Berry1984,Volovik1987,Xiao2010,Nagaosa2012-1,Nagaosa2012-2,Nagaosa2013}. 
They are fictitious electromagnetic fields acting on electrons coupled to the spin textures, and thus, give rise to unusual 
quantum transport and optical phenomena, such as the topological Hall effect~\cite{Loss1992, Ye1999, Bruno2004, Onoda2004, Binz2008, Nakazawa2019}, the Nernst effect~\cite{Shiomi2013, Mizuta2016, Hirschberger2020TNE}, the magneto-optical Kerr effect~\cite{Feng2020,Hayashi2021}, and the emergent inductance~\cite{Nagaosa2019, Yokouchi2020, Kurebayashi2021, Ieda2021, Kitaori2021emergent}. 
Owing to these distinguishing properties, the topological spin textures have attracted a lot of attention for not only fundamental physics but also applications to next-generation electronic devices. 

\begin{figure}[tb]
\includegraphics[width=0.95\columnwidth]{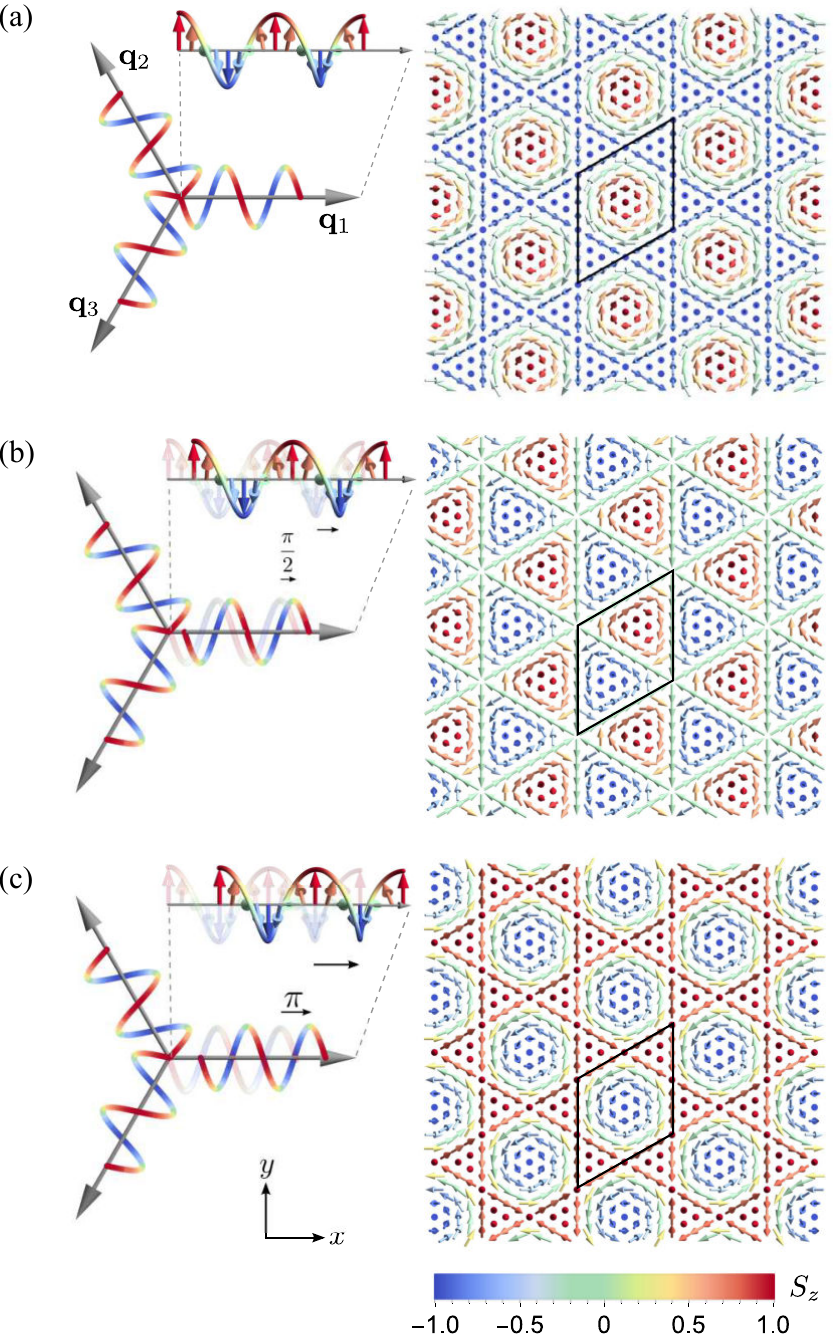}
\caption{
\label{fig:phase_schematic}
Variations of spin textures while changing the phase in the superpositions of three proper screws with the wave vectors ${\bf q}_1$, ${\bf q}_2$, and ${\bf q}_3$. 
(b) and (c) are obtained from (a) by the phase shift of $\frac{\pi}{2}$ and $\pi$, respectively.
The left panels display the schematic pictures of the superposed waves  
and the right panels show the spin textures obtained by the superpositions. 
The color of the arrows in the left panels represents the out-of-plane component of spins, as indicated in the inset of (c). 
The skyrmion number changes from (a) $N_{\rm sk}=1$ to (b) $N_{\rm sk}=0$, and to  
(c) $N_{\rm sk}=-1$, and the symmetry of the spin texture changes from (a) sixfold, (b) threefold, and (c) sixfold rotational symmetry.
The black rhombus represents the magnetic unit cell. 
See Sec.~\ref{sec:3.2} for the details.
}
\end{figure}

In magnetic materials, the topological spin textures often appear in the form of a periodic array of the topological objects. 
For instance, the magnetic skyrmions appear by forming a periodic lattice called the 
skyrmion lattice (SkL)~\cite{Muhlbauer2009, Yu2010, Yu2011, Munzer2010, Seki2012, Adams2012}, and the magnetic hedgehogs (and the antihedgehogs) appear as the hedgehog lattice (HL)~\cite{Tanigaki2015, Kanazawa2016, Yang2016, Fujishiro2019, Ishiwata2020, Okumura2020, Aoyama2021}. 
These periodic structures can be represented by superpositions of multiple spin density waves, and hence, called multiple-$Q$ spin textures.
An example is shown in Fig.~\ref{fig:phase_schematic}(a), where a SkL is given by a superposition of three proper screws and called the $3Q$-SkL. 
As such superpositions yield superstructures as the interference patterns, the topological spin textures can be viewed as ``spin moir\'e''~\cite{Shimizu2021moire}. 
Analogous to moir\'e fringes in optics, there are many ways to modulate the spin moir\'e, such as the number of superposed waves~\cite{Binz2006-1,Binz2006-2,Park2011}, 
the amplitudes of each spin density waves~\cite{Shimizu2021anisotropy}, 
and the angles between the propagating directions of the superposed waves~\cite{Shimizu2021moire}. 
Such modulations bring about various topological phases with different topological invariants and topological phase transitions between them. 

Among such parameters in spin moir\'e, it was recently pointed out that 
the phase degree of freedom in the superposed waves is an important parameter 
to control not only the spin textures but also their symmetry and topological properties~\cite{Kurumaji2019,Hayami2021phase}.
The situation is illustrated for superpositions of three proper screws in Fig.~\ref{fig:phase_schematic}.
Figure~\ref{fig:phase_schematic}(a) shows a SkL by a superposition of three proper screws running in the $120^\circ$ directions of ${\bf q}_1$, ${\bf q}_2$, and ${\bf q}_3$. 
The spin texture comprises a hexagonal array of skyrmions with the skyrmion number $N_{\rm sk}=1$ per magnetic unit cell 
and has sixfold rotational symmetry; see Sec.~\ref{sec:3.2} for the details. 
Let us consider a phase shift in the ${\bf q}_1$ component from this state. 
The results obtained by $\frac{\pi}{2}$ and $\pi$ shifts are shown in Figs.~\ref{fig:phase_schematic}(b) and \ref{fig:phase_schematic}(c), respectively. 
The symmetry is reduced to threefold for the $\frac{\pi}{2}$ shift, but recovered to sixfold for the $\pi$ shift. 
Accordingly, the topological property is also changed: 
The $\frac{\pi}{2}$ shift gives a periodic array of half skyrmions called merons and antimerons, leading to $N_{\rm sk}=0$, while the $\pi$ shift leads to a SkL with $N_{\rm sk}=-1$. 
Thus, the phases of the superposed waves are relevant degrees of freedom, but their impact has not been fully elucidated thus far, for not only SkLs but also the other topological spin textures like HLs.

In this paper, we systematically clarify the effect of phase shifts on the typical multiple-$Q$ spin textures, two-dimensional (2D) SkLs and three-dimensional (3D) HLs, focusing on their topological properties and the emergent magnetic fields. 
We first establish a generic framework to deal with the phase shift by introducing the hyperspace with an additional dimension corresponding to the phase degree of freedom, 
inspired by the description of the phason degree of freedom in quasicrystals~\cite{Levine1984, Levine1986, Socolar1986, Steinhardt1987}. 
In the hyperspace representation, the 2D SkLs composed of the three spin density waves with the phase degree of freedom are mapped to 3D HLs in which the Dirac strings connecting the hedgehogs and antihedgehogs correspond to the skymion and antiskyrmion cores in the original 2D SkLs. 
Similarly, the 3D HLs composed of four spin density waves are mapped to four-dimensional (4D) loop lattices in which intersections of the membranes defined by the loops, which we call ``the Dirac planes'', by 3D hyperplanes give hedgehog-antihedgehog pairs connected by the Dirac strings in the original 3D HLs. 
Analyzing the topological objects in the hyperspace representation, we systematically elucidate the evolution of the multiple-$Q$ spin structures for the phase shift as well as the magnetization change. 

In the 2D case, considering the superpositions of three proper screws or sinusoidal waves, we obtain various $3Q$-SkLs with $N_{\rm sk}$ ranging from $-2$ to $2$ depending on the phase and magnetization. 
We find that the phase diagram is dominated by the SkLs with $N_{\rm sk}=\pm 1$ in the case of the proper screw superpositions, whereas the $N_{\rm sk}=\pm 2$ regions become wider in the sinusoidal case. 
Interestingly, at zero magnetization, we always obtain the $N_{\rm sk}=\pm 1$ ($\pm 2$) SkLs for any phase shifts in the screw (sinusoidal) case; namely, the $N_{\rm sk}=\pm 2$ ($\pm 1$) SkLs are obtained only with nonzero magnetization in the screw (sinusoidal) case. 

On the other hand, in the 3D case, we clarify the topological phase diagrams for the superpositions of four proper screws or sinusoidal waves. 
We find various $4Q$-HLs classified by the number of the hedgehogs and antihedgehogs per unit cube, $N_{\rm m}$: $N_{\rm m}=8$, $16$, $32$, and $48$ for the screw case, and $N_{\rm m}=8$ $16$, $24$, $32$, and $48$ for the sinusoidal case. 
For the former case, the emergent magnetic field is always negative, while for the latter, it takes both positive and negative values. 
Notably, we find unusual Dirac strings running on the horizontal planes perpendicular to the magnetization direction. 
In the case of the screw superpositions, they give rise to pair creation of the hedgehogs and antihedgehogs and accordingly the increase of $N_{\rm m}$ from $16$ to $48$ while increasing the magnetization. 
This is highly unusual since the increase of the magnetization usually results in pair annihilation and the reduction of $N_{\rm m}$. 
In this screw case, $N_{\rm m}$ is always $16$ for any phases at zero magnetization, and the HLs with larger $N_{\rm m}$ appear only for nonzero magnetization. 
In contrast, in the case of the sinusoidal superpositions, the zero magnetization state has always the largest $N_{\rm m}=48$, and $N_{\rm m}$ decreases monotonically while increasing the magnetization. 
We also show that, in both cases, the amplitude of the emergent magnetic field is maximally enhanced by fusion of three hedgehogs and antihedgehogs on the horizontal Dirac strings where $N_{\rm m}$ changes from $48$ to $16$.

Finally, we study how the phases evolve in the actual multiple-$Q$ spin textures in microscopic models. 
Specifically, analyzing the numerical data for the 2D Kondo lattice model~\cite{Ozawa2017} and the 3D effective spin model~\cite{Okumura2020}, we extract the sum of phases in the superposed waves by fitting the spin configurations obtained by the numerical simulations. 
We show that phase shifts indeed take place in both cases around the topological phase transitions caused by an external magnetic field: 
For  the SkL, the sum of phases jumps from $\sim 0$ to $\sim \frac{\pi}{4}$ accompanied by the reduction of $|N_{\rm sk}|$ from $2$ to $1$, while for the HLs, it rapidly decreases from $\sim \frac{\pi}{3}$ to $\sim 0$ accompanied by the reduction of $N_{\rm m}$ from $16$ to $8$.

Our results establish the generic and systematic way to investigate the effect of phase shifts in the multiple-$Q$ spin textures. Moreover, they open a way for unexplored topological magnetic states and phase transitions, which may bring about nontrivial electronic structures and quantum transport properties through the emergent electromagnetic fields.  
Thus, our findings would shed light on the engineering of the multiple-$Q$ spin textures and related physics through the phase degree of freedom which has been overlooked thus far. 

The rest of the paper is organized as follows.
In Sec.~\ref{sec:2}, we introduce the hyperspace representation for general multiple-$Q$ spin textures. 
In Sec.~\ref{sec:3}, applying the framework to 2D $3Q$ states (Sec.~\ref{sec:3.1}), 
we elucidate the effect of phase shifts and magnetization changes on the spin textures, the symmetry, and 
the topological properties of the $3Q$ states composed of three proper screws (Sec.~\ref{sec:3.2}) and 
three sinusoidal waves (Sec.~\ref{sec:3.3}). 
In Sec.~\ref{sec:4}, we present the results for the 3D $4Q$ states: the hyperspace representation (Sec.~\ref{sec:4.1}), and the effect of phase shifts and magnetization changes on the $4Q$ states composed four proper screws (Sec.~\ref{sec:4.2}) and four sinusoidal waves (Sec.~\ref{sec:4.3}).
In Sec.~\ref{sec:5}, we present the analysis of the actual numerical data for the 2D Kondo lattice model (Sec.~\ref{sec:5.1}) and the 3D effective spin model (Sec.~\ref{sec:5.2}).
We discuss the results in Sec.~\ref{sec:6}. 
Section~\ref{sec:7} is devoted to the summary of this paper.

\section{Phase degree of freedom and hyperspace representation \label{sec:2}}

In this section, we propose a theoretical framework to systematically analyze the phase degree  
of freedom in multiple-$Q$ spin structures. 
We consider a generic form of the multiple-$Q$ spin structures in $d$-dimensional continuous space, 
which is given by the function of the real-space position ${\bf r}=(x_1, \ldots, x_d)$ as
\begin{eqnarray}
\Sr \propto 
\sum_{\eta=1}^{N_{Q}}~
\left(
\psi_{\eta}^{\rm c}{\bf e}_{\eta}^1\cos\QQ_{\eta} 
+\psi_{\eta}^{\rm s} {\bf e}_{\eta}^2\sin\QQ_{\eta}
\right)
+ m\hat{\bf z}, 
\label{eq:general_ansatz}
\end{eqnarray}
where $N_{Q}$ is the number of superposed waves, $\psi_{\eta}^{\rm c}$ and $\psi_{\eta}^{\rm s}$ are the amplitudes of cosinusoidal and  
sinusoidal waves with the wave vector ${\bf q}_{\eta}=(q_{\eta}^1, \ldots, q_{\eta}^{d})$, respectively, ${\bf e}_{\eta}^1$ and ${\bf e}_{\eta}^2$ are unit vectors,
$\QQ_{\eta}={\bf q}_{\eta}\cdot{\bf r}+\vp_{\eta}$, and $m$ represents the uniform magnetization and $\hat{\bf z}$ is the unit vector along the $z$ direction.
The spin length is normalized as $|\Sr|=1$ for any ${\bf r}$.
Note that $m$ is not the net magnetization because of the normalization.

\subsection{Phase degree of freedom \label{sec:2.1}}

Let us consider how the spin structures are modulated by changing the phases $\vp_{\eta}$. 
When $N_Q \leq d$ and ${\bf q}_{\eta}$ are linearly independent, the periodic spin textures are described by the set of $N_Q$ linearly-independent magnetic translation vectors ${\bf a}_{\eta}$, 
which satisfy 
\begin{eqnarray}
{\bf a}_{\eta}\cdot{\bf q}_{\eta'}=2\pi\delta_{\eta\eta'},
\label{eq:orthogonal_relation}
\end{eqnarray}
where $\delta_{\eta\eta'}$ is the Kronecker delta.
In this case, a phase shift from $\vp_\eta$ to $\vp_\eta+\Delta\vp_{\eta}$ is reduced to a spatial translation from ${\bf r}$ to ${\bf r}+\Delta{\bf r}$ with 
\begin{eqnarray}
\Delta{\bf r} = \sum_{\eta=1}^{N_Q} \frac{\Delta\vp_{\eta}}{2\pi} {\bf a}_{\eta},
\label{eq:phasishift_lin_indep}
\end{eqnarray} 
since the following relation holds: 
\begin{eqnarray}
{\bf q}_{\eta}\cdot\Delta{\bf r}
=\sum_{\eta'}\frac{\Delta\vp_{\eta'}}{2\pi}{\bf q}_{\eta}\cdot{\bf a}_{\eta'} 
=\Delta\vp_{\eta}.
\end{eqnarray} 
Hence, the phase degree of freedom is irrelevant when $N_Q \leq d$~\cite{2-1-1_note}.

The situation is, however, different for $N_Q>d$. 
In this case, the wave vectors ${\bf q}_{\eta}$ are not linearly independent of each other, as exemplified for $N_Q=3$ and $d=2$ in 
Fig.~\ref{fig:phase_schematic}. 
This means that a phase shift cannot be reduced to a spatial translation, as ${\bf a}_{\eta}$ defined 
by Eq.~(\ref{eq:orthogonal_relation}) with $d$ out of $N_Q$ wave vectors ${\bf q}_{\eta}$ do not satisfy 
the relation in Eq.~(\ref{eq:phasishift_lin_indep}).

\subsection{Hyperspace representation \label{sec:2.2}}
To discuss the effect of the phase shift in the case of $N_Q>d$ systematically, it is convenient to introduce 
$N_Q$-dimensional ($N_Q$D) hyperspace. 
In the hyperspace, we can introduce a position vector ${\bf R}=(X_1,X_2, \ldots, X_{N_Q})$ and $N_Q$ linearly-independent wave vectors 
${\bf Q}_{\eta}=(Q_{\eta}^{1}, Q_{\eta}^{2}, \ldots, Q_{\eta}^{N_Q})$ ($\eta=1,2,\ldots,N_Q$) so that they satisfy the relations~\cite{22_note}
\begin{eqnarray}
\QQ_{\eta}={\bf q}_{\eta}\cdot{\bf r}+\vp_{\eta}={\bf Q}_{\eta}\cdot{\bf R}. 
\label{eq:Q_r_R}
\end{eqnarray}
In this hyperspace, we can define the $N_Q$ linearly-independent magnetic translation vectors ${\bf A}_{\eta}$ ($\eta=1,2,\ldots, N_Q$) satisfying 
\begin{eqnarray}
{\bf A}_{\eta}\cdot{\bf Q}_{\eta'} = 2\pi\delta_{\eta\eta'}.    
\label{eq:orthogonal_hyper}
\end{eqnarray}
By using Eqs.~(\ref{eq:Q_r_R}) and (\ref{eq:orthogonal_hyper}), the hyperspace position ${\bf R}$ and 
the real-space one ${\bf r}$ are related with each other as 
\begin{eqnarray}
\left(\begin{array}{c} 
X_1 \\ X_2 \\ \vdots \\ X_{N_Q}
\end{array}\right)
=
M_Q^{-1}M_q
\left(\begin{array}{c} 
x_1 \\ x_2 \\ \vdots \\ x_d \\ 1 
\end{array}\right),
\label{eq:r2R}
\end{eqnarray}
where 
\begin{eqnarray}
M_Q
&=&
\left(\begin{array}{c} 
{\bf Q}_1 \\ {\bf Q}_2 \\ \vdots \\ {\bf Q}_{N_Q}
\end{array}\right)
=
\left(\begin{array}{cccc} 
Q_1^{1} & Q_1^{2} & \ldots & Q_1^{N_Q}  \\ 
Q_2^{1} & Q_2^{2} & \ldots & Q_2^{N_Q} \\
\vdots & \vdots & \ddots & \vdots \\
Q_{N_Q}^{1} & Q_{N_Q}^{2} & \ldots & Q_{N_Q}^{N_Q}
\end{array}\right), \\
M_q
&=&
\left(\begin{array}{c | c} 
{\bf q}_1 & \vp_1 \\ 
{\bf q}_2 & \vp_2 \\ 
\vdots & \vdots \\ 
{\bf q}_{N_Q} & \vp_{N_Q} 
\end{array}\right)
=
\left(\begin{array}{cccc} 
q_1^{1} & \ldots & q_1^{d} & \vp_1 \\ 
q_2^{1} & \ldots & q_2^{d} & \vp_2 \\
\vdots & \ddots & \vdots & \vdots \\
q_{N_Q}^{1} & \ldots & q_{N_Q}^{d} & \vp_{N_Q} \\
\end{array}\right).  
\end{eqnarray}
Note that $M_Q^{-1}$ is given as
\begin{eqnarray}
M_Q^{-1}
&=&
\frac{1}{2\pi}
\left(\begin{array}{cccc} 
{\bf A}_1 & {\bf A}_2 & \ldots & {\bf A}_{N_Q}
\end{array}\right) \notag \\
&=&
\frac{1}{2\pi}
\left(\begin{array}{cccc} 
A_1^{1} & A_2^{1} & \ldots & A_{N_Q}^{1} \\ 
A_1^{2} & A_2^{2} & \ldots & A_{N_Q}^{2} \\
\vdots & \vdots & \ddots & \vdots \\
A_1^{N_Q} & A_2^{{N_Q}} & \ldots & A_{N_Q}^{{N_Q}}
\end{array}\right).
\end{eqnarray}
The relation in Eq.~(\ref{eq:r2R}) can be regarded as a surjective mapping from the set of the real-space position ${\bf r}$ 
and the phases $\vp_{\eta}$ onto the hyperspace position ${\bf R}$~\cite{1_note}.
In this setting, Eq.~(\ref{eq:Q_r_R}) gives one-to-one correspondence between a multiple-$Q$ spin configuration 
with the phase variables $\vp_{\eta}$ in the original $d$-dimensional real space and that in the $N_Q$D hyperspace with fixed phases 
(in other words, without the phase degrees of freedom). 
In this representation, a phase shift by $\Delta\vp_{\eta}$ in the original real space corresponds to a translation in the hyperspace by 
\begin{eqnarray}
\Delta{\bf R} = \sum_{\eta=1}^{N_Q} \frac{\Delta\vp_{\eta}}{2\pi} {\bf A}_{\eta}. 
\label{eq:phase_translation_3D}
\end{eqnarray}

Consequently, the hyperspace representation enables us to treat the phase degree of freedom as additional coordinates in the hyperspace. 
We note that the situation is analogous to the hyperspace introduced to understand the structures of quasiperiodic crystals, where the number of translation vectors are in general larger than the system dimension and 
the quasycrystals are obtained by a ``slice" of a periodic structure in the hyperspace with additional dimensions spanned by the same number of vectors~\cite{Levine1984, Levine1986, Socolar1986, Steinhardt1987}. 
In the quasicrystals, the additional variables in the hyperspace are called phasons~\cite{Levine1985, Bak1985, Kalugin1985, Hu2000}, 
which also supports the analogy.

\section{$3Q$ skyrmion lattices \label{sec:3}}

In this section, we discuss the effect of phase shifts on the spin structures composed of three wave vectors 
in two dimensions, i.e., $N_{Q}=3$ and $d=2$, by using the hyperspace representation introduced in Sec.~\ref{sec:2}. 
Specifically, we consider Eq.~(\ref{eq:general_ansatz}) with three ${\bf q}_{\eta}$ given by
\begin{align}
&&{\bf q}_1 = \left( q, 0 \right),\ 
{\bf q}_2 = \left( -\frac{q}{2}, \frac{\sqrt{3}q}{2} \right),\ 
{\bf q}_3 = \left( -\frac{q}{2}, -\frac{\sqrt{3}q}{2} \right),
\label{eq:3Q_q_eta}
\end{align}
where $|{\bf q}_{\eta}|=q$.
For this $3Q$ spin structure, the 2D magnetic translation vectors are defined as
\begin{align}
{\bf a}_1 = \frac{2\pi}{q}\left(1,\frac{1}{\sqrt{3}}\right),\ \ 
{\bf a}_2 = \frac{2\pi}{q}\left(0,\frac{2}{\sqrt{3}}\right).
\label{eq:3Q_a_eta}
\end{align}

In the following, we focus on the two types of the $3Q$ spin structures.
One is the superposition of proper screws, which includes a $3Q$-SkL found in a wide range of materials, as introduced in Sec.~\ref{sec:1}.
We call this type the screw $3Q$ state; see Sec.~\ref{sec:3.2}.
The other type is the superposition of sinusoidal waves, which includes a $3Q$-SkL with the skyrmion number of two found in the Kondo lattice system on the 
triangular lattice~\cite{Ozawa2017} and its effective spin model~\cite{Hayami2017}.
We call this type the sinusoidal $3Q$ state; see Sec.~\ref{sec:3.3}. 
Before going into the analyses, we present the hyperspace representation in Sec.~\ref{sec:3.1}, which is commonly used for these two types.

\subsection{Hyperspace representation of the $3Q$ states \label{sec:3.1}}

\begin{figure}[tb]
	\includegraphics[width=1.0\columnwidth]{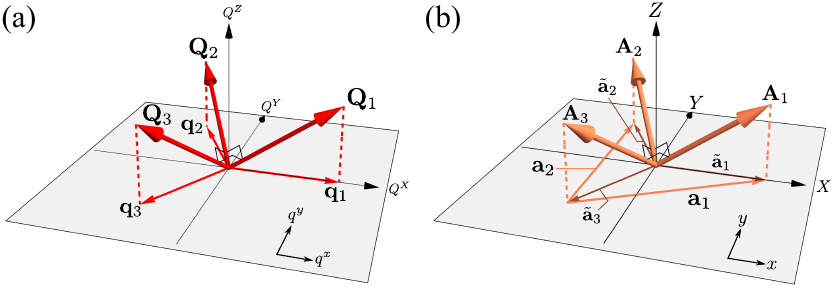}
	\caption{
	\label{fig:3Q_setup}
	(a) Schematic pictures of the relations between the wave vectors in the 3D reciprocal hyperspace, ${\bf Q}_{\eta}$, 
	and those in the 2D reciprocal space, ${\bf q}_{\eta}$.  
	(b) Corresponding relations between the magnetic translation vectors in the 3D hyperspace, 
	${\bf A}_{\eta}$, and those in the 2D real space, ${\bf a}_{\eta}$. 
	The projections of ${\bf A}_{\eta}$ onto the $xy$ plane denoted by $\tilde{\bf a}_{\eta}$ are also shown.
	}
\end{figure}

Using the framework introduced in Sec.~\ref{sec:2.2}, we construct the hyperspace representation of the $3Q$ spin structures. 
First, we set the 3D wave vectors ${\bf Q}_{\eta}=(Q_{\eta}^X, Q_{\eta}^Y, Q_{\eta}^Z)$ in Eq.~(\ref{eq:Q_r_R}) as
\begin{eqnarray}
&&{\bf Q}_1=q\left( 1 , 0 , \frac{1}{\sqrt{2}} \right), 
\label{eq:Q_1}\\ 
&&{\bf Q}_2=q\left( -\frac{1}{2} , \frac{\sqrt{3}}{2} , \frac{1}{\sqrt{2}} \right),
\label{eq:Q_2}\\
&&{\bf Q}_3=q\left( -\frac{1}{2}, -\frac{\sqrt{3}}{2}, \frac{1}{\sqrt{2}} \right),
\label{eq:Q_3} 
\end{eqnarray}
so that the projection of ${\bf Q}_\eta$ onto the $q^xq^y$ plane is ${\bf q}_{\eta}$, 
and ${\bf Q}_1$, ${\bf Q}_2$, and ${\bf Q}_3$ are orthogonal to each other as shown in Fig.~\ref{fig:3Q_setup}(a), without loss of generality.
Then, we obtain the corresponding magnetic translation vectors ${\bf A}_{\eta}$ in the 3D hyperspace as 
\begin{eqnarray}
&&{\bf A}_1=\frac{4\pi}{3q}\left( 1 , 0 , \frac{1}{\sqrt{2}} \right),
\label{eq:A_1}\\
&&{\bf A}_2=\frac{4\pi}{3q}
\left( -\frac{1}{2} , \frac{\sqrt{3}}{2}, \frac{1}{\sqrt{2}} \right),
\label{eq:A_2}\\ 
&&{\bf A}_3=\frac{4\pi}{3q}
\left( -\frac{1}{2}, -\frac{\sqrt{3}}{2}, \frac{1}{\sqrt{2}} \right). 
\label{eq:A_3}
\end{eqnarray}
Figure~\ref{fig:3Q_setup}(b) illustrates the relation between ${\bf A}_{\eta}$ and ${\bf a}_{\eta}$. 
Note that ${\bf a}_{\eta}$ is given by ${\bf a}_{\eta}=\tilde{\bf a}_{\eta}-\tilde{\bf a}_3$, where $\tilde{\bf a}_{\eta}$ is the projection of ${\bf A}_{\eta}$ onto the $xy$ plane.

By using Eq.~(\ref{eq:r2R}), the hyperspace positions are related with the real-space positions as 
\begin{eqnarray}
\left(\begin{array}{c} 
X \\ Y \\ Z
\end{array}\right)
=
V_{3}
\left(\begin{array}{c} 
x \\ y \\ 1
\end{array}\right),
\label{eq:r2R_3Q}
\end{eqnarray}
where $V$ includes the phases as 
\begin{align}
V_{3}=
\left(\begin{array}{ccc} 
{\bf v}_1 & {\bf v}_2 & {\bf v}_3
\end{array}\right)
=
\left(\begin{array}{ccc} 
1 & 0 & \frac{2}{3q}\left( \vp_1 - \frac{1}{2}(\vp_2+\vp_3) \right) \\
0 & 1 & \frac{1}{\sqrt{3}q}\left(\vp_2 - \vp_3 \right) \\
0 & 0 & \frac{\sqrt{2}}{3q}\left( \vp_1 + \vp_2+\vp_3 \right)
\end{array}\right). 
\end{align}
It is worth noting that Eq.~(\ref{eq:r2R_3Q}) can be rewritten as 
\begin{equation}
{\bf R} = x{\bf v}_1 + y{\bf v}_2 + {\bf v}_3.
\end{equation} 
This means that the spin configuration on the original 2D $xy$ plane is the same as the one on a slice of the hyperspace spin configuration spanned by 
${\bf v}_1$ and ${\bf v}_2$ including the point at ${\bf v}_3$, namely, the horizontal plane with 
\begin{equation}
Z=\frac{\sqrt{2}}{3q}\tilde{\vp}, 
\label{eq:Z_cubic}
\end{equation}
where 
\begin{align}
\tilde{\vp} = \sum_{\eta} \vp_\eta = \vp_1 + \vp_2 + \vp_3. 
\label{eq:3q_phitilde}
\end{align} 
Thus, only the summation of the phases $\vp_\eta$ is relevant for the present $3Q$ spin textures, instead of each value of $\vp_\eta$. 
Note that $\tvp$ has $2\pi$ periodicity; namely, the spin configuration in the 3D hyperspace becomes equivalent with period of $\frac{2\sqrt{2}\pi}{3q}$ in the $Z$ direction. 
In the original 2D space, the $2\pi$ phase shift in $\tvp$ ($\Delta\tvp = \sum_{\eta} \Delta\vp_{\eta} = 2\pi$) 
corresponds to a spatial translation by 
\begin{eqnarray}
\Delta{\bf r}_{\eta}=\tilde{\bf a}_{\eta}-\sum_{\eta'=1}^{3}\frac{\Delta\vp_{\eta'}}{2\pi}\tilde{\bf a}_{\eta'}, 
\label{eq:3q_2pi_translation}
\end{eqnarray}
where $\eta$ may take any of $1$, $2$, and $3$. 

In the following, we apply the above hyperspace representation to analyze the effect of phase shifts on the magnetic and topological properties of two types of 2D $3Q$ states.

\subsection{Screw $3Q$ state \label{sec:3.2}}

In this subsection, we analyze the effect of phase shifts on the $3Q$ state 
composed of three proper screws. 
The spin texture is given by Eq.~(\ref{eq:general_ansatz}) with $N_Q=3$ and
\begin{eqnarray}
&&\psi_{\eta}^{\rm c}=\psi_{\eta}^{\rm s}=
\frac{1}{\sqrt{3}}
, \\ 
&&{\bf e}_{\eta}^1=\hat{\bf z}, \ 
{\bf e}_{\eta}^2={\bf e}_{\eta}^0\times{\bf e}_{\eta}^1, \ 
{\bf e}_{\eta}^0=\left(\frac{q^x_{\eta}}{q}, \frac{q^y_{\eta}}{q}, 0 \right)^{\mathsf{T}}, 
\end{eqnarray}
where $\mathsf{T}$ denotes the transpose of the vector. The explicit form is given as
\begin{eqnarray}
{\bf S}({\bf r})
&\propto&
\left(\begin{array}{c}  
\frac{\sqrt{3}}{2}(\sin\QQ_{2} - \sin\QQ_{3}) \\
-\sin\QQ_{1}+\frac{1}{2}(\sin\QQ_{2} + \sin\QQ_{3}) \\
\cos\QQ_1+\cos\QQ_2+\cos\QQ_3+\sqrt{3}m
\end{array} \right).  
\label{eq:chiral_3Q_ansatz}
\end{eqnarray}

Before going into the analyses of the topological properties, let us discuss the symmetry of the screw $3Q$ state in Eq.~(\ref{eq:chiral_3Q_ansatz}). 
Table~\ref{tab:3q_scr_sym} summarizes the symmetry operations on the spin texture, together with the decomposition of each operation into the change of the sum of phases $\tvp$, spatial translation, and magnetization change. 
We here consider the point group operations and time-reversal operation that do not change the wave vectors of the proper screws. 
Other symmetry operations are expressed by the combinations of those in Table~\ref{tab:3q_scr_sym}, e.g., $C_{2y}=C_{2z}C_{2x}=C_{6z}^3C_{2x}$. 
From Table~\ref{tab:3q_scr_sym}, we can obtain the symmetry operations which do not change both $\tvp$ and $m$, i.e., the spin texture. 
Specifically, we find that the screw $3Q$ state is symmetric under $C_{6z}$, $\mathcal{T}C_{2x}$, and their combinations for $\tvp=0$ and $\pi$, otherwise 
$C_{3z}$, $\mathcal{T}C_{2x}$, and their combinations. 
We note that spatial translation combined with the $C_{6z}$ rotation can be represented by shifting the rotation axis. 

\begin{table}
\caption{\label{tab:3q_scr_sym}
Symmetry operations for the screw $3Q$ state in Eq.~(\ref{eq:chiral_3Q_ansatz}) and their decompositions into the change in the sum of phases, spatial translation in the original 2D space, and the magnetization change: 
$C_{n\alpha}$ represents an $n$-fold rotation about the axis in the $\alpha$ direction and $\mathcal{T}$ represents the time-reversal operation. 
}
\begin{ruledtabular}
\begin{tabular}{c|ccc}
operation & sum of phases & translation  & magnetization   \\
\hline 
$C_{6z}$ & $\tvp \rightarrow 2\pi - \tvp$ & $\tilde{\bf a}_{\eta}+\tilde{\bf a}_{\eta'}$  & $m \rightarrow m$ \\
$C_{2x}$ &  $\tvp \rightarrow \pi + \tvp$ & $\tilde{\bf a}_{\eta}$  & $m \rightarrow -m$ \\
$\mathcal{T}$ & $\tvp \rightarrow \pi + \tvp$ & $\tilde{\bf a}_{\eta}$ & $m \rightarrow -m$ 
\end{tabular}
\end{ruledtabular}
\end{table}

\begin{figure}[tb]
	\includegraphics[width=1.0\columnwidth]{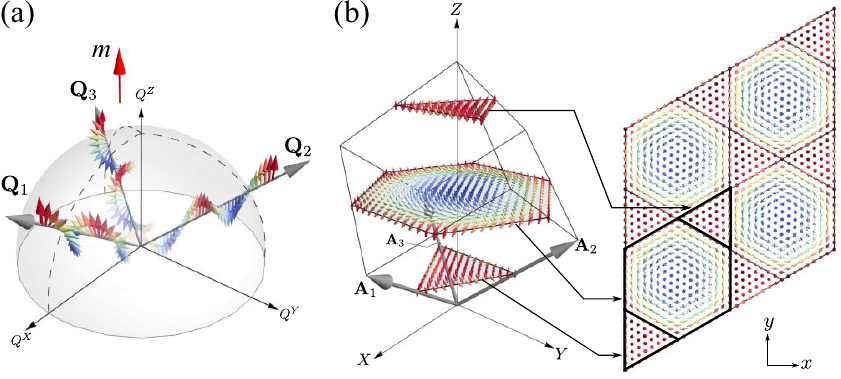}
	\caption{
	\label{fig:3qscr_spin_hyper}
	(a) Schematic of the three proper screws in the 3D reciprocal hyperspace, whose wave vectors ${\bf Q}_\eta$ are defined by Eqs.~(\ref{eq:Q_1}), (\ref{eq:Q_2}), and (\ref{eq:Q_3}) for the screw $3Q$ state in Eq.~(\ref{eq:chiral_3Q_ansatz}). 
	(b) Corresponding cubic MUC in the hyperspace and 
	spin structures on the three horizontal planes with $Z=\frac{\sqrt{2}}{3q}(2n+1)\pi$ ($n=0,1,2$). 
	The right panel in (b) shows the 2D spin texture on the three slices, which corresponds to 
	the spin structure in Eq.~(\ref{eq:chiral_3Q_ansatz}) with $\tilde{\vp}=\vp_1+\vp_2+\vp_3=(2n+1)\pi$, 
	where $n$ is an integer.  
	The rhombus consisting of the two triangles and one hexagon denotes the 2D MUC in the original 2D plane. 
	}
\end{figure}

\subsubsection{Hedgehogs in hyperspace \label{sec:3.2.1}}

To discuss the effect of phase shifts on the screw $3Q$ state in Eq.~(\ref{eq:chiral_3Q_ansatz}), 
we study the corresponding spin texture in the 3D hyperspace whose reciprocal space is spanned by the wave vectors ${\bf Q}_\eta$ 
in Eqs.~(\ref{eq:Q_1}), (\ref{eq:Q_2}), and (\ref{eq:Q_3}).
The schematic picture is shown in Fig.~\ref{fig:3qscr_spin_hyper}(a). 
As discussed in the previous subsection, a real-space spin configuration for a given phase summation $\tilde{\vp}$ on the original 2D plane corresponds to 
a hyperspace spin configuration on the horizontal plane with $Z=\frac{\sqrt{2}}{3q}\tilde{\vp}$ [see Eq.~(\ref{eq:Z_cubic})]. 
Such a correspondence is exemplified in Fig.~\ref{fig:3qscr_spin_hyper}(b) for $\tilde{\vp}=(2n+1)\pi$. 
In this case, the intersections of the 3D cubic magnetic unit cell (MUC) and the horizontal planes with $Z=\frac{\sqrt{2}}{3q}(2n+1)\pi$ $(n=0,1,2)$ give the hexagon and the two triangles, which comprise the rhombus MUC in the original 2D plane.

In the 3D hyperspace, the spin structure composed of three proper screws may comprise a 3D topological spin structure called 
$3Q$-HL~\cite{Kanazawa2016, Zhang2016, Okumura2020, Shimizu2021moire}. 
It has a periodic array of topological defects called the hedgehogs and antihedgehogs, whose cores are the singular points where the spin length vanishes (see Sec.~\ref{sec:3.2.2}). 
Indeed, by solving the equation ${\bf S}({\bf r})=0$ for Eq.~(\ref{eq:chiral_3Q_ansatz}), we obtain the following eight solutions: 
\begin{eqnarray}
(\QQ_1^{*},\QQ_2^{*},\QQ_3^{*}) &=& 
\left(\pi+p_1^{\rm scr}(m), \pi+p_1^{\rm scr}(m), \pi+p_1^{\rm scr}(m) \right), \notag \\ 
&&
\left(\pi-p_1^{\rm scr}(m), \pi-p_1^{\rm scr}(m), \pi-p_1^{\rm scr}(m) \right), \notag \\
&&
\left(\pi-p_2^{\rm scr}(m), \pi-p_2^{\rm scr}(m), p_2^{\rm scr}(m) \right)  \notag \\
&&\qquad\qquad\quad
\mbox{and cyclic permutations}, \notag \\
&&
\left(\pi+p_2^{\rm scr}(m), \pi+p_2^{\rm scr}(m), 2\pi-p_2^{\rm scr}(m) \right) \notag \\
&&\qquad\qquad\quad
\mbox{and cyclic permutations}, 
\label{eq:chiral_3Q_sol}
\end{eqnarray}
where 
\begin{eqnarray}
p_1^{\rm scr}(m) = \arccos\left(\frac{m}{\sqrt{3}}\right), \quad
p_2^{\rm scr}(m) = \arccos\left(\sqrt{3}m\right).
\end{eqnarray}
By using the relation 
\begin{eqnarray}
{\bf R}^{*} = \sum_{\eta} \frac{\QQ_{\eta}^{*}}{2\pi} {\bf A}_{\eta}, 
\label{eq:R^*}
\end{eqnarray}
we obtain the positions of the eight singular points in the hyperspace as 
\begin{eqnarray}
&&
(X^{*}, Y^{*}, Z^{*}) = \notag \\
&&\quad
\frac{\sqrt{2}}{3q}\left(0, 0, 3(\pi+p_1^{\rm scr}(m)) \right), 
\ 
\frac{\sqrt{2}}{3q}\left(0, 0, 3(\pi-p_1^{\rm scr}(m)) \right), \notag \\
&& \quad
\left( \frac{\pi-2p_2^{\rm scr}(m)}{3q}, \frac{\pi-2p_2^{\rm scr}(m)}{\sqrt{3}q}, \frac{\sqrt{2}(2\pi-p_2^{\rm scr}(m))}{3q} \right)\notag \\ 
&&\quad
\qquad \qquad \qquad \qquad \mbox{ and $C_3^{Z}$ symmetric points}, \notag \\
&&\quad
\left( -\frac{\pi-2p_2^{\rm scr}(m)}{3q}, -\frac{\pi-2p_2^{\rm scr}(m)}{\sqrt{3}q}, \frac{\sqrt{2}(4\pi+p_2^{\rm scr}(m))}{3q} \right)\notag \\
&& \quad
\qquad \qquad \qquad \qquad \mbox{ and $C_3^{Z}$ symmetric points}, 
\label{eq:chiral_3Q_sol_R}
\end{eqnarray}
where the $C_3^{Z}$ symmetric points are obtained by $2\pi/3$ and $4\pi/3$ rotations about the $Z$ axis.

\subsubsection{Topological transition in 3D hyperspace \label{sec:3.2.2}}

\begin{figure}[tb]
	\centering
	\includegraphics[width=0.95\columnwidth]{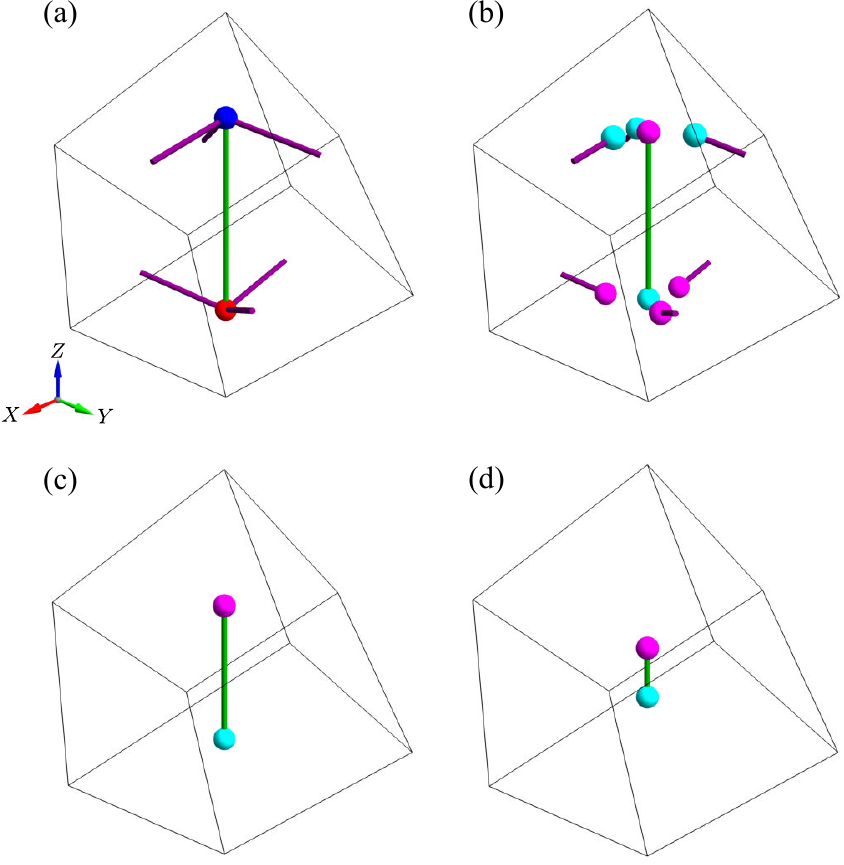}
	\caption{
	\label{fig:3qscr_defect}
	Change of the positions of the hedgehogs and antihedgehogs, and the Dirac strings connecting them in the spin structure 	in the 3D hyperspace corresponding to the screw $3Q$ state in Eq.~(\ref{eq:chiral_3Q_ansatz}) while changing $m$: (a) $m=0$, (b) $m=0.4$, (c) $m=0.8$, and (d) $m=1.6$. 
	The red, blue, magenta, and cyan spheres are the topological defects with 
	$Q_{\rm m}=2$, $-2$, $1$, and $-1$, respectively.  
	The green and purple lines denote the Dirac strings with the vorticity $\zeta=+1$ and $-1$, respectively.
	}
\end{figure}

Figure~\ref{fig:3qscr_defect} illustrates the systematic change of the topological defects while changing 
the magnetization along the $Z$ direction, $m$. 
When $m=0$, four out of the eight solutions in Eq.~(\ref{eq:chiral_3Q_sol_R}) become identical to 
$\frac{\sqrt{2}}{3q} \left( 0,0,\frac{9\pi}{2} \right)$ and the rest four become $\frac{\sqrt{2}}{3q} \left( 0,0,\frac{3\pi}{2} \right)$, 
since $p_1^{\rm scr}(0)=\frac{\pi}{2}$ and $p_2^{\rm scr}(0)=\frac{\pi}{2}$. 
Hence, there are only two topological defects located on the $Z$ axis, as shown in Fig.~\ref{fig:3qscr_defect}(a).
Following the arguments in Refs.~\cite{Park2011,Zhang2016,Kanazawa2016,Okumura2020,Shimizu2021moire}, we compute 
the monopole charge for these defects, which is defined by
\begin{eqnarray}
Q_{\rm m}=\frac{1}{4\pi}\int d\tilde{{\bf S}} \cdot {\bf b}({\bf R}),
\label{eq:monopole_charge}
\end{eqnarray}
where ${\bf b}({\bf R})=(b_X({\bf R}), b_Y({\bf R}), b_Z({\bf R}))$ is the scalar spin chirality defined in the hyperspace as 
\begin{eqnarray}
b_{i}({\bf R})=\frac{1}{2}\varepsilon^{ijk} {\bf S}({\bf R})\cdot\left(\frac{\partial {\bf S}({\bf R})}{\partial X_j} \times \frac{\partial {\bf S}({\bf R})}{\partial X_k} \right), 
\end{eqnarray}
where $\varepsilon^{ijk}$ is the Levi-Civita Symbol and ${\bf S}({\bf R})$ is the spin at ${\bf R}$ in the 3D hyperspace; the integral in Eq.~(\ref{eq:monopole_charge}) is taken on a closed surface surrounding the defect in the hyperspace. 
Here, $Q_{\rm m}$ takes an integer, which defines the topological nature of the defects; 
a defect with positive (negative) $Q_{\rm m}$ is a hedgehog (an antihedghog) which is regarded as a source (sink) of 
the emergent magnetic fields~\cite{Zhang2016,Kanazawa2016,Okumura2020,Shimizu2021moire}.
We find that the topological defect at $\frac{\sqrt{2}}{3q} \left( 0,0,\frac{3\pi}{2} \right)$ in Fig.~\ref{fig:3qscr_defect}(a) is a hedgehog with $Q_{\rm m}=+2$ (red sphere) 
and the other one at $\frac{\sqrt{2}}{3q} \left( 0,0,\frac{9\pi}{2} \right)$ is an antihedgehog with $Q_{\rm m}=-2$ (blue sphere). 
Following the procedures in Ref.~\cite{Shimizu2021moire}, we also identify four Dirac strings, which is the lines connecting the hedgehog and antihedgehog by the spins pointing downward (antiparallel to the direction of the magnetization).
The Dirac strings are distinguished by their vorticity given by
\begin{eqnarray}
\zeta = \frac{1}{2\pi}\oint d{\bf l} \cdot \boldsymbol{\nabla}\phi({\bf R}), 
\label{eq:vorticity}
\end{eqnarray}
where $\phi({\bf r})$ is the azimuthal angle of ${\bf S}({\bf R})$ and the integral is taken along a closed path surrounding
the string at ${\bf R}$ on a plane perpendicular to the $Z$ axis~\cite{Tatara2019, Shimizu2021moire}.
We find four Dirac strings, as shown in Fig.~\ref{fig:3qscr_defect}(a): 
One is the line along the $Z$ axis (green line) and the other three run through the 
MUC boundaries and connect the topological defects in the neighboring MUCs (purple lines). 
The former has the vorticity of $\zeta=+1$, while the latter three have $\zeta=-1$. 

When introducing $m$, each topological defect splits into four; 
the hedgehog with $Q_{\rm m}=+2$ splits into three hedgehogs with $Q_{\rm m}=+1$ (magenta spheres) and 
one antihedgehog with $Q_{\rm m}=-1$ (cyan sphere), while the antihedgehog with $Q_{\rm m}=-2$ splits into 
three antihedgehogs with $Q_{\rm m}=-1$ and one hedgehog with $Q_{\rm m}=+1$, as shown in Fig.~\ref{fig:3qscr_defect}(b). 
Note that the total monopole charge is conserved in each splitting. 
Thus, we have totally four hedgehogs with $Q_{\rm m}=+1$ and 
four antihedgehogs with $Q_{\rm m}=-1$ by introducing $m$.
With a further increase of $m$, three pairs connected by the Dirac strings with $\zeta=-1$ disappear with pair annihilation 
at $m=1/\sqrt{3}$, leaving one pair connected by the Dirac string with $\zeta=+1$, 
as shown in Fig.~\ref{fig:3qscr_defect}(c).
The remaining hedgehog and antihedgehog move toward each other while further increasing $m$, as shown in Fig.~\ref{fig:3qscr_defect}(d), and they also vanish with pair annihilation at $m=\sqrt{3}$.
Consequently, while increasing $m$, we have two topological transitions caused by pair annihilation of the hedgehogs and antihedgehogs at $m=1/\sqrt{3}$ and $m=\sqrt{3}$.

\subsubsection{Topological transition on 2D plane \label{sec:3.2.3}}

\begin{figure}[tb]
	\centering
	\includegraphics[width=1.0\columnwidth]{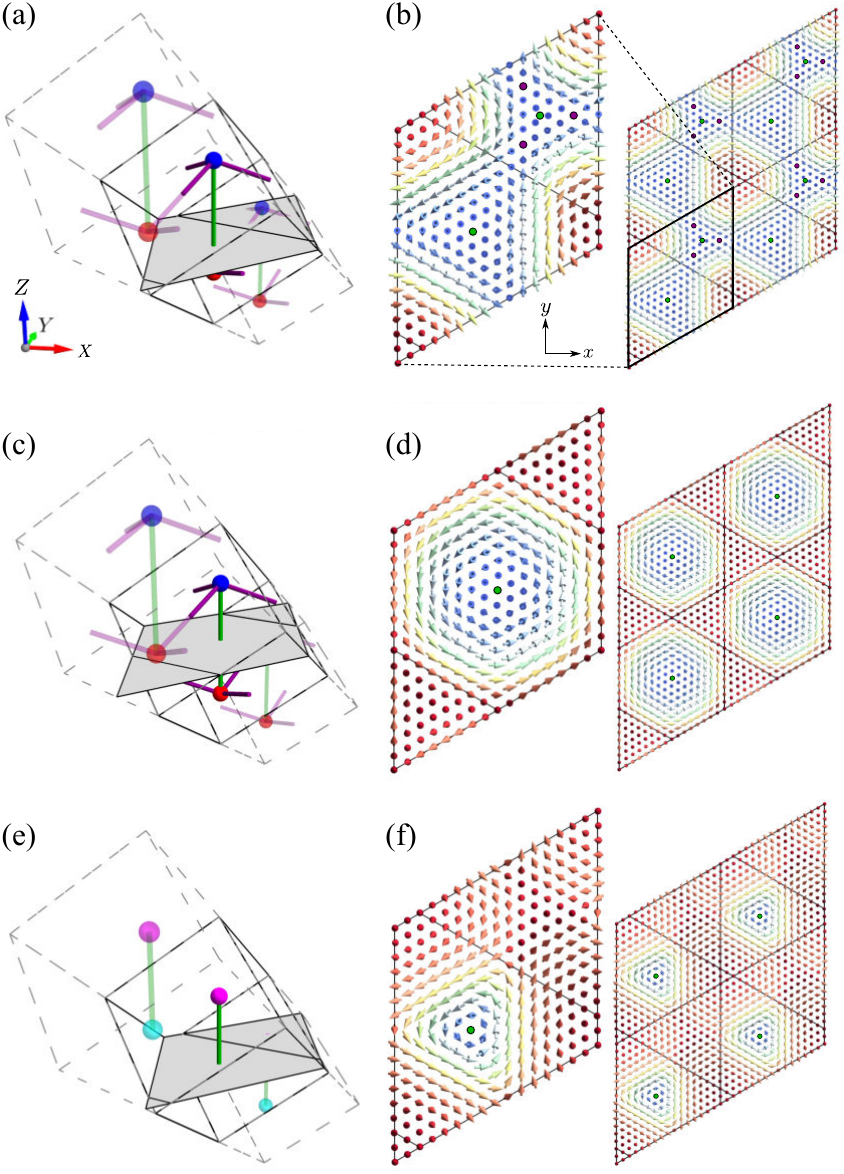}
	\caption{
	\label{fig:3qscr_intersection}
	(a) Hedgehogs and antihedgehogs, and Dirac strings for $m=0$ in the 3D hyperspace. 
	The pale colors are used for the objects in the neighboring MUCs denoted by the dashed cubes. 
	The gray plane represents the 2D MUC on the intersection of the plane with $Z=\frac{\sqrt{2}}{3q}\frac{9}{4}\pi$. 
	(b) The spin configuration on the gray plane in (a), which corresponds to the spin structure in 
	Eq.~(\ref{eq:chiral_3Q_ansatz}) with $\tilde{\vp}=\frac{9}{4}\pi$. 
	The green and purple circles denote the intersections of the Dirac strings with $\zeta=+1$ and $-1$, respectively. 
	The right panel displays the 2D real-space spin configuration by repeating the MUC. 
	Similar figures for (c),(d) $m=0$ and $\tilde{\vp}=3\pi$, and (e),(f) $m=0.9$ and $\tilde{\vp}=\frac{9}{4}\pi$. 
	}
\end{figure}

Using the results on the topological defects and the Dirac strings in the 3D hyperspace, we can discuss in a systematic way 
the topological properties of the 2D spin texture in Eq.~(\ref{eq:chiral_3Q_ansatz}) while changing $\tilde{\vp}$ and $m$. 
Figure~\ref{fig:3qscr_intersection} displays the relation between the horizontal slices in the 3D hyperspace at $Z=\frac{\sqrt{2}}{3q}\tilde{\vp}$ and the 2D spin textures with $\tilde{\vp}$. 
In Fig.~\ref{fig:3qscr_intersection}(a), we show the configurations of the hedgehogs and antihedgehogs, and the Dirac strings in the hyperspace at $m=0$, and the horizontal plane with $Z=\frac{\sqrt{2}}{3q}\frac{9}{4}\pi$. 
The spin configuration on the plane is shown in Fig.~\ref{fig:3qscr_intersection}(b), which corresponds to 
the 2D spin texture in Eq.~(\ref{eq:chiral_3Q_ansatz}) with $\tilde{\vp}=\frac94\pi$. 
The topological property of the 2D spin texture is characterized by the skyrmion number~\cite{Rajaraman1987, Braun2012,Nagaosa2013}
\begin{eqnarray}
N_{\rm sk} = \frac{1}{4\pi} \int_{\rm 2D~MUC} dXdY~b_Z({\bf R}), 
\label{eq:Nsk}
\end{eqnarray}
where the integral is taken within the 2D rhombic MUC.
The skyrmion number is also obtained by counting the vorticities of 
the Dirac strings as~\cite{Shimizu2021moire} 
\begin{eqnarray}
N_{\rm sk} = -\sum_{k} \zeta_k(Z), 
\label{eq:Nsk_vorticity}
\end{eqnarray}
where $\zeta_k(Z)$ denotes the vorticity of the $k$th Dirac string intersecting the 2D MUC~\cite{3_note}. 
For example, in the case of Fig.~\ref{fig:3qscr_intersection}(b), there are two intersections by the Dirac strings 
with $\zeta=+1$ and three with $\zeta=-1$, and hence, $N_{\rm sk}=+1$. 
Thus, this simple counting in the hyperspace representation enables us to identify the 2D spin texture at 
$m=0$ with $\tilde{\vp}=\frac94\pi$ as the $3Q$-SkL with $N_{\rm sk}=+1$, 
while the direct integration by Eq.~(\ref{eq:Nsk}) gives the same conclusion. 

Figures~\ref{fig:3qscr_intersection}(c) and \ref{fig:3qscr_intersection}(d) illustrate the situations with $\tilde{\vp}=3\pi$ at $m=0$.  
In this case, the 2D slice has a single intersection by the Dirac string with $\zeta=+1$, and hence, the spin texture with $\tilde{\vp}=3\pi$ is the $3Q$-SkL with $N_{\rm sk}=-1$. 
This demonstrates a switching of the topological property by the phase shift. 
In the hyperspace representation, such topological transitions occur when the gray horizontal plane crosses 
the hedgehogs or antihedgehogs at the end points of the Dirac strings. 

Since the topological defects change their positions with $m$ as shown in Fig.~\ref{fig:3qscr_defect}, 
the topological properties of the 2D spin structures change also with $m$. 
A demonstration is shown in Figs.~\ref{fig:3qscr_intersection}(e) and \ref{fig:3qscr_intersection}(f) for $m=0.9$.  
At this value of $m$, three pairs of the hedgehogs and antihedgehogs already vanish by pair annihilation, and only a single pair remains on the $Z$ axis, as shown in Fig.~\ref{fig:3qscr_intersection}(e). 
In this case, when the 2D slice intersects the Dirac string connecting the hedgehog-antihedgehog pair, the 2D spin structures becomes a $3Q$-SkL with $N_{\rm sk}=-1$, as exemplified in Fig.~\ref{fig:3qscr_intersection}(f) for $\tilde{\vp}=\frac94\pi$. 

In this way, we can systematically investigate the changes of the magnetic textures and the topological properties 
of the 2D $3Q$ spin structures while changing $\tilde{\vp}$ and $m$. 
The procedure is summarized as follows: 
(i) Define the 3D spin texture in the hyperspace by using Eqs.~(\ref{eq:Q_r_R}) and (\ref{eq:orthogonal_hyper}), 
in the present case, Eqs.~(\ref{eq:Q_1})-(\ref{eq:A_3}), 
(ii) identify the hedgehogs and antihedgehogs, and the Dirac strings connecting them for the 3D spin structure in the hyperspace, 
(iii) consider the 2D slice of the 3D spin structure at the horizontal plane with $Z=\frac{\sqrt{2}}{3q}\tilde{\vp}$, which gives the 2D spin texture with the phase summation $\tilde{\vp}$, and 
(iv) take the sum of the vorticities of the Dirac strings intersecting the plane, which gives the skyrmion number $N_{\rm sk}$ of the 2D spin structure through Eq.~(\ref{eq:Nsk_vorticity}).

\subsubsection{Topological phase diagram \label{sec:3.2.4}}

\begin{figure}[tb]
	\includegraphics[width=1.0\columnwidth]{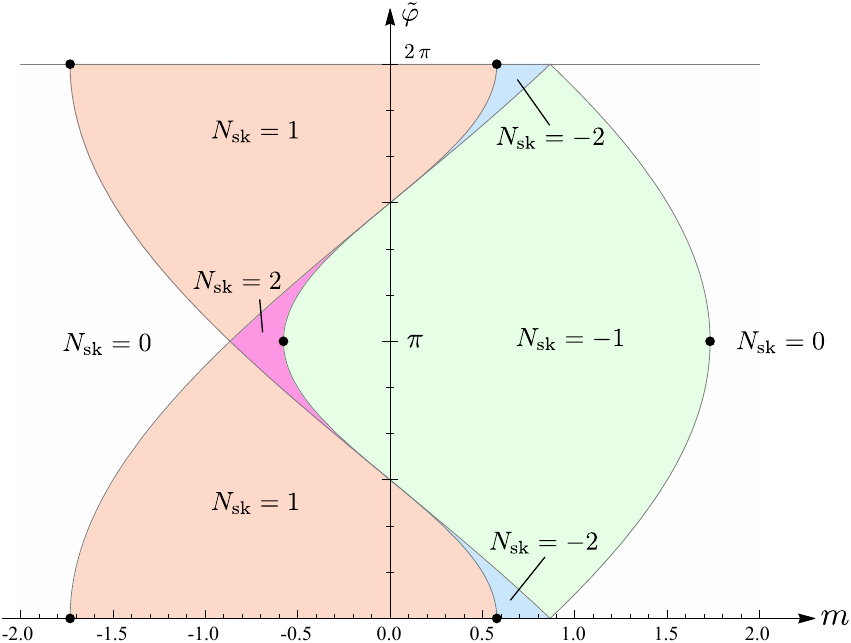}
	\caption{
	\label{fig:3qscr_Nsk}
	Topological phase diagram for the screw $3Q$ state in Eq.~(\ref{eq:chiral_3Q_ansatz}) 
	determined by the skyrmion number $N_{\rm sk}$ while changing $m$ and $\tilde{\vp}$. 
	The colored regions are topologically nontrivial phases with nonzero $N_{\rm sk}$, and 
	the white areas in the left and right hand sides denote topologically trivial phases with $N_{\rm sk}=0$.
	The black dots represent the pair annihilation of the hedgehogs and antihedgehogs in the 3D hyperspace, which are located at $(m,\tilde{\vp}) = (-\sqrt{3}, 0)$, $(-1/\sqrt{3}, \pi)$, $(1/\sqrt{3}, 0 )$, and $(\sqrt{3}, \pi)$.
	The phase diagram is periodic in the $\tilde{\vp}$ direction with period of $2\pi$. 
	}
\end{figure}

\begin{figure}[tbh]
	\centering
	\includegraphics[width=0.9\columnwidth]{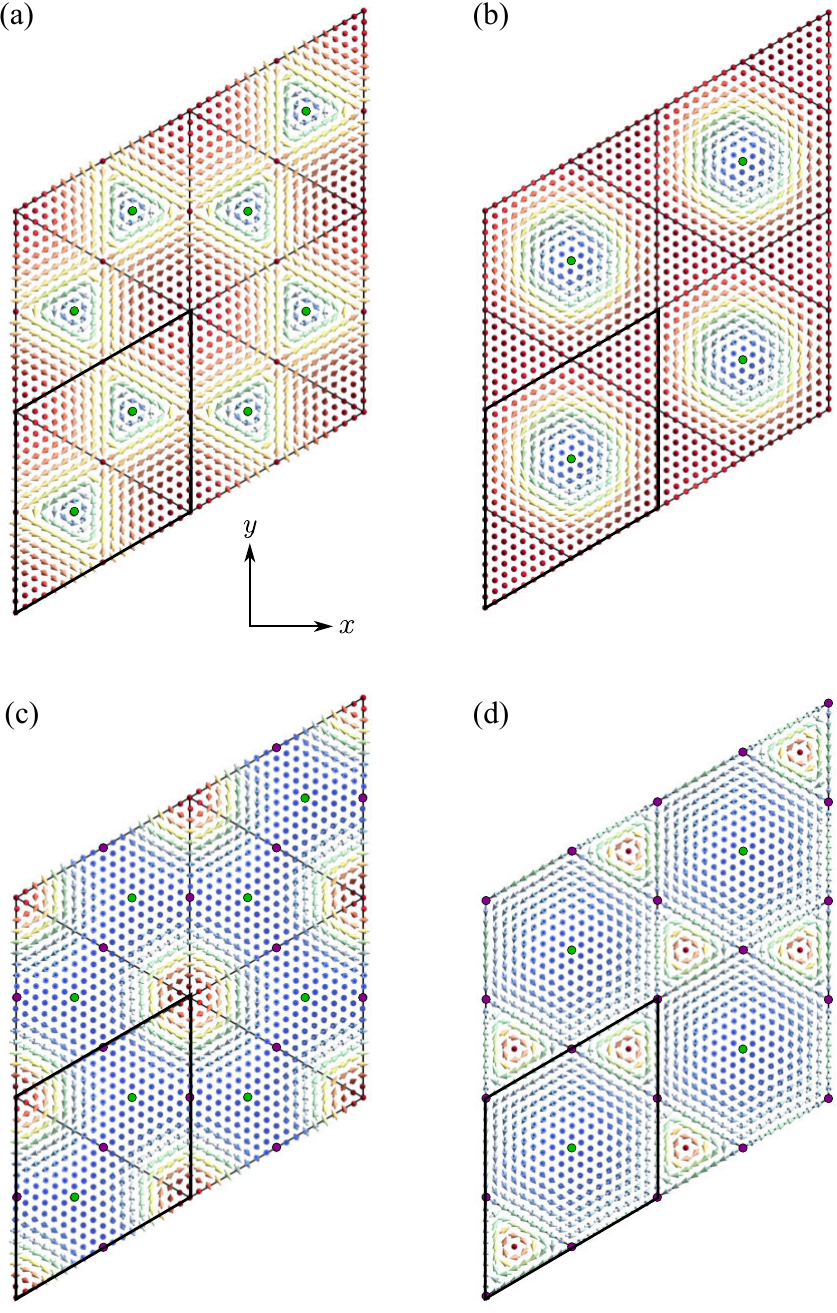}
	\caption{
	\label{fig:3qscr_spin}
	Real-space spin configurations of the screw $3Q$ states: 
	(a) $m=0.7$ and $\tilde{\vp}=0$, (b) $m=0.7$ and $\tilde{\vp}=\pi$, 
	(c) $m=-0.7$ and $\tilde{\vp}=0$, and (d) $m=-0.7$ and $\tilde{\vp}=\pi$.
	Each spin configuration is topologically nontrivial with 
	(a) $N_{\rm sk}=-2$, (b) $N_{\rm sk}=-1$, (c) $N_{\rm sk}=1$, and (d) $N_{\rm sk}=2$. 
	The notations are common to those in Fig.~\ref{fig:3qscr_intersection}. 
	}
\end{figure}

Figure~\ref{fig:3qscr_Nsk} summarizes the topological phase diagram on the plane of $m$ and $\tilde{\vp}$ for Eq.~(\ref{eq:chiral_3Q_ansatz}) obtained by the above procedure. 
The result is periodic in the $\tvp$ direction with period of $2\pi$ and symmetric with respect to $\tvp=\pi$. 
Note that the phase diagram for $m<0$ is obtained by mirroring that for $m>0$ with $\pi$ shift of $\tilde{\vp}$ and 
the sign inversion of $N_{\rm sk}$ since the spin texture with ($\varphi_\eta, m$) is obtained by time-reversal operation on that with ($\varphi_\eta+\pi, -m$). 
We find four topologically nontrivial phases with $N_{\rm sk}=-2$, $-1$, $1$, and $2$. 
The major portions of the phase diagram are occupied by the SkLs with $N_{\rm sk}=\pm1$, 
while the SkL with $N_{\rm sk}=\pm2$ appear in the small areas in between them only for $m\neq 0$; the SkL at $m=0$ always has $N_{\rm sk}=\pm 1$. 
The black dots appearing at the ends of the $N_{\rm sk}=\pm 1$ domes at $m=\pm 1/\sqrt{3}$ and $m=\pm \sqrt{3}$ correspond to the topological transitions by 
the pair annihilation of the hedgehogs and antihedgehogs in the hyperspace; see Sec.~\ref{sec:3.2.2}. 

Typical spin configurations of the screw $3Q$-SkLs with different $N_{\rm sk}$ are shown in Fig.~\ref{fig:3qscr_spin}. 
Figure~\ref{fig:3qscr_spin}(a) shows the spin configuration of the SkL with $N_{\rm sk}=-2$ at $m=0.7$ and $\tilde{\vp}=0$.  
In this state, small two skyrmions exist in the MUC and constitute a honeycomb lattice structure. 
There are two Dirac strings with $\zeta=1$ crossing the 2D plane, resulting in $N_{\rm sk}=-2$. 
Figure~\ref{fig:3qscr_spin}(b) is for the SkL with $N_{\rm sk}=-1$ at $m=0.7$ and $\tilde{\vp}=\pi$. 
In this case, a single skyrmion exists at the center of the MUC. 
The Dirac string with $\zeta=1$ through the skyrmion core leads to $N_{\rm sk}=-1$. 
Figures~\ref{fig:3qscr_spin}(c) and \ref{fig:3qscr_spin}(d) show the spin configurations for $m=-0.7$ at $\tilde{\vp}=0$ and 
$\tilde{\vp}=\pi$, respectively. 
These are obtained by flipping all the spins in Figs.~\ref{fig:3qscr_spin}(b) and \ref{fig:3qscr_spin}(a), 
and hence, $N_{\rm sk}=1$ and $2$, respectively. 
We note that the spin configurations with $\tilde{\vp}=n\pi$ ($n$ is an integer) have sixfold rotational symmetry, 
while the others with $\tilde{\vp}\neq n\pi$ are threefold, consistent with the symmetry arguments in Table~\ref{tab:3q_scr_sym}. 

Let us conclude this section by discussing some implications of our topological phase diagram to the phase control.  
In the previous experimental and theoretical studies for the chiral magnets~\cite{Binz2006-1, Muhlbauer2009, Yu2010, Yu2011, Han2010, Buhrandt2013}, 
the SkLs with $N_{\rm sk}=-1$ ($+1$) were observed in an external magnetic field applied to the ($-$)$\hat{\bf z}$ direction. 
This corresponds to the state with $\tilde{\vp}\sim\pi$ for $m>0$ 
and $\tilde{\vp}\sim 0$ for $m<0$ in the phase diagram in Fig.~\ref{fig:3qscr_Nsk}. 
The other SkLs with $N_{\rm sk}=\pm2$, however, have not been reported thus far. 
Our topological phase diagram indicates that it is necessary to cause the phase shift by $\simeq \pi$ in a magnetic field for reaching the $N_{\rm sk}=\pm2$ states. 
This is an interesting issue to be addressed since the emergent magnetic field in the $N_{\rm sk}=\pm2$ states becomes twice as large as that in the $N_{\rm sk}=\pm1$ ones. 
In addition, in the previous studies, the SkLs with $N_{\rm sk}=\pm1$ turn into a $1Q$ conical state or a uniformly polarized state 
while increasing the magnetic field, not into the topologically trivial $3Q$ state with $N_{\rm sk}=0$ shown in the phase diagram in Fig.~\ref{fig:3qscr_Nsk}. 
This suggests that it is difficult to access the points where the hedgehogs and antihedgehogs cause pair annihilation in the hyperspace (the black dots in Fig.~\ref{fig:3qscr_Nsk}). 
Once one can avoid the transition to the $1Q$ conical state, it might be possible to find novel topological phenomena arising from the singularity in the emergent electromagnetic fields due to the pair annihilation. 
It is worth noting that some possible ways to control the phase degree of freedom were recently proposed~\cite{Hayami2021phase}. 
We will discuss this issue in Sec.~\ref{sec:6}.

\subsection{ Sinusoidal 3$Q$ state \label{sec:3.3}}

We next discuss the phase degree of freedom for the sinusoidal $3Q$ state given by
\begin{align}
{\bf S}({\bf r})\propto
\left(\begin{array}{c}  
\frac{\sqrt{3}}{2}\sin\theta(-\cos\QQ_2 + \cos \QQ_3 ) \\
(-1)^{\Gamma}\frac{1}{2}\sin\theta\left( 2 \cos\QQ_1 - \cos \QQ_2 - \cos \QQ_3 \right) \\
\cos\theta(\cos\QQ_1 + \cos\QQ_2 + \cos\QQ_3 + 3\tilde{m})
\end{array} \right), 
\label{eq:nonchiral_3Q_ansatz}
\end{align}
which is obtained from Eq.~(\ref{eq:general_ansatz}) by taking $N_Q=3$ and
\begin{eqnarray}
&&\psi_{\eta}^{\rm c} = \frac{1}{\sqrt{3}}
, \ \ \psi_{\eta}^{\rm s} = 0, \\
&&{\bf e}_1^1=
\left( 0 , (-1)^{\Gamma} \sin\theta , \cos\theta \right)^{\mathsf{T}}, 
\ \ {\bf e}_2^1 =
R_{\Gamma}{\bf e}_{1}^1, 
\ \ {\bf e}_3^1=
R_{\Gamma}^2{\bf e}_{1}^1, 
\label{eq:evecs_nonchiral3Q} \\
&&
\tilde{m} = \frac{m}{\sqrt{3}\cos\theta}, 
\label{eq:mtilde}
\end{eqnarray}
where $\Gamma$ takes 0 or 1, $0<\theta<\frac{\pi}{2}$, and $R_{\Gamma}$ represents a $(-1)^{\Gamma}\frac{2\pi}{3}$ rotation about the $z$ axis given by 
\begin{eqnarray}
R_{\Gamma}=\left(\begin{array}{ccc}
\cos\left((-1)^{\Gamma}\frac{2\pi}{3}\right) & -\sin\left((-1)^{\Gamma}\frac{2\pi}{3}\right) & 0 \\
\sin\left((-1)^{\Gamma}\frac{2\pi}{3}\right) & \cos\left((-1)^{\Gamma}\frac{2\pi}{3}\right) & 0 \\
0 & 0 & 1
\end{array}\right).
\end{eqnarray}
In Eq.~(\ref{eq:nonchiral_3Q_ansatz}), $\Gamma$ is a parameter to describe the chirality of the spin texture; the spin texture with $\Gamma=1$ is obtained by flipping the $S_y$ component of that with $\Gamma=0$. 
Note that the spin texture with $\Gamma=0$ has threefold rotational symmetry, whereas that with $\Gamma=1$ does not. 
Meanwhile, $\theta$ describes the angle of the sinusoidal plane in the constituent waves.
In the following, we mainly focus on the spin textures with $\Gamma=0$, while we touch on those with $\Gamma=1$ in Sec.~\ref{sec:5.1}. 

Following the arguments in Sec.~\ref{sec:3.2}, we summarize the symmetry operations and their decompositions
for the sinusoidal $3Q$ state in Eq.~(\ref{eq:nonchiral_3Q_ansatz}) 
with $\Gamma=0$ and $1$ in Tables~\ref{tab:3q_sin_sym} and \ref{tab:3q_sin_sym_G=1}, respectively. 
Other symmetry operations are expressed by the combinations of those in the tables.  
From Table~\ref{tab:3q_sin_sym}, we find that the sinusoidal $3Q$ state with $\Gamma=0$ 
is symmetric under $C_{3z}$, $\mathcal{I}$, $\mathcal{T}C_{2x}$, and their combinations for $\tvp=0$ and $\pi$, otherwise $C_{3z}$, $\mathcal{T}C_{2x}$, and their combinations. 
Meanwhile, from Table~\ref{tab:3q_sin_sym_G=1}, we find that the state with $\Gamma=1$ is symmetric under $\mathcal{I}$, $\mathcal{T}C_{2x}$, and their combinations for $\tvp=0$, $\pi$, otherwise $\mathcal{T}C_{2x}$.

\begin{table}
\caption{\label{tab:3q_sin_sym}
Similar table to Table~\ref{tab:3q_scr_sym} for the sinusoidal $3Q$ state with $\Gamma=0$ in Eq.~(\ref{eq:nonchiral_3Q_ansatz}): 
$\mathcal{I}$ represents the spatial-inversion operation and the other notations are common to those in Table~\ref{tab:3q_scr_sym}.
}
\begin{ruledtabular}
\begin{tabular}{c|ccc}
operation & sum of phases & translation & magnetization  \\
\hline 
$C_{3z}$ & $\tvp \rightarrow \tvp$ & $0$ & $m \rightarrow m$  \\
$C_{2x}$ & $\tvp \rightarrow \pi + \tvp$ & $\tilde{\bf a}_{\eta}$ & $m \rightarrow -m$ \\
$\mathcal{I}$ & $\tvp \rightarrow 2\pi - \tvp$ & $\tilde{\bf a}_{\eta}+\tilde{\bf a}_{\eta'}$ & $m \rightarrow m$ \\
$\mathcal{T}$ & $\tvp \rightarrow \pi + \tvp$ & $\tilde{\bf a}_{\eta}$ & $m \rightarrow -m$
\end{tabular}
\end{ruledtabular}
\end{table}

\begin{table}
\caption{\label{tab:3q_sin_sym_G=1}
Similar table for the sinusoidal $3Q$ state with $\Gamma=1$ in Eq.~(\ref{eq:nonchiral_3Q_ansatz}). 
}
\begin{ruledtabular}
\begin{tabular}{c|ccc}
operation & sum of phases & translation & magnetization  \\
\hline 
$C_{2x}$ & $\tvp \rightarrow \pi + \tvp$ & $\tilde{\bf a}_{\eta}$ & $m \rightarrow -m$  \\
$\mathcal{I}$ & $\tvp \rightarrow 2\pi - \tvp$ & $\tilde{\bf a}_{\eta}+\tilde{\bf a}_{\eta'}$ & $m \rightarrow m$ \\
$\mathcal{T}$ & $\tvp \rightarrow \pi + \tvp$ & $\tilde{\bf a}_{\eta}$ & $m \rightarrow -m$
\end{tabular}
\end{ruledtabular}
\end{table}

\subsubsection{Hedgehogs in hyperspace \label{sec:3.3.1}}

Following the procedure in Sec.~\ref{sec:3.2.1}, we can identify the hyperspace positions of the topological defects 
in the 3D spin texture corresponding to Eq.~(\ref{eq:nonchiral_3Q_ansatz}). 
Solving ${\bf S}({\bf r})=0$, we obtain the following eight solutions: 
\begin{eqnarray}
(\QQ_1^{*},\QQ_2^{*},\QQ_3^{*}) &=& 
\left(\pi+p^{\rm sin}(\tilde{m}), \pi+p^{\rm sin}(\tilde{m}), \pi+p^{\rm sin}(\tilde{m}) \right), \notag \\ 
&&
\left(\pi-p^{\rm sin}(\tilde{m}), \pi-p^{\rm sin}(\tilde{m}), \pi-p^{\rm sin}(\tilde{m}) \right), \notag \\
&&
\left(\pi+p^{\rm sin}(\tilde{m}), \pi-p^{\rm sin}(\tilde{m}), \pi-p^{\rm sin}(\tilde{m}) \right) \notag \\ 
&&\qquad\qquad\quad
\mbox{and cyclic permutations}, \notag \\
&&
\left(\pi-p^{\rm sin}(\tilde{m}), \pi+p^{\rm sin}(\tilde{m}), \pi+p^{\rm sin}(\tilde{m}) \right) \notag \\
&&\qquad\qquad\quad 
\mbox{and cyclic permutations},
\label{eq:nonchiral_3Q_sol}
\end{eqnarray}
where 
\begin{eqnarray}
p^{\rm sin}(\tilde{m})=\arccos(\tilde{m}).
\end{eqnarray}
By using the relation in Eq.~(\ref{eq:R^*}), we obtain the positions of the eight singular points as 
\begin{eqnarray}
(X^{*}, Y^{*}, Z^{*}) &=&
\frac{\sqrt{2}}{3q}\left(0, 0, 3(\pi+p^{\rm sin}(\tilde{m})) \right), \notag \\ 
&&
\frac{\sqrt{2}}{3q}\left(0, 0, 3(\pi-p^{\rm sin}(\tilde{m})) \right), \notag \\
&&
\left( \frac{4p^{\rm sin}(\tilde{m})}{3q}, 0, \frac{\sqrt{2}}{3q}(3\pi-p^{\rm sin}(\tilde{m})) \right) \notag \\
&&\qquad\qquad\quad
\mbox{and $C_3^{Z}$ symmetric points}, \notag \\
&&
\left( -\frac{4p^{\rm sin}(\tilde{m})}{3q}, 0, \frac{\sqrt{2}}{3q}(3\pi+p^{\rm sin}(\tilde{m})) \right) \notag \\
&&\qquad\qquad\quad
\mbox{and $C_3^{Z}$ symmetric points}. 
\label{eq:nonchiral_3Q_sol_R}
\end{eqnarray}

\begin{figure}[tb]
\centering
\includegraphics[width=0.95\columnwidth]{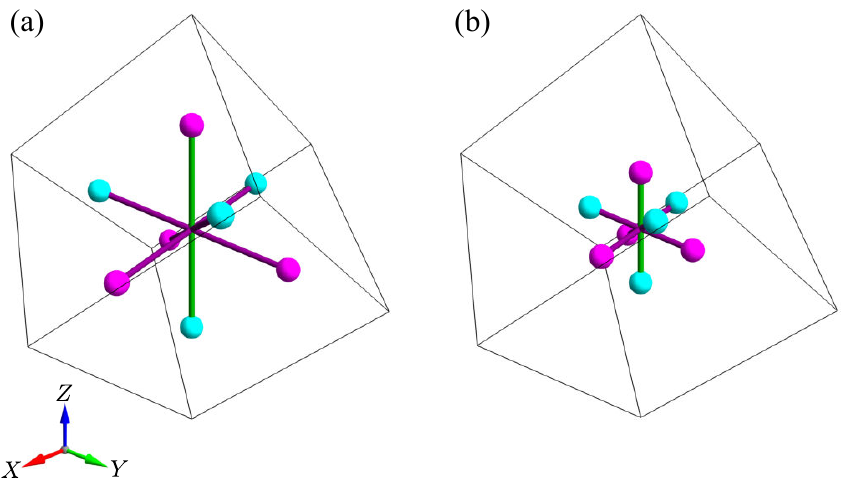}
\caption{
\label{fig:3qsin_defect}
Hedgehogs and antihedgehogs, and Dirac strings in the spin structure in the 3D hyperspace corresponding to the sinusoidal $3Q$ state in Eq.~(\ref{eq:nonchiral_3Q_ansatz}) for (a) $\tilde{m}=0$ and (b) $\tilde{m}=0.8$.
The notations are common to those in Fig.~\ref{fig:3qscr_defect}. 
The results are obtained for $\Gamma=0$ in Eq.~(\ref{eq:nonchiral_3Q_ansatz}); in the case of $\Gamma=1$, all the monopole charges of the hedgehogs and antihedgehogs, and the vorticities of the Dirac strings reverse their signs. 
}
\end{figure}

\subsubsection{Topological transition in 3D hyperspace \label{sec:3.3.2}}

Figure~\ref{fig:3qsin_defect} illustrates the evolution of the topological defects and 
the Dirac strings in the hyperspace while changing $\tilde{m}$ in Eq.~(\ref{eq:nonchiral_3Q_ansatz}) with $\Gamma=0$. 
The monopole charges, the Dirac strings, and their vorticities are obtained by the same procedure as 
in Sec.~\ref{sec:3.2.2}; see Eqs.~(\ref{eq:monopole_charge}) and (\ref{eq:vorticity}). 
Note that $\theta$ in Eq.~(\ref{eq:nonchiral_3Q_ansatz}) is irrelevant to the positions of the topological objects in the hyperspace. 
In the absence of the magnetization ($\tilde{m}=0$), four out of the eight defects in Eq.~(\ref{eq:nonchiral_3Q_sol_R}) 
are classified into the hedgehogs with $Q_{\rm m}=1$ and the other fours are antihedgehogs with $Q_{\rm m}=-1$. 
The eight topological defects form the NaCl-like structure, as shown in Fig.~\ref{fig:3qsin_defect}(a). 
We find that four pairs of the hedgehogs and antihedgehogs are connected by 
four Dirac strings with $\zeta=\pm 1$, which cross each other at the center of the MUC. 
When introducing $\tilde{m}$, the hedgehog and antihedgehog pairs move toward each other along the Dirac strings, 
and the cube defined by the eight defects shrinks, as shown in Fig.~\ref{fig:3qsin_defect}(b).
All of the eight defects come to the center of the MUC and vanish with pair annihilatation simultaneously at 
$\tilde{m}=1$. 
In the case of $\Gamma=1$, the positions of the topological defects remain the same, but the signs of 
all $Q_{\rm m}$ and $\zeta$ are reversed. 

\begin{figure}[tb]
\includegraphics[width=1.0\columnwidth]{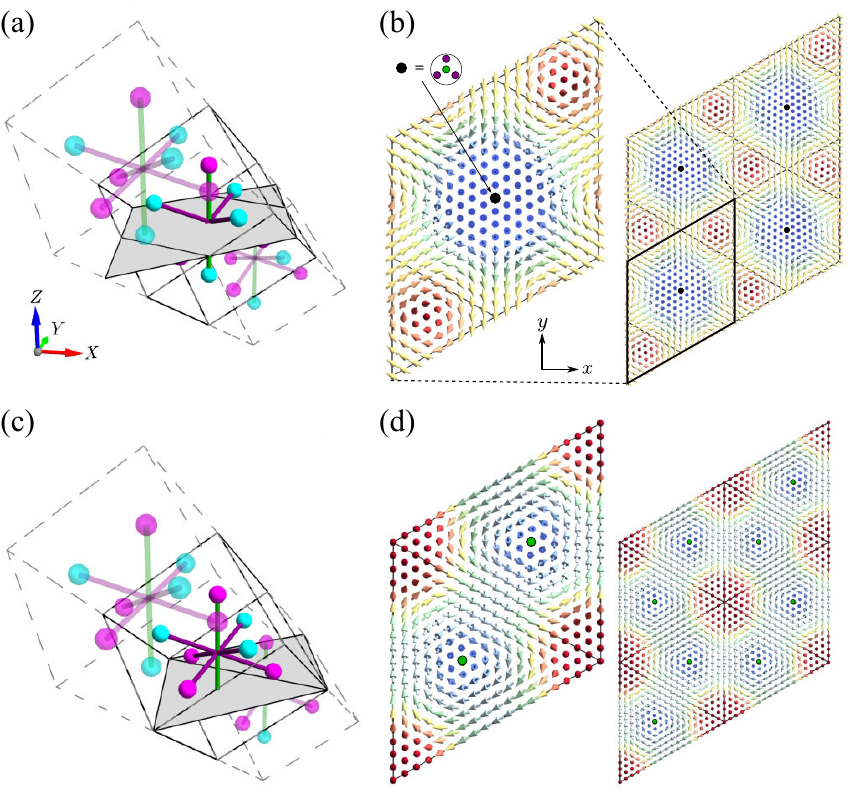}
\caption{
\label{fig:3qsin_intersection}
(a) Hedgehogs and antihedgehogs, and Dirac strings for $\tilde{m}=0$ 
in the 3D hyperspace for the sinusoidal $3Q$ state with $\Gamma$=0. 
The gray plane represents the intersection of the plane with $Z=\frac{\sqrt{2}}{3q}3\pi$ ($\tvp=3\pi$).
(b) The spin configuration on the gray plane in (a), which corresponds with the spin structure in 
Eq.~(\ref{eq:nonchiral_3Q_ansatz}) with $\tvp=3\pi$ and $\Gamma=0$. 
We take $\theta=\arccos\frac{1}{\sqrt{3}}$ in Eq.~(\ref{eq:evecs_nonchiral3Q}).
The vorticity at the black circle takes $\zeta=-2$ (see the text for details). 
(c) and (d) Similar figures for $\tilde{m}=0$ and $\tvp=2\pi$. 
The notations are common to those in Fig.~\ref{fig:3qscr_intersection}.
}
\end{figure}

\subsubsection{Topological transition on 2D plane \label{sec:3.3.3}}

As in Sec.~\ref{sec:3.2.3}, the 2D spin structure with phase $\tvp$ in Eq.~(\ref{eq:nonchiral_3Q_ansatz}) is obtained as the slice of the 3D hedgehog lattice at $Z=\frac{\sqrt{2}}{3q}\tvp$, and the skyrmion number $N_{\rm sk}$ is given by 
the sum of the vorticity of the Dirac strings as Eq.~(\ref{eq:Nsk_vorticity}). 
Figure~\ref{fig:3qsin_intersection}(a) shows the configurations of the hedgehogs, antihedgehogs, and the Dirac strings 
for the spin structure of Eq.~(\ref{eq:nonchiral_3Q_ansatz}) with $\Gamma=0$ 
in the hyperspace at $\tilde{m}=0$, and the slice at $Z=\frac{\sqrt{2}}{3q}3\pi$. 
The spin configuration on the slice is shown in Fig.~\ref{fig:3qsin_intersection}(b), which corresponds to the 2D spin texture in Eq.~(\ref{eq:nonchiral_3Q_ansatz}) with $\tvp=(2n+1)\pi$ ($n$ is an integer) and $\Gamma=0$. 
While the topological objects in the hyperspace are independent of the value of $\theta$ as mentioned in Sec.~\ref{sec:3.3.2}, the spin configuration in the original 2D plane depends on $\theta$; we take $\theta=\arccos\frac{1}{\sqrt{3}}$ in Fig.~\ref{fig:3qsin_intersection}(b). 
In the hyperspace, as the four Dirac strings cross each other at the center of the MUC and the slice includes the crossing point, 
the sum of the vorticities at the intersection is given by $\zeta=3\times(-1)+1=-2$, as depicted by the black circles in the figure.
Thus, we can identify the 2D spin texture at $\tilde{m}=0$ and $\tvp=3\pi$ as the $3Q$-SkL with $N_{\rm sk}=2$.

Figures~\ref{fig:3qsin_intersection}(c) and \ref{fig:3qsin_intersection}(d) illustrate the situation with $\tvp=2\pi$ at $\tilde{m}=0$. 
In this case, the 2D slice has two intersections of the Dirac strings with $\zeta=+1$, and hence, the spin texture 
with $\tvp=2\pi$ is the $3Q$-SkL with $N_{\rm sk}=-2$, which is a time-reversal counterpart of the spin texture in Fig.~\ref{fig:3qsin_intersection}(b).
Thus, similar to the screw $3Q$ spin structures in Sec.~\ref{sec:3.2.3}, the phase shift can cause topological transitions in the sinusoidal 
ones.

\subsubsection{Topological phase diagram \label{sec:3.3.4}}

\begin{figure}[tb]
\includegraphics[width=1.0\columnwidth]{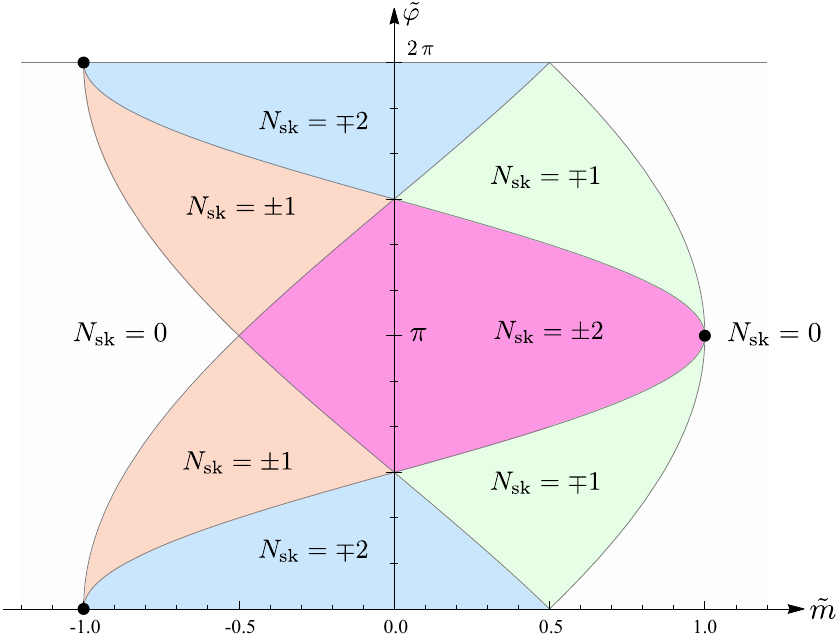}
\caption{
\label{fig:3qsin_Nsk}
Topological phase diagram for the sinusoidal $3Q$ state in Eq.~(\ref{eq:nonchiral_3Q_ansatz}) 
determined by the skyrmion number $N_{\rm sk}$ on the plane of $\tilde{m}$ and $\tilde{\vp}$. 
The upper (lower) signs of $N_{\rm sk}$ are for $\Gamma=0$ ($1$). 
The notations are common to those in Fig.~\ref{fig:3qscr_Nsk}. 
The black points represent the simultaneous annihilation of the four hedgehogs and the four antihedgehogs in the hyperspace, which are located at $(\tilde{m},\tilde{\vp})=(-1, 0 )$ and $(1, \pi)$.
}
\end{figure}

\begin{figure*}[tb]
\centering
\includegraphics[width=1.9\columnwidth]{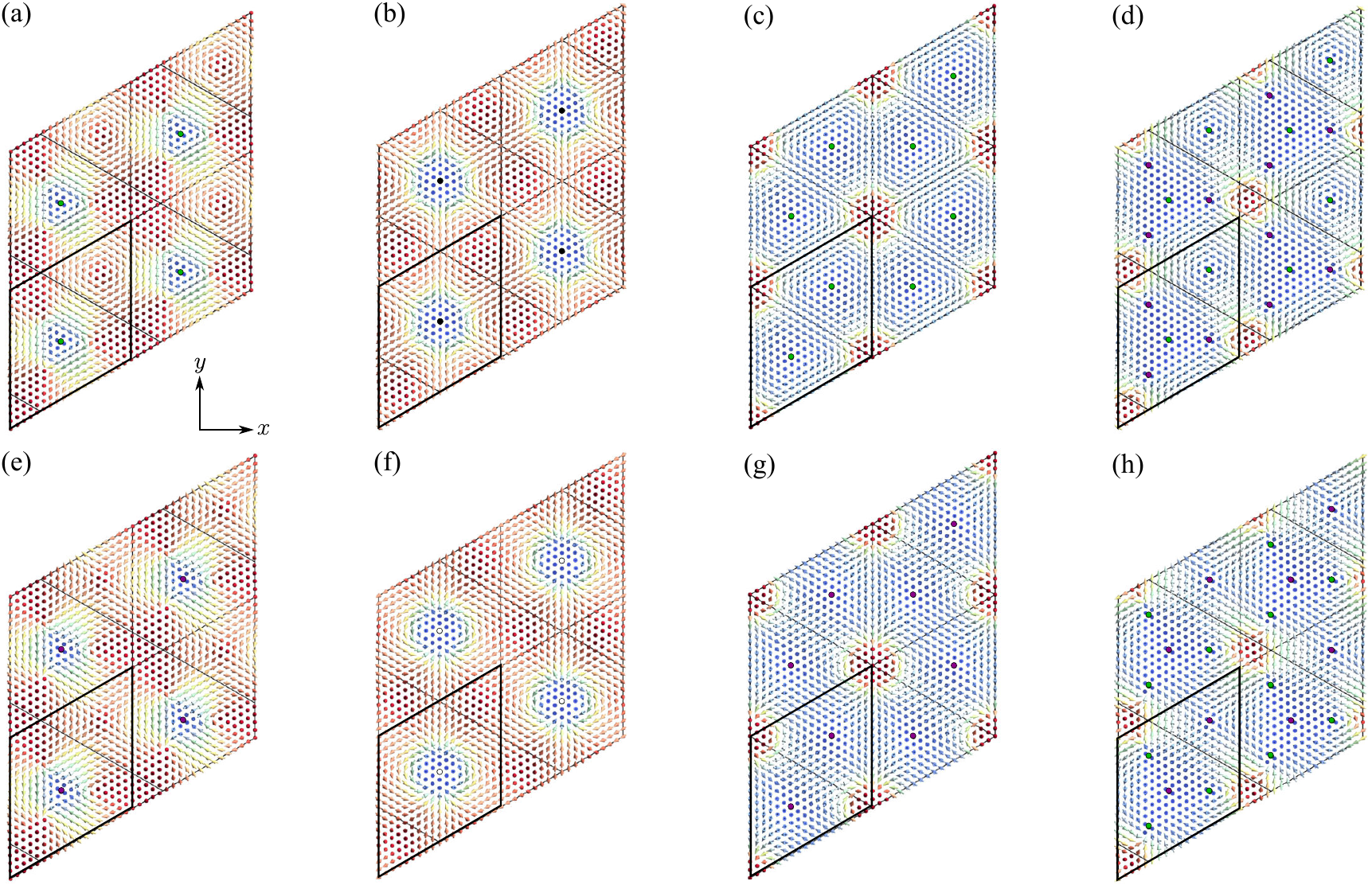}
\caption{
\label{fig:3qsin_spin}
Real-space spin configurations of the sinusoidal $3Q$ state in Eq.~(\ref{eq:nonchiral_3Q_ansatz}) 
with (a)-(d) $\Gamma=0$ and (e)-(h) $\Gamma=1$, and $\theta=\arccos\frac{1}{\sqrt{3}}$: 
(a)(e) $\tilde{m}=0.5$ and $\tilde{\vp}=\frac{\pi}{2}$, (b)(f) $\tilde{m}=0.5$ and $\tilde{\vp}=\pi$, 
(c)(g) $\tilde{m}=-0.5$ and $\tilde{\vp}=0$, (d)(h) $\tilde{m}=-0.5$ and $\tilde{\vp}=\frac{\pi}{2}$.
Each spin configuration is topologically nontrivial and has
(a) $N_{\rm sk}=-1$, (b) $N_{\rm sk}=2$, (c) $N_{\rm sk}=-2$, (d) $N_{\rm sk}=1$, 
(e) $N_{\rm sk}=1$, (f) $N_{\rm sk}=-2$, (g) $N_{\rm sk}=2$, and (h) $N_{\rm sk}=-1$. 
The white circles in (f) denote the vorticity $\zeta=2$. 
Other notations are common to those in Fig.~\ref{fig:3qsin_intersection}. 
In (b) and (f), the four Dirac strings cross at the center of the MUC; see Fig.~\ref{fig:3qsin_defect} and the text for details.
}
\end{figure*}

Figure~\ref{fig:3qsin_Nsk} summarizes the topological phase diagram on the plane of 
$\tilde{m}$ and $\tvp$ for the 2D spin texture in Eq.~(\ref{eq:nonchiral_3Q_ansatz}). 
The result is common to $\Gamma=0$ and $1$, while the sign of the skyrmion number $N_{\rm sk}$ in each phase is opposite: The upper (lower) signs are for $\Gamma=0$ ($1$). 
Similar to Fig.~\ref{fig:3qscr_Nsk}, the phase diagram has $2\pi$ periodicity and symmetric with respect to $\tvp=\pi$, and the result for $m<0$ is obtained by mirroring that for $m>0$ with $\pi$ shift of $\tvp$ and the sign inversion of $N_{\rm sk}$.
We find four topologically nontrivial phases with $N_{\rm sk}=-2$, $-1$, $1$, and $2$ as the proper screw case in Fig.~\ref{fig:3qscr_Nsk}, but with different distributions of each phase. 
In the present sinusoidal case, the large portions of the phase diagram are occupied by the SkL with $N_{\rm sk}=\pm2$, 
while the SkL with $N_{\rm sk}=\pm1$ appear in between them only for $\tilde{m}\neq 0$; the state at $\tilde{m}=0$ always has $N_{\rm sk}=\pm 2$, in contrast to Fig.~\ref{fig:3qscr_Nsk}.
Both $N_{\rm sk}=\pm2$ ($\mp2$) and $\mp1$ ($\pm1$) regions end at $\tilde{m}=1$ ($-1$) and $\tilde{\vp}=\pi$ ($0$) with the simultaneous 
pair annihilation of all the hedgehogs and antihedgehogs in the hyperspace, 
which are denoted by the black dots in Fig.~\ref{fig:3qsin_Nsk}. 

Figure~\ref{fig:3qsin_spin} showcases typical spin configurations of the sinusoidal $3Q$ state with $\theta=\arccos\frac{1}{\sqrt{3}}$. 
Here, we take $\Gamma=0$ in Figs.~\ref{fig:3qsin_spin}(a)-\ref{fig:3qsin_spin}(d) and $\Gamma=1$ in Figs.~\ref{fig:3qsin_spin}(e)-\ref{fig:3qsin_spin}(h). 
Figure~\ref{fig:3qsin_spin}(a) shows the spin configuration of the $N_{\rm sk}=-1$ state at $\tilde{m}=0.5$ and 
$\tilde{\vp}=\frac{\pi}{2}$.
In this state, there is a single Bloch type skyrmion with $N_{\rm sk}=-1$ per MUC. 
The Dirac string with $\zeta=1$ through the skyrmion core contributes to $N_{\rm sk}=-1$. 
Figure~\ref{fig:3qsin_spin}(b) is for the $N_{\rm sk}=2$ state at $\tilde{m}=0.5$ and $\tilde{\vp}=\pi$. 
This spin structure has a skyrmion with $N_{\rm sk}=2$ at the center of the MUC.
In this state, the 2D plane in the hyperspace intersects the crossing point of the four Dirac strings, 
which gives the total vorticity as $-2$; see Sec.~\ref{sec:3.3.3}.
Figures~\ref{fig:3qsin_spin}(c) and \ref{fig:3qsin_spin}(d) show the spin configurations with $\tilde{m}=-0.5$ at $\tilde{\vp}=0$ 
and $\tilde{\vp}=\frac{\pi}{2}$; the former is obtained by time-reversal operation on Fig.~\ref{fig:3qsin_spin}(b), 
while the latter is obtained by time-reversal operation combined with sixfold rotation operation about the $z$ axis on Fig.~\ref{fig:3qsin_spin}(a).
The corresponding results for $\Gamma=1$ are shown in Figs.~\ref{fig:3qsin_spin}(e)-\ref{fig:3qsin_spin}(h).
In contrast to the screw $3Q$ case in Sec.~\ref{sec:3.2.4}, all these sinusoidal $3Q$ cases  
with $\Gamma=0$ have threefold rotational symmetry independent of $\tilde{\vp}$ and $\tilde{m}$, 
while those with $\Gamma=1$ do not; see Sec.~\ref{sec:3.3}. 
In addition, the spin textures with $\tvp=n\pi$ ($n$ is an integer) has inversion symmetry independent of $\Gamma$, which is consistent with the symmetry arguments in Tables~\ref{tab:3q_sin_sym} and \ref{tab:3q_sin_sym_G=1}.

The phase diagram in Fig.~\ref{fig:3qsin_Nsk} indicates that the system undergoes a topological phase transition from $N_{\rm sk}=\pm 2$ to $\pm 1$, and finally to $N_{\rm sk}=0$ while increasing $\tilde{m}$. 
Such transitions were found in the previous numerical study of the Kondo lattice model on a triangular lattice while increasing the magnetic field~\cite{Ozawa2016}. 
Since the $N_{\rm sk}=\pm 2$ and $\pm 1$ phases appear predominantly in the different $\tvp$ regions in our phase diagram, the topological phase transition between them might be accompanied by a phase shift, but the phase degree of freedom was not studied in the previous study. 
We will discuss this issue by analyzing  the phases in the spin structures obtained by the previous study in Sec.~\ref{sec:5.1}.


\section{$4Q$ hedgehog lattices \label{sec:4} }

\begin{figure}[tb]
\centering
\includegraphics[width=1.0\columnwidth]{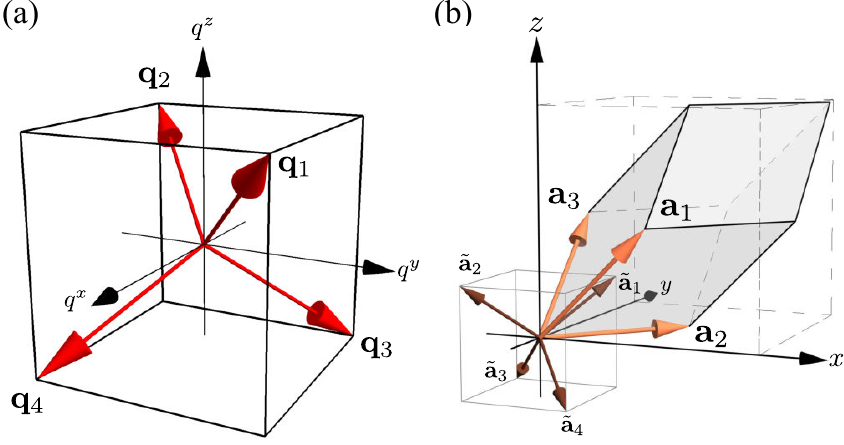}
\caption{
\label{fig:4q_setup}
(a) Schematic picture of the wave vectors in 3D reciprocal space, ${\bf q}_{\eta}$.
(b) Corresponding magnetic translation vectors in the 3D real space, ${\bf a}_{\eta}$, represented by orange arrows. 
The brown arrows represent the projections of ${\bf A}_{\eta}$ onto the $xyz$ space denoted by $\tilde{\bf a}_{\eta}$.  
The gray rhombohedron is the MUC, and the dashed cube with the side length of 
$L=\frac{2\sqrt{3}\pi}{q}$ includes four MUCs. 
The small gray cube has the side length of $\frac{L}{2}$.
}
\end{figure}

In this section, we elucidate the effect of phase shifts on the spin textures composed of four wave vectors 
in three dimensions, i.e., $N_{Q}=4$ and $d=3$, by using the hyperspace representation. 
Specifically, we consider Eq.~(\ref{eq:general_ansatz}) with four ${\bf q}_{\eta}$ given by
\begin{eqnarray}
&&{\bf q}_1 = \frac{q}{\sqrt{3}}\left( 1, 1, 1 \right),\ \ 
{\bf q}_2 = \frac{q}{\sqrt{3}}\left( -1, -1, 1 \right), \nonumber \\ 
&& 
{\bf q}_3 = \frac{q}{\sqrt{3}}\left( -1, 1,- 1 \right),\ \ 
{\bf q}_4 = \frac{q}{\sqrt{3}}\left( 1, -1, -1 \right).
\label{eq:4Q_q_eta}
\end{eqnarray}
For this $4Q$ spin structure, the 3D magnetic translation vectors are defined as
\begin{eqnarray}
&&
{\bf a}_1 = \frac{\sqrt{3}\pi}{q}\left(1, 0, 1\right),\ \ 
{\bf a}_2 = \frac{\sqrt{3}\pi}{q}\left(1, 1, 0\right), \nonumber \\
&&
{\bf a}_3 = \frac{\sqrt{3}\pi}{q}\left(0, 1, 1\right).
\label{eq:4Q_a_eta}
\end{eqnarray}
The four wave vectors and the three magnetic translation vectors are depicted in 
Figs.~\ref{fig:4q_setup}(a) and \ref{fig:4q_setup}(b), respectively.
While the 3D MUC is given by the gray rhombohedron in Fig.~\ref{fig:4q_setup}(b), 
we compute the topological properties for the dashed cube with the side length of $L=\frac{2\sqrt{3}\pi}{q}$, 
which includes four MUCs, in the following analyses.

In parallel with the arguments in Sec.~\ref{sec:3}, we focus on the two types of the $4Q$ spin structures in the following: 
the superposition of four proper screws and that of four sinusoidal waves. 
The former screw $4Q$ state breaks spatial-inversion symmetry and possesses the chirality; 
see Sec.~\ref{sec:4.2}. 
This type of spin structure is found in a noncentrosymmetric material MnSi$_{1-x}$Ge$_{x}$ and a centrosymmetric material SrFeO$_3$, as introduced in Sec.~\ref{sec:1}.
On the other hand, the latter sinusoidal $4Q$ state retains spatial-inversion symmetry or $S_4$ improper rotational symmetry about the $z$ axis; see Sec.~\ref{sec:4.3}. 
In the next subsection, we present the hyperspace representation applicable to these two types of spin configurations.

\subsection{Hyperspace representation of the $4Q$ states \label{sec:4.1}}

Following the same procedure as for the $3Q$ states in Sec.~\ref{sec:3.1}, 
we construct the hyperspace representation of the $4Q$ spin structures.
Considering 4D reciprocal hyperspace, we set the wave vectors 
${\bf Q}_{\eta}=( Q_{\eta}^{X}, Q_{\eta}^{Y}, Q_{\eta}^{Z}, Q_{\eta}^{W})$ in Eq.~(\ref{eq:Q_r_R}) without loss of generality: 
\begin{eqnarray}
&&
{\bf Q}_1=q\left(
\frac{1}{\sqrt{3}} , \frac{1}{\sqrt{3}} , \frac{1}{\sqrt{3}} , \frac{1}{q} 
\right), \label{eq:4Q_Q1}\\
&& 
{\bf Q}_2=q\left(
-\frac{1}{\sqrt{3}} , -\frac{1}{\sqrt{3}} , \frac{1}{\sqrt{3}} , \frac{1}{q} 
\right),\\
&&
{\bf Q}_3=q\left(
-\frac{1}{\sqrt{3}} , \frac{1}{\sqrt{3}} , -\frac{1}{\sqrt{3}} , \frac{1}{q} 
\right),\\
&&
{\bf Q}_4=q\left(
\frac{1}{\sqrt{3}} , -\frac{1}{\sqrt{3}} , -\frac{1}{\sqrt{3}} , \frac{1}{q} 
\right),
\label{eq:4Q_Q4}
\end{eqnarray}
where $Q_{\eta}^{X}$, $Q_{\eta}^{Y}$, and $Q_{\eta}^{Z}$ are taken to be proportional to $q_{\eta}^{x}$, $q_{\eta}^{y}$, and $q_{\eta}^{z}$ in Eq.~(\ref{eq:4Q_q_eta}), respectively. 
In this setting, ${\bf q}_{\eta}$ is a projection of ${\bf Q}_{\eta}$ onto the $q^x q^y q^z$ space. 
Then, we obtain the corresponding magnetic translation vectors ${\bf A}_{\eta}$ in the 4D hyperspace as
\begin{eqnarray}
&&
{\bf A}_1=\frac{\pi}{2q}\left( 
\sqrt{3} , \sqrt{3} , \sqrt{3} , q 
\right),\\
&& 
{\bf A}_2=\frac{\pi}{2q}\left(
-\sqrt{3} , -\sqrt{3} , \sqrt{3} , q 
\right),\\
&& 
{\bf A}_3=\frac{\pi}{2q}\left(
-\sqrt{3} , \sqrt{3} , -\sqrt{3} , q 
\right),\\
&&
{\bf A}_4=\frac{\pi}{2q}\left(
\sqrt{3} , -\sqrt{3} , -\sqrt{3} , q 
\right).
\end{eqnarray}
Note that ${\bf a}_1$, ${\bf a}_2$, and ${\bf a}_3$ are given by $\tilde{\bf a}_1-\tilde{\bf a}_3$, $\tilde{\bf a}_1-\tilde{\bf a}_2$, and $\tilde{\bf a}_1-\tilde{\bf a}_4$, respectively, where $\tilde{\bf a}_{\eta}$ are the projections of ${\bf A}_{\eta}$ onto the $xyz$ space; 
see Fig.~\ref{fig:4q_setup}(b). 

By using Eq.~(\ref{eq:r2R}), the hyperspace positions ${\bf R}=(X, Y, Z, W)$ are related with the real-space positions ${\bf r}$ as 
\begin{eqnarray}
\left(\begin{array}{c} 
X \\ Y \\ Z \\ W
\end{array}\right)
=
V_{4}
\left(\begin{array}{c} 
x \\ y \\ z \\ 1
\end{array}\right),
\label{eq:r2R_4Q}
\end{eqnarray}
where $V$ includes the phases as 
\begin{eqnarray}
V_{4}=
\left(\begin{array}{cccc} 
1 & 0 & 0 & \frac{\sqrt{3}}{4q}\left( \vp_1 - \vp_2 - \vp_3 + \vp_4 \right) \\
0 & 1 & 0 & \frac{\sqrt{3}}{4q}\left( \vp_1 - \vp_2 + \vp_3 - \vp_4 \right) \\
0 & 0 & 1 & \frac{\sqrt{3}}{4q}\left( \vp_1 + \vp_2 - \vp_3 - \vp_4 \right) \\
0 & 0 & 0 & \frac{1}{4}\left( \vp_1 + \vp_2 + \vp_3 + \vp_4 \right)
\end{array}\right). 
\end{eqnarray}
Equation~(\ref{eq:r2R_4Q}) tells that the spin configuration in the original 3D $xyz$ space is the same as the one on a hyperplane in the 4D hyperspace with 
\begin{eqnarray}
W=\frac{1}{4}\tvp,  
\end{eqnarray}
where 
\begin{eqnarray}
\tvp = \sum_{\eta} \vp_{\eta} = \vp_1 + \vp_2 + \vp_3 + \vp_4.
\end{eqnarray}
Similar to the $3Q$ case in Sec.~\ref{sec:3.1}, only the summation of the phases $\vp_{\eta}$ is relevant, 
instead of each value of $\vp_{\eta}$, and $\tvp$ has $2\pi$ periodicity.
Due to this periodicity, the $2\pi$ phase shift in $\tvp$ ($\Delta\tvp=\sum_{\eta} \vp_{\eta} = 2\pi$) corresponds to a spatial translation by 
\begin{eqnarray}
\Delta{\bf r}_{\eta}=\tilde{\bf a}_{\eta}-\sum_{\eta'=1}^{4}\frac{\Delta\vp_{\eta'}}{2\pi}\tilde{\bf a}_{\eta'},
\end{eqnarray}
where $\eta$ may take any of 1, 2, 3, and 4.

\subsection{Screw $4Q$ state \label{sec:4.2}}

In this subsection, we analyze the effect of phase shifts on the screw $4Q$ state 
composed of four proper screws given by  
\begin{widetext}
\begin{eqnarray}
\Sr
&\propto&
\left(
\begin{array}{c}
\frac{1}{\sqrt{2}}\left(
(-\cos {\mathcal {\mathcal Q}}_1+\cos {\mathcal Q}_2-\cos {\mathcal Q}_3+\cos {\mathcal Q}_4)
+\frac{1}{\sqrt{3}}(-\sin {\mathcal Q}_1+\sin {\mathcal Q}_2-\sin {\mathcal Q}_3+\sin {\mathcal Q}_4)
\right) \\
\frac{1}{\sqrt{2}}\left(
(\cos {\mathcal Q}_1-\cos {\mathcal Q}_2-\cos {\mathcal Q}_3+\cos {\mathcal Q}_4)
-\frac{1}{\sqrt{3}}(\sin {\mathcal Q}_1-\sin {\mathcal Q}_2-\sin {\mathcal Q}_3+\sin {\mathcal Q}_4)
\right) \\
\sqrt{\frac{2}{3}}(\sin {\mathcal Q}_1+\sin {\mathcal Q}_2+\sin {\mathcal Q}_3+\sin {\mathcal Q}_4) + 2m
\end{array}
\right),
\label{eq:4qchiral_ansatz}
\end{eqnarray}
\end{widetext}
which is obtained from Eq.~(\ref{eq:general_ansatz}) by taking $N_Q=4$ and 
\begin{eqnarray}
&&\psi_{\eta}^{\rm c}=\psi_{\eta}^{\rm s}=\frac{1}{2}, \\ 
&&{\bf e}_{\eta}^1=\frac{\hat{\bf z}\times{\bf e}_{\eta}^0}{|\hat{\bf z}\times{\bf e}_{\eta}^0|}, \ 
{\bf e}_{\eta}^2={\bf e}_{\eta}^0\times{\bf e}_{\eta}^1, \ 
{\bf e}_{\eta}^0=\frac{{\bf q}_{\eta}}{q}.
\end{eqnarray} 

With the same manner to Tables~\ref{tab:3q_scr_sym}--\ref{tab:3q_sin_sym_G=1}, we summarize the symmetry operations and their decompositions for the screw $4Q$ state in Table~\ref{tab:4q_scr_sym}. 
Similar to the previous arguments, from Table~\ref{tab:4q_scr_sym}, we can obtain the symmetry operations which do not change the spin texture. 
In the 3D system, however, some of the symmetry operations are nonsymmorphic. 
For instance, the $C_{4z}$ operation at $\tvp=\pi$  is reduced to the spatial translation by $\tilde{\bf a}_{\eta}$ as shown in Table~\ref{tab:4q_scr_sym}, and hence, the operations of $\{C_{4z}|-\tilde{\bf a}_{\eta}\}$ does not change the spin texture, where $\{\mathcal{O}|{\bf t}\}$ denotes the translation by ${\bf t}$ after operating $\mathcal{O}$. 
This corresponds to a screw operation. 
Note that the situation is different from the $3Q$ cases, where a translation combined with rotation can be represented by a shift of the rotation axis.
By considering such relations, we find that the screw $4Q$ state is symmetric under 
$C_{4z}$, $\mathcal{T}C_{2x}$, and their combinations for $\tvp=0$, 
$\{C_{4z}|-\tilde{\bf a}_{\eta}\}$, $\mathcal{T}C_{2x}$, and their combinations for $\tvp=\pi$, 
otherwise $C_{2z}$, $\mathcal{T}C_{2x}$, and their combinations.

\begin{table}
\caption{\label{tab:4q_scr_sym}
Similar table to Tables~\ref{tab:3q_scr_sym}--\ref{tab:3q_sin_sym_G=1} for the screw $4Q$ state in Eq.~(\ref{eq:4qchiral_ansatz}). 
The notations are common to those in Table~\ref{tab:3q_scr_sym}.
We take $\vp_{\eta}=\frac{\tvp}{4}$ without loss of generality.
}
\begin{ruledtabular}
\begin{tabular}{c|ccc}
operation & sum of phases & translation & magnetization   \\
\hline 
$C_{4z}$ & $\tvp \rightarrow 2\pi - \tvp$ & $\tilde{\bf a}_{\eta}$ & $m \rightarrow m$ \\
$C_{2x}$ & $\tvp \rightarrow \tvp$ & $\tilde{\bf a}_{\eta}+\tilde{\bf a}_{\eta'}$ & $m \rightarrow -m$ \\
$\mathcal{T}$ & $\tvp \rightarrow \tvp$ & $\tilde{\bf a}_{\eta}+\tilde{\bf a}_{\eta'}$ & $m \rightarrow -m$
\end{tabular}
\end{ruledtabular}
\end{table}

\subsubsection{Hedgehogs in hyperspace \label{sec:4.2.1}}

\begin{figure*}[tb]
\centering
\includegraphics[width=1.0\textwidth]{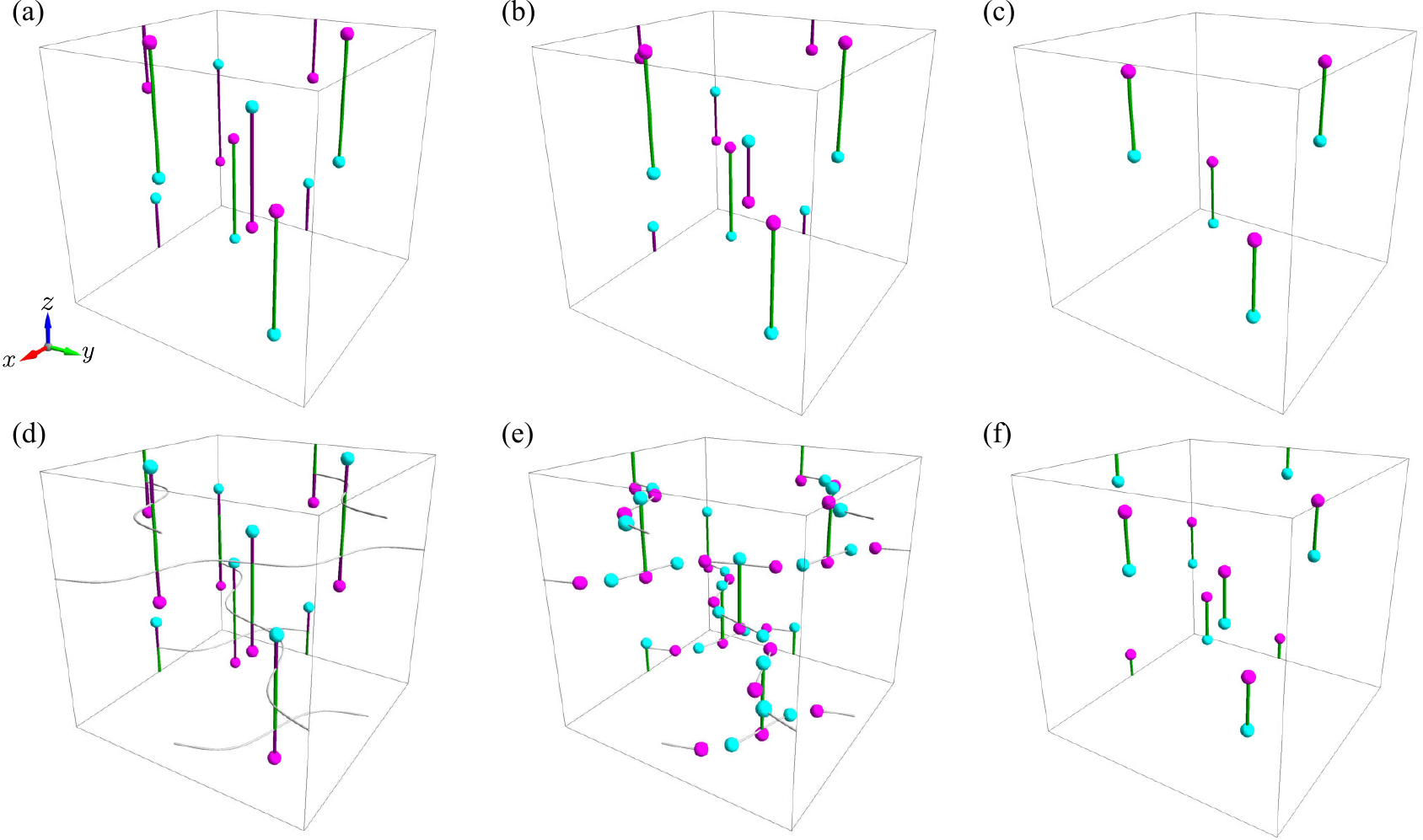}
\caption{
\label{fig:4qchiral_hedgehogs}
Real-space distribution of the hedgehogs and antihedgehogs, and the Dirac strings 
within the $L^3$ cube [see Fig.~\ref{fig:4q_setup}(b)] for the screw $4Q$ state in Eq.~(\ref{eq:4qchiral_ansatz}) 
while changing $m$ and $\tvp$: 
(a) $m=0$, (b) $m=0.3$, and (c) $m=0.9$ at $\tvp=\pi/3$, and 
(d) $m=0$, (e) $m=0.7$, and (f) $m=0.9$ at $\tvp=\pi$. 
The notations are common to those in Fig. \ref{fig:3qscr_defect}, except for the Dirac strings running on the horizontal planes, 
whose vorticities are ill-defined, denoted by the white lines. 
}
\end{figure*}

To discuss the phase shift in the screw $4Q$ state in Eq.~(\ref{eq:4qchiral_ansatz}), we study the corresponding spin texture 
in the 4D hyperspace whose reciprocal space is 
spanned by the wave vectors ${\bf Q}_{\eta}$ in Eqs.~(\ref{eq:4Q_Q1})--(\ref{eq:4Q_Q4}). 
As discussed in the previous subsection, a real-space spin configuration for a given phase summation $\tvp$ in the original 3D space corresponds to a hyperspace spin configuration on the hyperplane with $W=\frac{\tvp}{4}$. 
Following the procedure in Sec.~\ref{sec:3.2.1}, we first compute the positions of the topological defects in the 4D hyperspace. 
In the present case, however, the solutions for ${\bf S}({\bf R})=0$ are given by lines rather than points in the 4D hyperspace.
When $|m|<\frac{4}{\sqrt{6}}\sin W$, we obtain two solutions analytically:
\begin{eqnarray}
(X^*, Y^*, Z^*)=
L\left( 0, 0, \frac12 \pm \frac{1}{2\pi}r_1^{\rm scr}(m, W) \right), 
\label{eq:4qchiral_sol1}
\end{eqnarray}
where 
\begin{eqnarray}
r_1^{\rm scr}(m, W)=\arccos\left(\frac{\sqrt{6}m}{4\sin W}\right).
\end{eqnarray} 
Here, we show the solutions within $0 \leq X,Y < \frac{L}{2}$ and $0 \leq Z < L$, but the spatial translations of them with ${\bf a}_{\eta}$ in Eqs.~(\ref{eq:4Q_a_eta}) also satisfy ${\bf S}({\bf R})=0$. 
Meanwhile, when $|m|<\frac{4}{\sqrt{6}}\cos W$, we obtain other two solutions as 
\begin{eqnarray}
(X^*, Y^*, Z^*)&=&
L\left( \frac14, \frac14,  \frac{1}{2\pi}r_2^{\rm scr}(m, W) \right), \notag\\
&&
L\left( \frac14, \frac14,  \frac12 - \frac{1}{2\pi}r_2^{\rm scr}(m, W) \right), 
\label{eq:4qchiral_sol2}
\end{eqnarray}
where 
\begin{eqnarray}
r_2^{\rm scr}(m, W)=\arcsin\left(\frac{\sqrt{6}m}{4\cos W}\right).
\end{eqnarray}
The spatial translations also apply to this case. 
These solutions do not change their $XY$ coordinates with $m$ and $\tvp$. 
In addition to the above analytical solutions, we also find the other solutions by numerically solving the equation 
${\bf S}({\bf R})=0$. 
The two conditions $S_x({\bf R})=0$ and $S_y({\bf R})=0$ can be reduced to 
\begin{eqnarray}
\tan\left(\frac{q}{\sqrt{3}}Z^*\right) = \pm\sqrt{\frac{3\tan^2W-1}{3-\tan^2W}}
\label{eq:4qchiral_sol3_1}
\end{eqnarray}
and
\begin{align}
\tan \left(\frac{q}{\sqrt{3}}
Y^{*}\right)=\mp\sqrt{
\frac{\left(\sqrt{3}\tan W+1\right)\left(\sqrt{3}+\tan W\right)}
{\left(\sqrt{3}\tan W-1\right)\left(\sqrt{3}-\tan W\right)}
}\tan \left(\frac{q}{\sqrt{3}}X^{*}\right)
\label{eq:4qchiral_sol3_2}
\end{align}
Meanwhile, $S_z({\bf R})=0$ is reduced to 
\begin{eqnarray}
&&\cos \left(\frac{q}{\sqrt{3}}X^{*}\right)\cos \left(\frac{q}{\sqrt{3}}Y^{*}\right)\cos \left(\frac{q}{\sqrt{3}}Z^{*}\right)\sin W \notag \\
&&- \sin \left(\frac{q}{\sqrt{3}}X^{*}\right)\sin \left(\frac{q}{\sqrt{3}}Y^{*}\right)\sin \left(\frac{q}{\sqrt{3}}Z^{*}\right)\cos W \notag \\
&&+\frac{\sqrt{6}}{4}m=0.
\label{eq:4qchiral_sol3_3}
\end{eqnarray} 
We find that the numerical solutions exist when $\frac{1}{\sqrt{3}}<\tan W<\sqrt{3}$, namely $\frac{2\pi}{3}<\tvp<\frac{4\pi}{3}$. 
We note that the solutions in Eqs.~(\ref{eq:4qchiral_sol3_1})-(\ref{eq:4qchiral_sol3_3}) were not mentioned in the previous study for the screw $4Q$ state with $\tvp=\pi$~\cite{Park2011}.
While above solutions are points in the original 3D space, the line solutions in original 3D space are also obtained for only $m=0$. 
When $W=0$, $\frac{\pi}{6}$ and $\frac{\pi}{3}$, we find analytical solutions 
\begin{eqnarray}
(X^*,Y^*,Z^*) = \left(0,0,*\right), \left(*, \frac{L}{4}, 0\right), \left(0, *, \frac{L}{4}\right), 
\label{eq:4qchiral_line_node}
\end{eqnarray}
where $*$ takes any value, respectively. 

In the 4D hyperspace, the topological objects defined by the above solutions of ${\bf S}({\bf R})=0$ except for Eq.~(\ref{eq:4qchiral_line_node}) form closed loops, and 
the intersection of the loops by the hyperplane with $W=\frac{\tvp}{4}$ gives topological point defects in the original 3D space, 
which correspond to the hedgehogs and antihedgehogs discussed below. 
The hedgehog and antihedgehog always appear in pairs for each closed loop. 
The Dirac string connecting the hedgehog-antihedgehog pair in the 3D space is derived from the intersection of a 2D membrane in the 4D hyperspace whose edge and surface are defined by ${\bf S}({\bf R})=0$ and ${\bf S}({\bf R})=-\hat{\bf z}$, respectively. 
The 2D membrane can be regarded as an extension of the Dirac string in the higher dimension and hence, we may call it the Dirac plane. 
Thus, the hyperspace representation of the $4Q$ state is given by a 4D lattice composed of such closed loops, and the $4Q$ spin texture 
with the phase degree of freedom is defined as a 3D intersection of the 4D loop lattice. 

Since it is difficult to visualize the 4D hyperspace, we present the topological objects in the original 3D space which are derived from the hyperspace representation above. 
Figure~\ref{fig:4qchiral_hedgehogs} shows the systematic change of the topological defects in the 3D space while changing $m$ in Eq.~(\ref{eq:4qchiral_ansatz}) with $\tvp=\frac{\pi}{3}$ and $\pi$. 
Following the procedure in Sec.~\ref{sec:3.2.2}, we compute the monopole charge for the topological defects, $Q_{\rm m}$ in Eq.~(\ref{eq:monopole_charge}), and the vorticity of the Dirac strings which connect the hedgehogs and antihedgehogs, $\zeta$ in Eq.~(\ref{eq:vorticity}), by replacing ${\bf R}$ with ${\bf r}$. 
In order to distinguish the different topological phases, we also compute the total number of the topological defects, 
the hedgehogs and antihedgehogs, within the cube 
shown in Fig.~\ref{fig:4q_setup}(b), denoted by $N_{\rm m}$; the total number per MUC is given by $\frac{N_{\rm m}}{4}$. 

First, we discuss the case of $\tvp=\frac{\pi}{3}$ shown in Figs.~\ref{fig:4qchiral_hedgehogs}(a), 
\ref{fig:4qchiral_hedgehogs}(b), and \ref{fig:4qchiral_hedgehogs}(c). 
For $m=0$, there are 16 topological defects in total and half of them are hedgehogs with $Q_{\rm m}=+1$ and the others are antihedgehogs with $Q_{\rm m}=-1$, as shown in Fig.~\ref{fig:4qchiral_hedgehogs}(a).  
The hedgehogs and antihedgehogs derived from Eq.~(\ref{eq:4qchiral_sol1}) are connected by the Dirac strings with $\zeta=-1$, 
while the other topological defects from Eq.~(\ref{eq:4qchiral_sol2}) are connected by the Dirac strings with $\zeta=+1$.
All the Dirac strings run along the $z$ axis and have the same length of $\frac{L}{2}$.  
When introducing $m$, the hedgehogs and antihedgehogs move toward their counterparts along the Dirac strings, 
as exemplified in Fig.~\ref{fig:4qchiral_hedgehogs}(b) for $m=0.3$. 
The Dirac strings with $\zeta=-1$ become shorter than those with $\zeta=+1$: The lengths change as 
$L\left(\frac{1}{2}-\frac{1}{\pi}r_2^{\rm scr}\left(m, \frac{\tvp}{4}\right)\right)$ and 
$\frac{L}{\pi}r_1^{\rm scr}\left(m, \frac{\tvp}{4}\right)$ for $\zeta=+1$ and $-1$, respectively. 
This differentiation gives rise to a net emergent magnetic field, as will be discussed in Sec.~\ref{sec:4.2.2}. 
By further increasing $m$, the hedgehogs and antihedgehogs connected by the Dirac strings with $\zeta=-1$ disappear 
with pair annihilation at $m=\frac{4}{\sqrt{6}}\sin\frac{\tvp}{4}$; namely $N_{\rm m}$ is reduced to 8, which defines a 
topological transition between the phases with different $N_{\rm m}$.
The remaining hedgehogs and antihedgehogs move toward each other while further increasing $m$, 
as shown in Fig.~\ref{fig:4qchiral_hedgehogs}(c). 
They also vanish with pair annihilation at $m=\frac{4}{\sqrt{6}}\cos\frac{\tvp}{4}$, which defines the other topological transition into a topologically trivial state with $N_{\rm m}=0$.

Next, we discuss the case of $\tvp=\pi$ shown in Figs.~\ref{fig:4qchiral_hedgehogs}(d), 
\ref{fig:4qchiral_hedgehogs}(e), and \ref{fig:4qchiral_hedgehogs}(f).
When $m=0$, as shown in Fig.~\ref{fig:4qchiral_hedgehogs}(d), the system has eight pairs of the hedgehogs and antihedgehogs, similar to the case of $\tvp=\frac{\pi}{3}$.
However, while the positions of the topological defects as well as their total number are same as those in Fig.~\ref{fig:4qchiral_hedgehogs}(a), half of the hedgehog and antihedgehog pairs are exchanged; 
all the Dirac strings have the hedgehogs at their lower edges in Fig.~\ref{fig:4qchiral_hedgehogs}(d), whereas only half of them do in Fig.~\ref{fig:4qchiral_hedgehogs}(a). 
Moreover, we find additional Dirac strings running on the planes perpendicular to the $z$ axis, 
which are denoted by the white lines in the figure. 
Note that the vorticity in Eq.~(\ref{eq:vorticity}) is ill-defined for these horizontal Dirac strings. 
As shown in Fig.~\ref{fig:4qchiral_hedgehogs}(d), they intersect with 
the Dirac strings running along the $z$ axis, whose vorticities $\zeta$ change their signs at the crossing points. 
By introducing $m$, the hedgehogs and antihedgehogs move toward each other along the vertical Dirac strings as in the case of $\tvp=\frac{\pi}{3}$, while the horizontal Dirac strings are intact. 
When $m$ exceeds $\frac23$, however, $N_{\rm m}$ increases from 16 to 48, as depicted in Fig.~\ref{fig:4qchiral_hedgehogs}(e). 
This is caused by a peculiar topological transition with the increase of $N_{\rm m}$ discussed in detail below. 
The additional defects are obtained from the numerical solutions with Eqs.~(\ref{eq:4qchiral_sol3_1}), (\ref{eq:4qchiral_sol3_2}), and (\ref{eq:4qchiral_sol3_3}) for $m>\frac23$, which 
appear in pair on the horizontal Dirac strings; the horizontal Dirac strings are cut into pieces, both ends of which form hedgehogs or antihedgehogs. 
In other words, a cut results in pair creation of the hedgehog and antihedgehog. 
In this region, three hedgehogs and three antihedgehogs (a pair of hedgehog and antihedgehog with the vertical Dirac string, and hedgehog pair and antihedgehog pair with the horizontal Dirac strings) form a cluster like a twisted two-barred cross. 
While further increasing $m$, the hedgehogs and antihedgehogs move along the Dirac strings, and 
$N_{\rm m}$ decreases from 48 to 16 at $m=\sqrt{\frac{2}{3}}$. 
Here, two hedgehogs and one antihedgehog (or one hedgehog and two antihedgehogs) collide with each other at the same time in each cluster, 
leaving one (anti)hedgehog (see below). 
After this topological transition, all the Dirac strings have the hedgehogs at the upper edges, and hence, the vorticities are $\zeta=+1$, 
as shown in Fig.~\ref{fig:4qchiral_hedgehogs}(f). 
Finally, the remaining hedgehogs and antihedgehogs move toward their counterparts along the Dirac strings, and cause pair annihilation at 
$m=\frac{2}{\sqrt{3}}$, by which the system enters into a topologically trivial phase with $N_{\rm m}=0$.

\begin{figure}[tb]
\centering
\includegraphics[width=1.0\columnwidth]{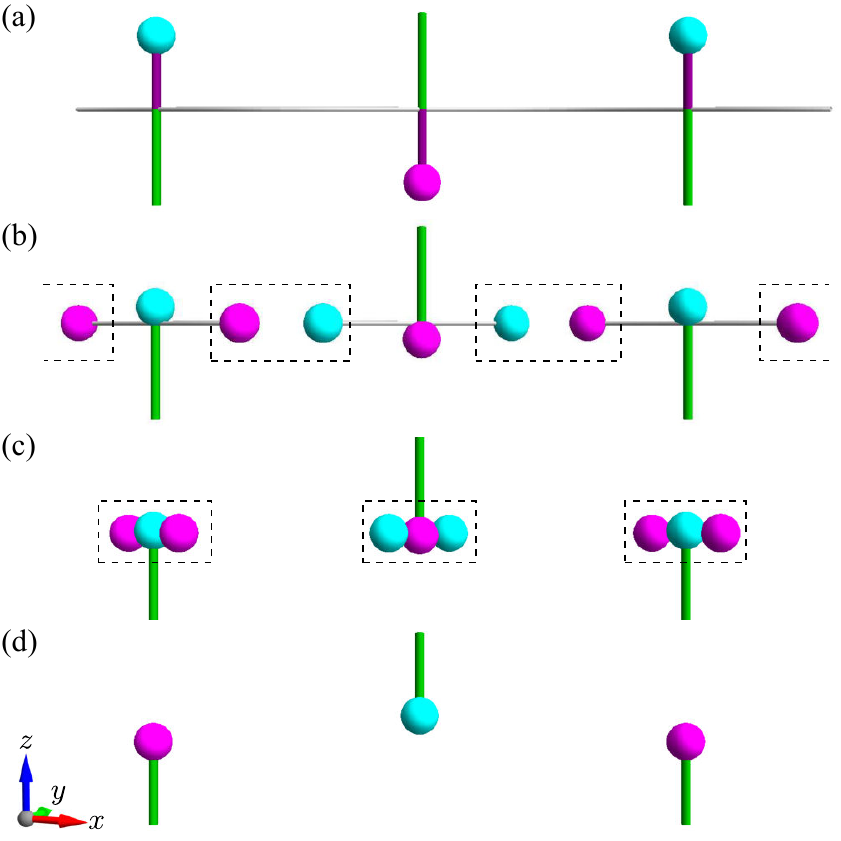}
\caption{
\label{fig:4qchiral_sign_change}
Enlarged figures for the systematic evolution of the hedgehogs and antihedgehogs while changing $m$ with $\tvp=\pi$: 
(a) $m=0$, (b) $m=0.7$, (c) $m=0.8$, and (d) $m=0.9$. 
The hedgehogs and antihedgehogs enclosed with the black dotted lines in (b) are generated by pair creation on the horizontal Dirac strings. 
On the other hand, the three topological defects enclosed with the black dotted lines in (c) collide 
with each other simultaneously and cause a fusion into a hedgehog or an antihedgehog in (d). 
A part of the topological defects and the Dirac strings are depicted for better visibility.
The notations are same with those in Fig.~\ref{fig:4qchiral_hedgehogs}.  
}
\end{figure}

The evolution of the topological objects for the case of $\tvp=\pi$ includes, at least, two striking features. 
One is the increase of the total number of the hedgehogs and antihedgehogs while increasing $m$. 
This is highly nontrivial since usually the hedgehogs and antihedgehogs move toward each other along the Dirac string and cause pair annihilation, and hence, their number does not increase while increasing $m$~\cite{Binz2006-1, Park2011, Zhang2016, Kanazawa2016, Shimizu2021moire}. 
It is worthy noting that the increase is caused by pair creation on the horizontal Dirac strings whose vorticities are ill-defined. 
The process is detailed in Figs.~\ref{fig:4qchiral_sign_change}(a) and \ref{fig:4qchiral_sign_change}(b). 
The other striking feature is the simultaneous fusion of three defects, 
as depicted in Figs.~\ref{fig:4qchiral_sign_change}(c) and \ref{fig:4qchiral_sign_change}(d). 
In this process, two hedgehogs (antihedgehogs) on the horizontal Dirac string and one antihedgehog (hedgehog) on the vertical Dirac string move toward the crossing point of the vertical and horizontal Dirac strings, and cause the fusion into a single hedgehog (antihedgehog).  
Note that the total monopole charge is conserved through the fusion. 
Hence, both striking features originate from the peculiar horizontal Dirac strings. 

Thus far, we show the results only for $\tvp=\pi/3$ and $\pi$, but we note that the other cases fall into either behavior qualitatively. 
Specifically, the cases for $\frac{2\pi}{3}<\tvp<\frac{4\pi}{3}$ belong to the latter, while the others to the former. 
In the latter class, we find the additional numerical solutions as described above, which lead to the peculiar pair creation and fusion of the topological defects. 
We note that the fusion occurs simultaneously for all the created pairs when $\tvp=\pi$, while it takes place successively half by half for other values in 
$\frac{2\pi}{3}<\tvp<\frac{4\pi}{3}$, as discussed below.

\subsubsection{Topological phase diagram \label{sec:4.2.2}}

\begin{figure}[tb]
\centering
\includegraphics[width=1.0\columnwidth]{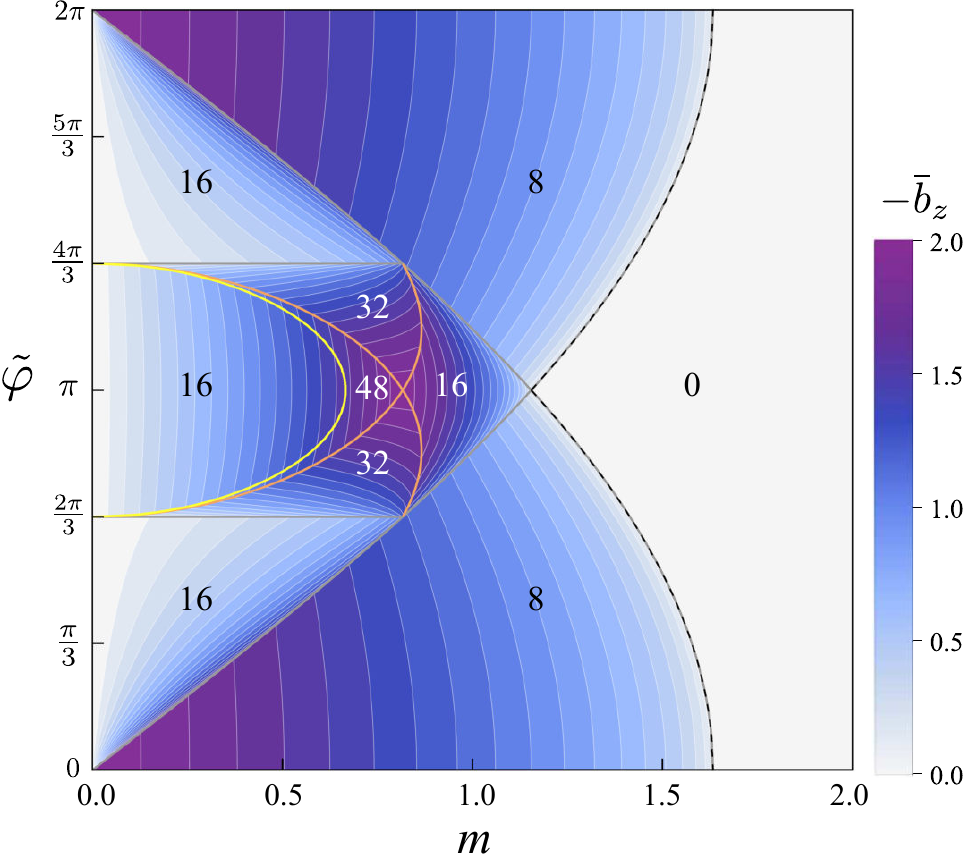}
\caption{
\label{fig:4qch_pd}	
Topological phase diagram for the screw $4Q$ state on the plane of $m$ and $\tvp$, determined by the number of 
hedgehogs and antihedgehogs within the cube, $N_{\rm m}$ [see Fig.~\ref{fig:4q_setup}(b)]. The contour plot indicates the emergent magnetic field 
$-\bar{b}_z$; the white lines denote the contours drawn every $0.1$, and the black dashed lines denote the boundary to $\bar{b}_z=0$.
The gray solid lines denote pair annihilation of hedgehogs and antihedgehogs. 
The yellow line and the orange lines 
denote pair creation of hedgehogs and antihedgehogs and fusion of three topological defects, respectively, while increasing $m$. 
}
\end{figure}

\begin{figure*}[tb]
\centering
\includegraphics[width=2.0\columnwidth]{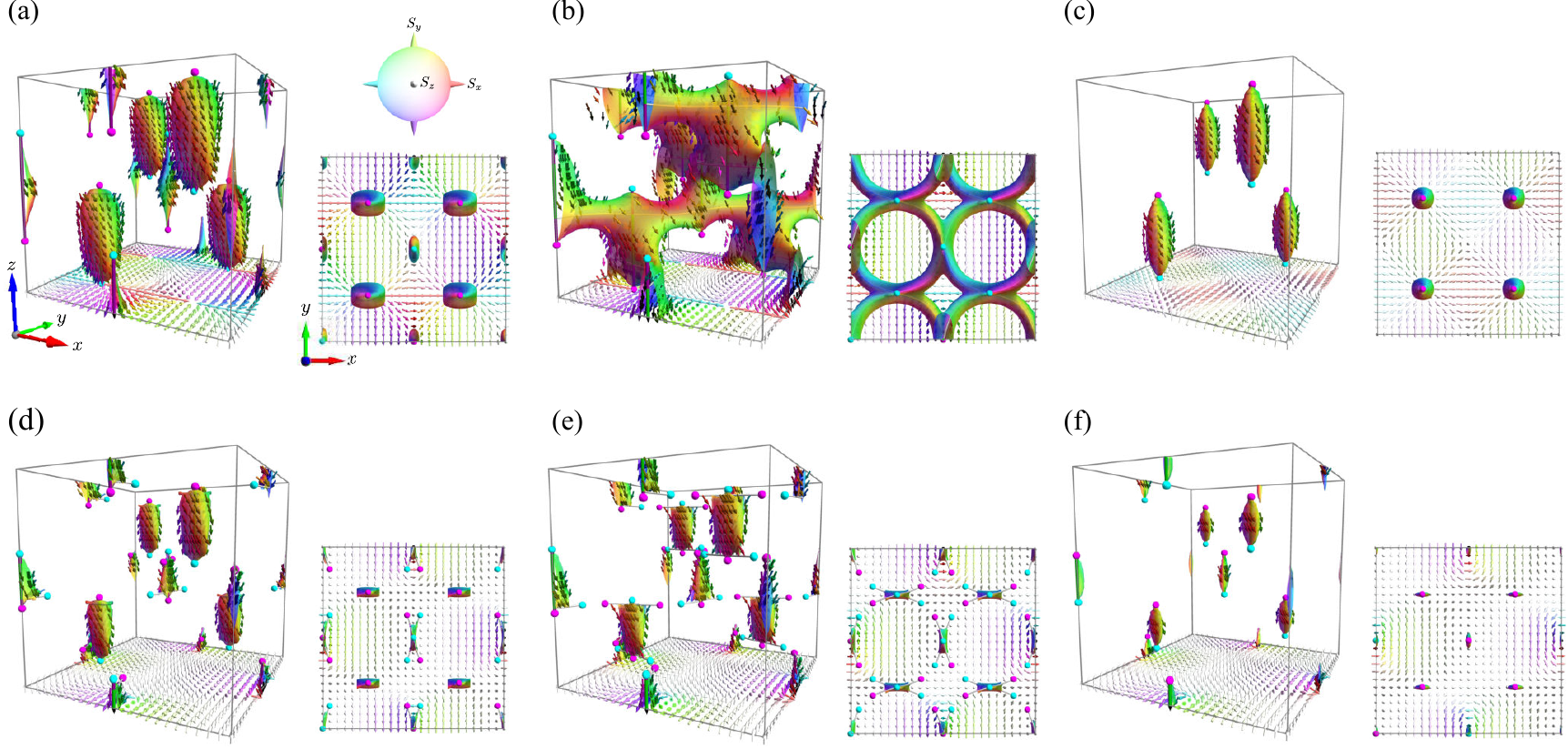}
\caption{
\label{fig:4qch_spin}
Real-space spin configurations on the isosurfaces with $S_z({\bf r})=-0.9$ within the $L^3$ cube for the screw 4$Q$ state in Eq.~(\ref{eq:4qchiral_ansatz}) for different topological phases in Fig.~\ref{fig:4qch_pd}: 
(a) $m=0$ and $\tvp=\frac{\pi}{3}$ ($N_{\rm m}=16$), (b) $m=0$ and $\tvp=\pi$ ($N_{\rm m}=16$), (c) $m=0.4$ and $\tvp=0$ ($N_{\rm m}=8$), (d) $m=0.7$ and $\tvp=\frac{5\pi}{6}$ ($N_{\rm m}=32$), (e) $m=0.7$ and $\tvp=\pi$ ($N_{\rm m}=48$), and (f) $m=0.9$ and $\tvp=\pi$ ($N_{\rm m}=16$). 
The spin configurations are also shown on the bottom plane of the cube.
The color of the arrows and the isosurfaces denote the $xyz$ and $xy$ components of $\Sr$, respectively; see the inset in (a). 
In each figure, the right panel shows the top view.
}
\end{figure*}

Figure~\ref{fig:4qch_pd} summarizes the topological phase diagram on the plane of $m$ and $\tvp$ 
for the screw $4Q$ state in Eq.~(\ref{eq:4qchiral_ansatz}), determined by $N_{\rm m}$. 
The result is periodic in the $\tvp$ direction with period of $2\pi$ and symmetric with respect to $\tvp=\pi$. 
We also plot the emergent magnetic field $\bar{b}_z$, which is defined as 
\begin{eqnarray}
\bar{b}_z = 
\frac{1}{4\pi L}\int d\mathcal{V} b_z({\bf r}),
\label{eq:bbar}
\end{eqnarray}
where the volume integration is taken within the $L^3$ cube in Fig.~\ref{fig:4q_setup}(b).
Following the arguments in Refs.~\cite{Park2011, Zhang2016, Kanazawa2016, Shimizu2021moire}, 
$\bar{b}_z$ is rewritten into 
\begin{eqnarray}
\bar{b}_z = -\frac{1}{L}\sum_{k}\left( l_k^{+} - l_{k}^{-} \right), 
\label{eq:bbar2}
\end{eqnarray}
where $l_{k}^{+}$ and $l_{k}^{-}$ denote the length of the $k$th Dirac strings with $\zeta=+1$ and $\zeta=-1$ 
projected onto the $z$ axis, respectively, and the sum is taken for all the Dirac strings involved in the cube.
Note that the values of $-\bar{b}_z$ are plotted by contour in Fig.~\ref{fig:4qch_pd}. 
The phase diagram for $m<0$ is obtained in the same form with sign reversal of $-\bar{b}_z$ since the spin texture with $(\vp_{\eta}, m)$ is obtained by time-reversal operation on that with $(\vp_{\eta}+\pi, -m)$. 

In the phase diagram, we find the topological phases with $N_{\rm m}=8$, $16$, $32$, and $48$, 
in addition to the trivial phase with $N_{\rm m}=0$ in the large $m$ region.
When $0 < \tvp < \frac{2\pi}{3}$ or $\frac{4\pi}{3} < \tvp < 2\pi$, the phases with $N_{\rm m}=8$ and 16 appear. 
In the phase with $N_{\rm m}=16$ for small $m$, 8 hedgehogs and 8 antihedgehogs are connected by the Dirac strings with $\zeta=+1$ and $-1$, and the Dirac strings with $\zeta=-1$ are shorter than those with $\zeta=+1$, as exemplified in Fig.~\ref{fig:4qchiral_hedgehogs}(b). 
While increasing $m$, the length difference between the long and short Dirac strings increases, leading to the increase of $-\bar{b}_z$. 
Specifically, the value of $-\bar{b}_z$ is given by 
\begin{equation}
-\bar{b}_z = \pm2\left[ 1-\frac{2}{\pi}\left(
r_1^{\rm scr}\left(m, \frac{\tvp}{4}\right) + r_2^{\rm scr}\left(m, \frac{\tvp}{4}\right) 
\right) \right],
\end{equation}
where the upper (lower) sign is for $0 < \tvp < \frac{2\pi}{3}$ ($\frac{2\pi}{3} < \tvp < 2\pi$).
At $m=\frac{4}{\sqrt{6}}\sin\frac{\tvp}{4}$ for $0<\tvp<\frac{2\pi}{3}$ and $m=\frac{4}{\sqrt{6}}\cos\frac{\tvp}{4}$ for 
$\frac{4\pi}{3}<\tvp<2\pi$, half of the topological defects connected by the Dirac strings with $\zeta=-1$ disappear with pair annihilation, leaving the defects connected by the 
Dirac stings with $\zeta=+1$, as exemplified in Fig.~\ref{fig:4qchiral_hedgehogs}(c).
The pair annihilation occurs at smaller $m$ when $\tvp$ approaches $0$ or $2\pi$, and $-\bar{b}_z$ is enhanced to $-\bar{b}_z\to 2$ when $m\to 0$ at $\tvp=0$ or $2\pi$ 
(in the limit, there are only four Dirac string with $\zeta=+1$ and length $\frac{L}{2}$). 
Meanwhile, in the phase with $N_{\rm m}=8$, the increase of $m$ reduces the length of the Dirac strings with $\zeta=+1$, leading to the decrease of $-\bar{b}_z$ as 
\begin{equation}
-\bar{b}_z =
\begin{cases}
2\left[ 1-\frac{2}{\pi}r_2^{\rm scr}\left(m, \frac{\tvp}{4}\right) \right]
& 
{\rm for} 
\ \ 0 < \tvp <
\frac{2\pi}{3} \\
\frac{4}{\pi}r_1^{\rm scr}\left(m, \frac{\tvp}{4}\right) 
&
{\rm for} 
\ \ \frac{4\pi}{3} < \tvp < 2\pi. 
\end{cases}
\end{equation}
Finally, $-\bar{b}_z$ vanishes by the pair annihilation of the remaining hedgehogs and antihedgehogs at 
$m=\frac{4}{\sqrt{6}}\cos\frac{\tvp}{4}$ for $0<\tvp<\frac{2\pi}{3}$ and $m=\frac{4}{\sqrt{6}}\sin\frac{\tvp}{4}$ for 
$\frac{4\pi}{3}<\tvp<2\pi$, and the system enters into the topologically trivial phase with $N_{\rm m}=0$ for larger $m$. 

On the other hand, when $\frac{2\pi}{3} < \tvp < \frac{4\pi}{3}$, the topological phases with $N_{\rm m}=32$ and 48 appear additionally 
in the intermediate region of $m$, as shown in Fig.~\ref{fig:4qch_pd}.  
In this range of $\tvp$, while increasing $m$ from the phase with $N_{\rm m}=16$, pair creation of the hedgehogs and antihedgehogs 
occurs on the phase boundary to the phase with $N_{\rm m}=48$, which is denoted by the yellow line in the figure. 
No anomaly is found in $-\bar{b}_z$ at the topological transition since the pair-created topological objects move along the horizontal Dirac strings and this evolution does not contribute to $l_{k}^{\pm}$ in Eq.~(\ref{eq:bbar2}); see Fig.~\ref{fig:4qchiral_hedgehogs}(e). 
In these phases, however, the increase of $m$ reduces the length of the Dirac strings with $\zeta=-1$, leading to the increase of 
$-\bar{b}_z$ as 
\begin{equation}
-\bar{b}_z = 2\left[ 1 - \frac{2}{\pi}\left( 
r_1^{\rm scr}\left(m, \frac{\tvp}{4}\right) - r_2^{\rm scr}\left(m, \frac{\tvp}{4}\right) 
\right) \right]. 
\end{equation}
While increasing $m$, half of the pair-created topological defects disappear through the fusion, which causes the topological transition from $N_{\rm m}=48$ to $32$ (orange lines in the figure). 
In the $N_{\rm m}=32$ region, the value of $-\bar{b}_z$ is given by 
\begin{eqnarray}
-\bar{b}_z=&&2\left[ 1 \mp \frac{2}{\pi}\left( 
r_1^{\rm scr}\left(m, \frac{\tvp}{4}\right) + r_2^{\rm scr}\left(m, \frac{\tvp}{4}\right) 
\right) \right. \notag \\
&& \qquad \left. \pm \frac{4}{\pi}\arccos\left(\sqrt{2\cos^2\frac{\tvp}{4}-\frac{1}{2}}\right) \right],  
\end{eqnarray}
where the upper (lower) signs are for $\frac{2\pi}{3} < \tvp \leq \pi$ ($\pi \leq \tvp < \frac{4\pi}{3}$).
With a further increase of $m$, the rest half of the pair-created topological defects cause the fusion and $N_{\rm m}$ is reduced from $32$ to $16$. 
In the $N_{\rm m}=16$ phase, all of the topological defects are connected by the Dirac strings with 
$\zeta=+1$, as exemplified in Fig.~\ref{fig:4qchiral_hedgehogs}(f), where $-\bar{b}_z$ is given as 
\begin{equation}
-\bar{b}_z=2\left[ 1 + \frac{2}{\pi}\left( 
r_1^{\rm scr}\left(m, \frac{\tvp}{4}\right) - r_2^{\rm scr}\left(m, \frac{\tvp}{4}\right) 
\right) \right]. 
\end{equation} 
Note that $-\bar{b}_z$ takes the maximum value of $-\bar{b}_z=2$ at $(m, \tvp) = \left(\sqrt{\frac{2}{3}}, \pi \right)$, 
where all the pair-created defects cause the fusion simultaneously.

Figure~\ref{fig:4qch_spin} showcases typical spin configurations of the screw $4Q$ states for all the topological phases in Fig.~\ref{fig:4qch_pd}, together with the hedgehogs and antihedgehogs, and the Dirac strings. 
The spin configurations are shown on the isosurfaces with  $S_z({\bf r})=-0.9$ as well as the bottom plane of the $L^3$ cube. 
The isosurfaces, by definition, extend from the hedgehogs and antihedgehogs, and involve the Dirac strings inside. 
Figure~\ref{fig:4qch_spin}(a) is for the $N_{\rm m}=16$ state at $m=0$ and $\tvp=\frac{\pi}{3}$. 
From the spin configurations on the isosurfaces, it is observed that the helicity of the spin texture gradually increases or decreases with the $z$ coordinates, 
which leads to $\pm \pi$ difference between the top and bottom of each Dirac string. 
Figure~\ref{fig:4qch_spin}(b) is for the different $N_{\rm m}=16$ state at $m=0$ and $\tvp=\pi$. 
The result demonstrates that the phase shift drastically changes the spin configurations as well as the real-space distributions of the topological objects, 
while $N_{\rm m}$ is same as in Fig.~\ref{fig:4qch_spin}(a). 
In this state, the isosurfaces have complicated 3D networks due to the existence of the Dirac strings running on the horizontal planes. 
Figure~\ref{fig:4qch_spin}(c) is for the $N_{\rm m}=8$ phase at $m=0.4$ and $\tvp=0$. 
In this state, the isosurfaces become much simpler. 
Figures~\ref{fig:4qch_spin}(d) and \ref{fig:4qch_spin}(e) are for the $N_{\rm m}=32$ state at $m=0.7$ and $\tvp=\frac{5\pi}{6}$ and $N_{\rm m}=48$ states at $m=0.7$ and $\tvp=\pi$, respectively. 
In both states, the spin configuration on a horizontal $xy$ plane comprise a SkL with Bloch type skyrmions, as exemplified on the bottom plane of the cube in the figure. 
We note that the skyrmions are deformed and elongated along the direction of the horizontal Dirac strings. 
Figure~\ref{fig:4qch_spin}(f) is for the $N_{\rm m}=16$ state at $m=0.9$ and $\tvp=\pi$. 
In this state, while $N_{\rm m}$ takes the same value as that in Figs.~\ref{fig:4qch_spin}(a) and \ref{fig:4qch_spin}(b), the spin configuration is completely different from them. 
In all the cases, the spin configurations have twofold rotational symmetry about each vertical Dirac string. 
We note that the spin configurations with $\tvp=0$ have fourfold rotational symmetry about each Dirac string, as exemplified in Fig.~\ref{fig:4qch_spin}(c), while those with $\tvp=\pi$ are symmetric for the screw operation $\{C_{4z}|-\tilde{\bf a}_{\eta}\}$, as exemplified in Figs.~\ref{fig:4qch_spin}(b), \ref{fig:4qch_spin}(e), and \ref{fig:4qch_spin}(f).
These results are consistent with the symmetry arguments in Table~\ref{tab:4q_scr_sym}.

Let us briefly discuss the present results in comparison with the previous studies. A $4Q$ HL was experimentally discovered in MnSi$_{1-x}$Ge$_x$~\cite{Fujishiro2018}. In this study, the spin texture at zero magnetic field was interpreted as the screw $4Q$ state with four hedgehogs and four antihedgehogs ($N_{\rm m}=8$), which corresponds to $\tvp=0$ in our results, although the value of $\tvp$ was not examined experimentally. Meanwhile, the topological properties and the emergent magnetic field in the screw $4Q$ state were theoretically studied at $\tvp=\pi$ while changing the external magnetic field~\cite{Park2011}, but the study was limited to the $N_{\rm m}=16$ state as the solutions in Eqs.~(\ref{eq:4qchiral_sol3_1})-(\ref{eq:4qchiral_sol3_3}) were not included. Recently, some of the authors studied the evolution of the screw $4Q$ state on a 3D simple cubic lattice by variational calculations and simulated annealing, and found various types of topological transitions depending on the direction of the magnetic field~\cite{Okumura2020}. We will discuss one of them, paying attention to the phase shift in Sec.~\ref{sec:5.2}.

\subsection{Sinusoidal $4Q$ state \label{sec:4.3}}

Next, we analyze the phase shift in the sinusoidal $4Q$ state given by 
\begin{eqnarray}
\Sr
\propto
\left(
\begin{array}{c}
\cos\QQ_1 - \cos\QQ_2 - \cos\QQ_3 + \cos\QQ_4 \\
\cos\QQ_1 - \cos\QQ_2 + \cos\QQ_3 - \cos\QQ_4 \\
\cos\QQ_1 + \cos\QQ_2 - \cos\QQ_3 - \cos\QQ_4  + 2\sqrt{3}m
\end{array}
\right),\notag \\
\label{eq:4qnc_ansatz}
\end{eqnarray}
which is obtained from Eq.~(\ref{eq:general_ansatz}) by taking $N_Q=4$ and
\begin{eqnarray}
&&\psi_{\eta}^{\rm c}=\frac{1}{2}
,\ \ \psi_{\eta}^{\rm s}=0, \ \  
{\bf e}_{\eta}^1 = \frac{{\bf q}_{\eta}}{q}.
\end{eqnarray} 

Following the arguments in Sec.~\ref{sec:4.2}, we summarize the symmetry operations for the sinusoidal $4Q$ state in Table~\ref{tab:4q_sin_sym}.
We find that the spin texture is unchanged for $C_{4z}$, $\mathcal{I}$, $\mathcal{T}C_{2x}$, and their combinations for $\tvp=0$, 
$\{C_{4z}|-\tilde{\bf a}_{\eta}\}$, $\mathcal{I}$, $\mathcal{T}C_{2x}$, and their combinations for $\tvp=\pi$, 
otherwise $C_{2z}$, $S_{4z}$, $\mathcal{T}C_{2x}$, and their combinations, where $S_{4z}$ represents the fourfold improper rotation operation about the $z$ axis. 

\begin{table}
\caption{\label{tab:4q_sin_sym}
Similar table to Tables~\ref{tab:3q_scr_sym}--\ref{tab:4q_scr_sym} for the sinusoidal $4Q$ state in Eq.~(\ref{eq:4qnc_ansatz}).
}
\begin{ruledtabular}
\begin{tabular}{c|ccc}
operation & sum of phases & translation  & magnetization\\
\hline 
$C_{4z}$ &  $\tvp \rightarrow 2\pi - \tvp$ & $\tilde{\bf a}_{\eta}$ & $m \rightarrow m$ \\
$C_{2x}$ & $\tvp \rightarrow \tvp$ & $0$ & $m \rightarrow -m$ \\
$\mathcal{I}$ & $\tvp \rightarrow 2\pi - \tvp$ & $\tilde{\bf a}_{\eta}+\tilde{\bf a}_{\eta'}$ & $m \rightarrow m$  \\
$\mathcal{T}$ & $\tvp \rightarrow \tvp$ & $\tilde{\bf a}_{\eta}+\tilde{\bf a}_{\eta'}$ & $m \rightarrow -m$ 
\end{tabular}
\end{ruledtabular}
\end{table}

\subsubsection{Hedgehogs in hyperspace \label{sec:4.3.1}}

\begin{figure*}[tb]
\centering
\includegraphics[width=1.0\textwidth]{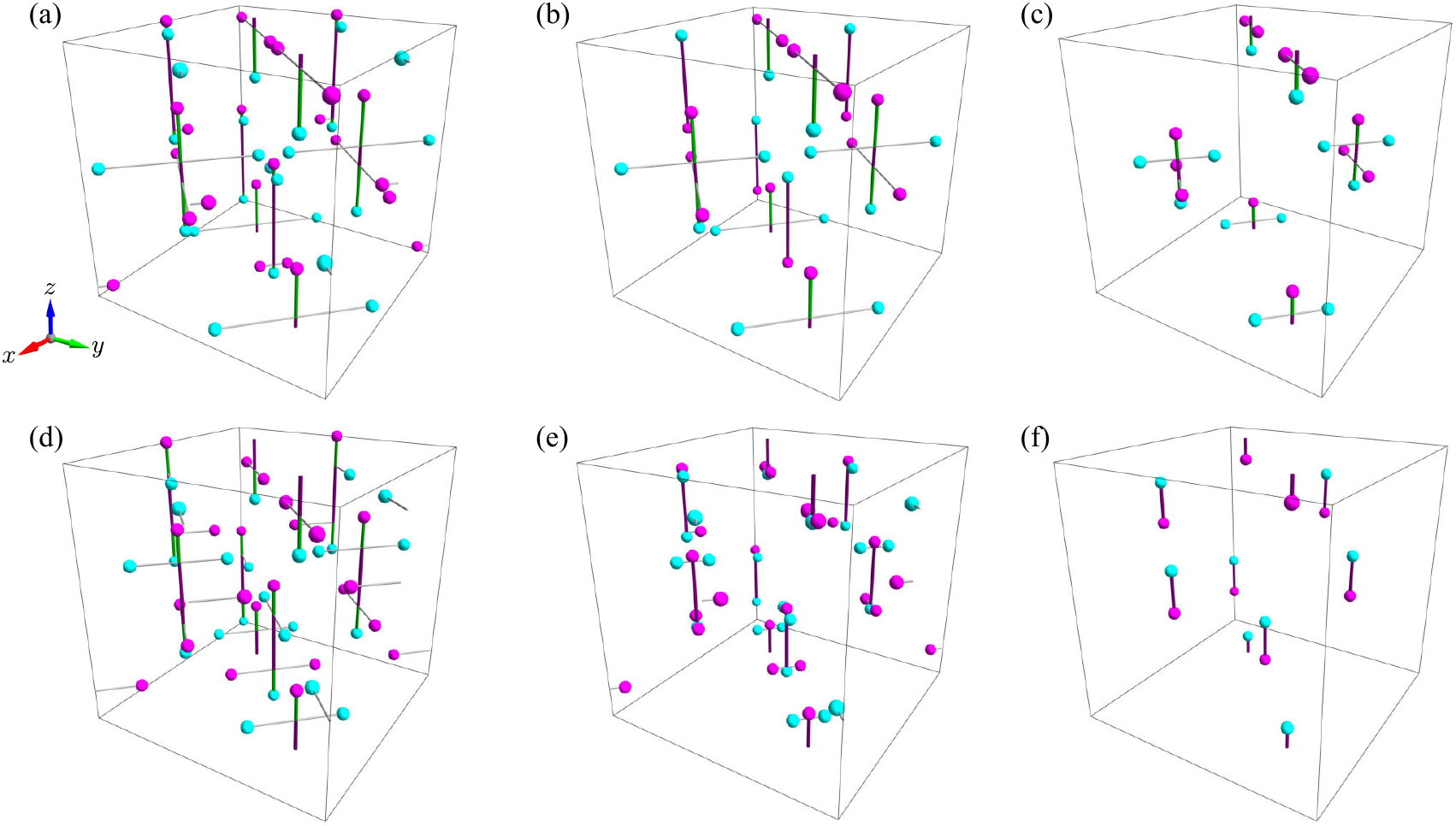}
\caption{
\label{fig:4qnc_hedgehogs}
Real-space distribution of the hedgehogs and antihedgehogs, and the Dirac strings 
within the $L^3$ cube [see Fig.~\ref{fig:4q_setup}(b)] for the sinusoidal $4Q$ state in Eq.~(\ref{eq:4qnc_ansatz}) 
while changing $m$ and $\tvp$: 
(a) $m=0$, (b) $m=0.1$, and (c) $m=0.7$ at $\tvp=\pi/3$, and 
(d) $m=0$, (e) $m=0.5$, and (g) $m=0.7$ at $\tvp=\pi$. 
The notations are common to those in Fig.~\ref{fig:4qchiral_hedgehogs}.
}
\end{figure*}

Following the same procedure in Sec.~\ref{sec:4.2.1}, we compute the positions of the topological defects in the 4D hyperspace 
for the spin texture corresponding to Eq.~(\ref{eq:4qnc_ansatz}). 
Here, we show the solutions ${\bf S}({\bf R})=0$ within $0 \leq X,Y < \frac{L}{2}$ and $0 \leq Z < L$, but the spatial translations of them 
with ${\bf a}_{\eta}$ in Eq.~(\ref{eq:4Q_a_eta}) also satisfy ${\bf S}({\bf R})=0$. 
When $|m| < \frac{2}{\sqrt{3}} \sin W$, we obtain two solutions analytically:
\begin{eqnarray}
(X^*, Y^*, Z^*)=L\left( 0, 0, \frac14 \pm \frac{1}{2\pi}r_1^{\rm sin}(m, W) \right), 
\label{eq:4qnc_sol1}
\end{eqnarray}
where 
\begin{eqnarray}
&&r_1^{\rm sin}(m,W)=\arccos\left(\frac{\sqrt{3}m}{2\sin W}\right).
\end{eqnarray}
Meanwhile, when $|m| < \frac{2}{\sqrt{3}}\cos W$, we obtain other two analytical solutions as 
\begin{eqnarray}
(X^*, Y^*, Z^*)&=&
L\left( \frac14, \frac14, \frac34 + \frac{1}{2\pi}r_2^{\rm sin}(m, W) \right), \notag\\
&&
L\left( \frac14, \frac14, \frac14 - \frac{1}{2\pi}r_2^{\rm sin}(m, W) \right), 
\label{eq:4qnc_sol2} 
\end{eqnarray}
where
\begin{eqnarray}
&&r_2^{\rm sin}(m,W)=\arcsin\left(\frac{\sqrt{3}m}{2\cos W}\right). 
\end{eqnarray}
The solutions in Eqs.~(\ref{eq:4qnc_sol1}) and (\ref{eq:4qnc_sol2}) do not change their $XY$ coordinates 
with $m$ and $\tvp$, as Eqs.~(\ref{eq:4qchiral_sol1}) and (\ref{eq:4qchiral_sol2}) for the screw $4Q$ case.
On the other hand, when $-\frac{2}{\sqrt{3}} \cos^2 W < m < \frac{2}{\sqrt{3}}\sin^2 W $, we obtain four solutions: 
\begin{eqnarray}
(X^*, Y^*, Z^*)&=&
L\left( \frac{1}{2\pi}r_{3+}^{\rm sin}(m,W), \frac12 - \frac{1}{2\pi}r_{3+}^{\rm sin}(m,W), \frac12 + \frac{W}{2\pi} \right), \notag\\
&&
L\left( \frac12 - \frac{1}{2\pi}r_{3+}^{\rm sin}(m,W), \frac{1}{2\pi}r_{3+}^{\rm sin}(m,W), \frac12 + \frac{W}{2\pi} \right), \notag\\
&&
L\left( \frac{1}{2\pi}r_{3+}^{\rm sin}(m,W), \frac{1}{2\pi}r_{3+}^{\rm sin}(m,W), \frac12 - \frac{W}{2\pi} \right), \notag\\
&&
L\left( \frac12 - \frac{1}{2\pi}r_{3+}^{\rm sin}(m,W), \frac12 - \frac{1}{2\pi}r_{3+}^{\rm sin}(m,W), \frac12 - \frac{W}{2\pi} \right), \notag \\
\label{eq:4qnc_sol3}
\end{eqnarray}
where 
\begin{eqnarray}
r_{3+}^{\rm sin}(m,W)=\frac{1}{2}\arccos\left( \cos \left(2W \right) + \sqrt{3}m \right).
\end{eqnarray}
For $-\frac{2}{\sqrt{3}} \sin^2 W < m < \frac{2}{\sqrt{3}}\cos^2 W $, we obtain four solutions: 
\begin{eqnarray}
(X^*, Y^*, Z^*)&=&
L\left( \frac{1}{2\pi}r_{3-}^{\rm sin}(m,W), \frac12 - \frac{1}{2\pi}r_{3-}^{\rm sin}(m,W), \frac{W}{2\pi} \right), \notag\\
&&
L\left( \frac12 - \frac{1}{2\pi}r_{3-}^{\rm sin}(m,W), \frac{1}{2\pi}r_{3-}^{\rm sin}(m,W), \frac{W}{2\pi} \right), \notag\\
&&
L\left( \frac{1}{2\pi}r_{3-}^{\rm sin}(m,W), \frac{1}{2\pi}r_{3-}^{\rm sin}(m,W), 1 - \frac{W}{2\pi} \right), \notag\\
&&
L\left( \frac12 - \frac{1}{2\pi}r_{3-}^{\rm sin}(m,W), \frac12 - \frac{1}{2\pi}r_{3-}^{\rm sin}(m,W), 1 - \frac{W}{2\pi} \right), \notag \\
\label{eq:4qnc_sol4}
\end{eqnarray}
where 
\begin{eqnarray}
&&r_{3-}^{\rm sin}(m,W)=\frac{1}{2}\arccos\left( \cos \left( 2W \right) - \sqrt{3}m \right).
\end{eqnarray}
In contrast to Eqs.~(\ref{eq:4qnc_sol1}) and (\ref{eq:4qnc_sol2}), the solutions in Eqs.~(\ref{eq:4qnc_sol3}) and (\ref{eq:4qnc_sol4}) 
change their $XY$ coordinates with $m$ and $\tvp$, while the $Z$ coordinates do not change.
These solutions correspond to those obtained numerically in the screw $4Q$ case. 

Figure~\ref{fig:4qnc_hedgehogs} showcases the systematic change of the topological defects in the original 3D space 
while changing $m$ in Eq.~(\ref{eq:4qnc_ansatz}) with $\tvp=\frac{\pi}{3}$ and $\pi$.
The monopole charge of the hedgehogs and antihedgehogs, $Q_{\rm m}$, and the vorticity of the Dirac strings, $\zeta$, are calculated 
by the same procedure in Sec.~\ref{sec:4.2.1}. 

First, we discuss the case of $\tvp=\frac{\pi}{3}$ shown in Figs.~\ref{fig:4qnc_hedgehogs}(a), 
\ref{fig:4qnc_hedgehogs}(b), and \ref{fig:4qnc_hedgehogs}(c).
When $m=0
$, there are 48 defects in total and half of them are hedgehogs with $Q_{\rm m}=+1$ and the others are antihedgehogs with $Q_{\rm m}=-1$, 
as shown in Fig.~\ref{fig:4qnc_hedgehogs}(a). 
These topological defects are derived from Eqs.~(\ref{eq:4qnc_sol1}), (\ref{eq:4qnc_sol2}), (\ref{eq:4qnc_sol3}), and (\ref{eq:4qnc_sol4}).
There are two types of the Dirac strings connecting the hedgehogs and antihedgehogs: 
One is the vertical strings for the hedgehog-antihedgehog pairs, and the other is the horizontal ones for the hedgehog-hedgehog pairs or the antihedgehog-antihedgehog pairs. 
Similar to the screw $4Q$ case, the vorticity is ill-defined for the latter horizontal ones, and they intersect the former vertical ones whose vorticities change their signs at the crossing points; 
three hedgehogs and three antihedgehogs form a cluster like a twisted two-barred cross. 
However, the horizontal ones are straight, in contrast to the curved ones in the screw $4Q$ case (see Fig.~\ref{fig:4qchiral_hedgehogs}).
Moreover, they are found for all $\tvp$, while they appear only for $\frac{2\pi}{3}<\tvp<\frac{4\pi}{3}$ in the screw $4Q$ case. 
All the vertical Dirac strings have the length of $\frac{L}{2}$ at $m=0
$, while the horizontal ones have a long or short length, resulting in the two types of clusters; 
the hedgehogs and antihedgehogs in the four clusters with longer horizontal strings are given by Eqs.~(\ref{eq:4qnc_sol2}) and (\ref{eq:4qnc_sol4}),
while those in the rest four with shorter ones are by Eqs.~(\ref{eq:4qnc_sol1}) and (\ref{eq:4qnc_sol3}). 
By introducing $m$, the topological defects move along the Dirac strings, and two hedgehogs and one antihedgehog 
(or one hedgehog and two antihedgehogs) collide with each other, leaving one (anti)hedgehog, at 
$m=\frac{2}{\sqrt{3}}\sin^2\frac{\tvp}{4}$ in the four clusters with shorter horizontal Dirac strings. 
The fusion process is similar to those in Fig.~\ref{fig:4qchiral_sign_change}.
At the topological transition by the fusion, $N_{\rm m}$ decreases from $48$ to $32$, and for larger $m$, the vertical Dirac strings left by the fusion, which have $\zeta=-1$, coexist with the clusters remaining, as shown in Fig.~\ref{fig:4qnc_hedgehogs}(b).
By further increasing $m$, the hedgehogs and antihedgehogs connected by the vertical Dirac strings cause the pair annihilation 
at $m=\frac{2}{\sqrt{3}}\sin\frac{\tvp}{4}$, leaving four clusters derived from Eqs.~(\ref{eq:4qnc_sol2}) and (\ref{eq:4qnc_sol4}), as shown in Fig.~\ref{fig:4qnc_hedgehogs}(c). 
At the topological transition by the pair annihilation, $N_{\rm m}$ is further reduced from $32$ to $24$. 
When $m=\frac{2}{\sqrt{3}}\cos^2\frac{\tvp}{4}$, the fusion takes place in the remaining four clusters, which reduces $N_{\rm m}$  
from 24 to 8, and leaves four vertical Dirac strings with $\zeta=-1$. 
Finally, the hedgehogs and antihedgehogs on the Dirac strings 
pair annihilate at $m=\frac{2}{\sqrt{3}}\cos\frac{\tvp}{4}$, where the system becomes topologically trivial with $N_{\rm m}=0$.

Next, we discuss the case of $\tvp=\pi$ shown in Figs.~\ref{fig:4qnc_hedgehogs}(d), 
\ref{fig:4qnc_hedgehogs}(e), and \ref{fig:4qnc_hedgehogs}(f).
When $m=0
$, $N_{\rm m}$ is 48 as in the case of $\tvp=\frac{\pi}{3}$; the positions of the hedgehogs and antihedgehogs 
connected by the vertical Dirac strings are the same, but those connected by the horizontal Dirac strings are different, as shown in Fig.~\ref{fig:4qnc_hedgehogs}(d). 
By introducing $m$, the topological defects move along the Dirac strings as shown in Fig.~\ref{fig:4qnc_hedgehogs}(e), and 
the fusion occurs at $m=\frac{1}{\sqrt{3}}$ simultaneously in all the clusters. 
This leaves eight pairs of the hedgehogs and antihedgehogs connected by the vertical Dirac strings, as shown in 
Fig.~\ref{fig:4qnc_hedgehogs}(f); $N_{\rm m}$ decreases from $48$ to $16$. 
In this state, all the Dirac strings have the length of $\frac{L}{\pi}r_1^{\rm sin}\left(m,\frac{\tvp}{4}\right)=L\left(\frac{1}{2}-\frac{1}{\pi}r_2^{\rm sin}\left(m,\frac{\tvp}{4}\right)\right)$ and the vorticity $\zeta=-1$. 
By further increasing $m$, the remaining pairs annihilate at $m=\sqrt{\frac{2}{3}}$.

\subsubsection{Topological phase diagram \label{sec:4.3.2}}

\begin{figure}[tb]
\centering
\includegraphics[width=1.0\columnwidth]{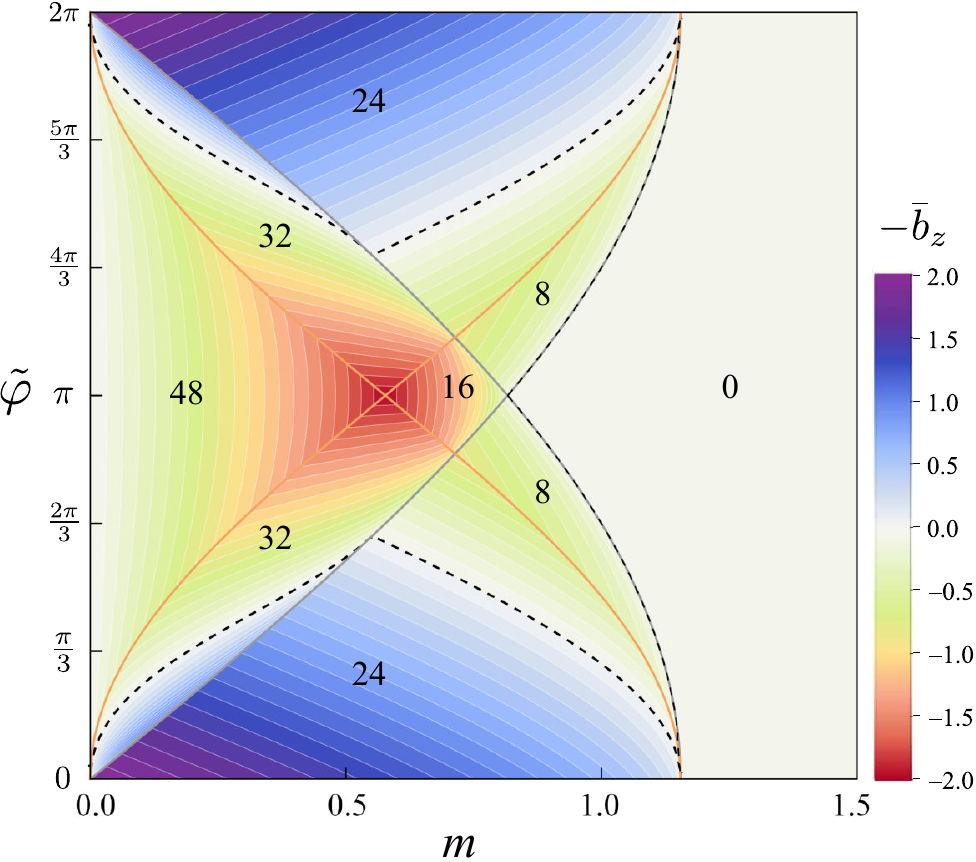}
\caption{
\label{fig:4qnc_pd}
Topological phase diagram for the sinusoidal $4Q$ state determined by $N_{\rm m}$, with the contour plot of $-\bar{b}_z$, on the plane of $m$ and $\tvp$.
$N_{\rm m}$ is indicated in each phase. 	
The white lines denote the contours drawn every 0.1, and the black dashed lines denote $\bar{b}_z=0$. 
The gray and orange lines are the phase boundaries between different $N_{\rm m}$ phases: The former denotes the pair annihilation of hedgehogs and antihedgehogs, while the latter denotes the fusion of three topological defects.
}
\end{figure}

Performing similar calculations to those in Sec.~\ref{sec:4.2.2} while changing $\tvp$ and $m$, we elaborate the phase diagram shown in Fig.~\ref{fig:4qnc_pd}. 
We also plot $-\bar{b}_z$ by the contour, as in Sec.~\ref{sec:4.2.2}. 
Similar to Fig.~\ref{fig:4qch_pd}, the result is again symmetric with respect to $\tvp=\pi$ and has $2\pi$ periodicity, and the phase diagram for $m<0$ is obtained in the same form with sign inversion of $-\bar{b}_z$. 
We find the topological phases with $N_{\rm m}=8$, 16, 24, 32, and 48 in the phase diagram. 
There are two interesting features, in comparison with the result for the screw $4Q$ case in Fig.~\ref{fig:4qch_pd}. 
One is that the topological phases with large values of $N_{\rm m}$, such as $N_{\rm m}=24$, 32, and 48, appear in the large portions of the phase diagram. 
In particular, the system has $N_{\rm m}=48$ for all $\tvp$ in the small $m$ limit, and turns into the $N_{\rm m}=32$ state by the fusion on the orange lines in the phase diagram. 
Near $\tvp=0$ or $2\pi$, the $N_{\rm m}=24$ regions extend widely in the intermediate $m$ region, and the $N_{\rm m}=32$ states appear in between. 
The value of $-\bar{b}_z$ in the phase with $N_{\rm m}=48$ is given by 
\begin{equation}
-\bar{b}_z=2\left[ 1+\frac{2}{\pi}\left( 
r_1^{\rm sin}\left(m, \frac{\tvp}{4}\right) - r_2^{\rm sin}\left(m, \frac{\tvp}{4}\right)
\right) \right].
\end{equation}
Meanwhile, the values of $-\bar{b}_z$ in the $N_{\rm m}=32$ and $24$ phases are given as 
\begin{equation}
-\bar{b}_z = 2\left[ 1 - \frac{2}{\pi}\left( 
r_1^{\rm sin}\left(m, \frac{\tvp}{4}\right) + r_2^{\rm sin}\left(m, \frac{\tvp}{4}\right)
\right) - \frac{\tvp}{\pi} \right],
\label{eq:bz_Nm=32}
\end{equation} 
and
\begin{equation}
-\bar{b}_z = 2\left[ 1 - \frac{2}{\pi}r_2^{\rm sin}\left(m, \frac{\tvp}{4}\right) - \frac{\tvp}{\pi} \right],
\label{eq:bz_Nm=24}
\end{equation}
respectively, when $0 \leq \tvp \leq \pi$; 
$-\bar{b}_z$ for $\pi \leq \tvp \leq 2\pi$ is obtained by the symmetry with respect to $\tvp=\pi$. 
These are in stark contrast to the screw $4Q$ case where the topological phases 
with $N_{\rm m}=32$ and 48 appear only in the limited region for $\frac{2\pi}{3} < \tvp < \frac{4\pi}{3}$ and $N_{\rm m}=24$ is not found. 
The other interesting feature is that the sign of $-\bar{b}_z$ is not limited to positive unlike the screw $4Q$ case; it can be  
negative while changing $m$ 
and $\tvp$ in the phases with $N_{\rm m}=32$ and $24$.
The sign changes are caused by the competition between the lengths of the Dirac strings with $\zeta=+1$ and $-1$; see also Eqs.~(\ref{eq:bz_Nm=32}) and (\ref{eq:bz_Nm=24}).

We find that $-\bar{b}_z$ takes the minimum value of $-2$ at $(m,\tvp)=\left(\frac{1}{\sqrt{3}}, \pi \right)$, where all the defects connected by the horizontal Dirac strings cause the fusion simultaneously. 
This state has the hedgehog-antihedgehog pairs given by the analytical solutions in Eqs.~(\ref{eq:4qnc_sol1}) and (\ref{eq:4qnc_sol2}) connected by the Dirac strings whose vorticities and  lengths are all $\zeta=-1$ and $\frac{L}{4}$, respectively. 
On the other hand, $-\bar{b}_z$ takes the maximum value of $2$ in the limit of $m\to 0$ at $\tvp=0$ ($2\pi$). 
This is understood as follows. 
When $\tvp=0$, the system has the hedgehogs and antihedgehogs given by Eqs.~(\ref{eq:4qnc_sol2}) and (\ref{eq:4qnc_sol4}) connected by the Dirac strings with the vorticities $\zeta=1$ and the lengths $L\left(\frac{1}{2} - \frac{1}{\pi}r_2^{\rm sin}\left(m, \frac{\tvp}{4}\right)\right)$. 
This leads to 
\begin{equation}
-\bar{b}_z=
2\left[
1 - \frac{2}{\pi}r_2^{\rm sin}\left(m, \frac{\tvp}{4}\right)\right]. 
\label{eq:4qnc_bbar_tvp=0}
\end{equation}
Meanwhile, when $\tvp=2\pi$, the hedgehogs and antihedgehogs are given by 
Eqs.~(\ref{eq:4qnc_sol1}) and (\ref{eq:4qnc_sol3}), and $-\bar{b}_z$ is given by 
\begin{eqnarray}
-\bar{b}_z=
\frac{4}{\pi}r_1^{\rm sin}\left(m, \frac{\tvp}{4}\right).
\label{eq:4qnc_bbar_tvp=2pi}
\end{eqnarray}
Note that Eqs.~(\ref{eq:4qnc_bbar_tvp=0}) and (\ref{eq:4qnc_bbar_tvp=2pi}) are related with each other by the symmetry with respect to $\tvp=\pi$. 
Hence, $-\bar{b}_z$ goes to $2$ in the limit of $m \rightarrow 0$ for both $\tvp=0$ and $2\pi$.

\begin{figure*}[tb]
\centering
\includegraphics[width=2.0\columnwidth]{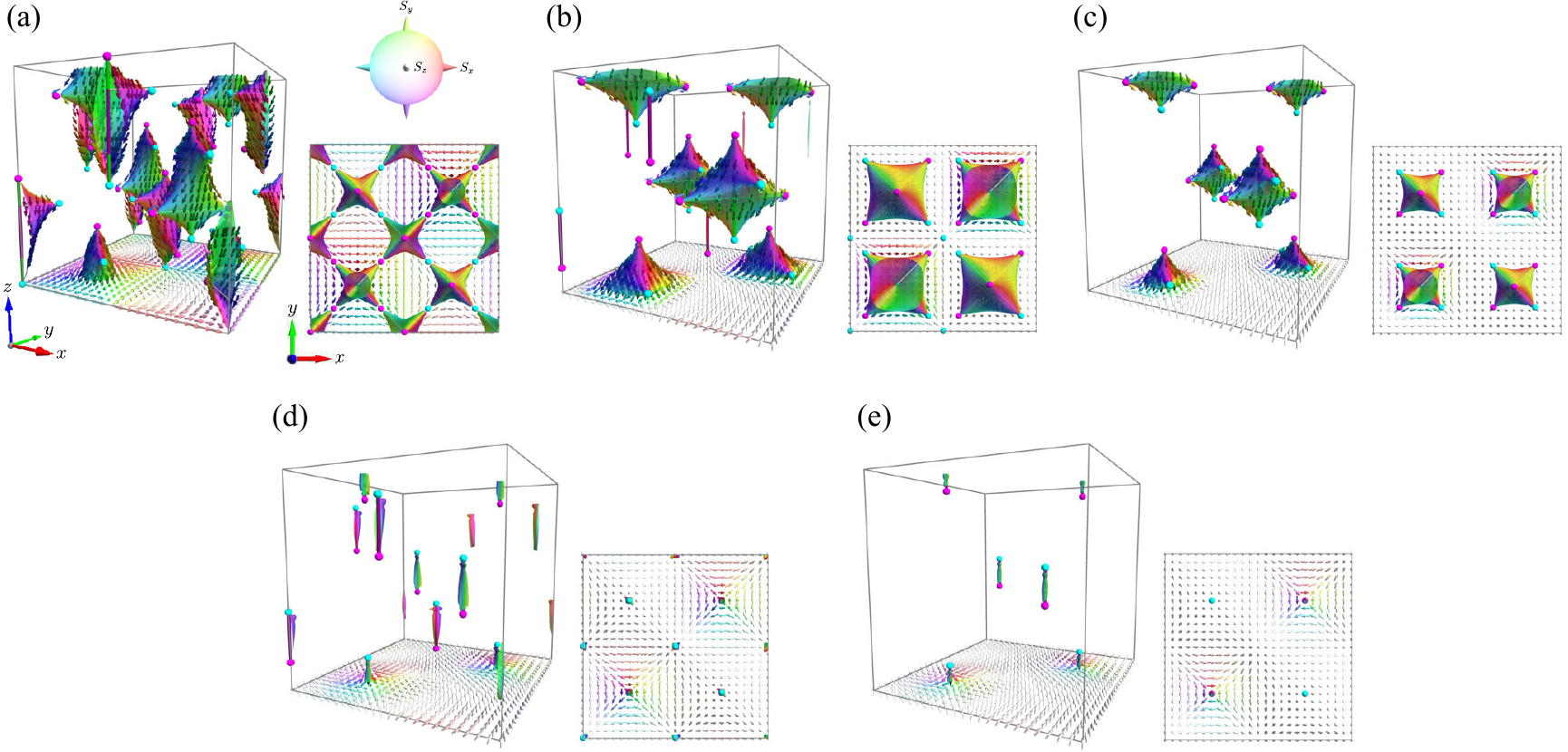}
\caption{
\label{fig:4qnc_spin}
Real-space spin configurations on the isosurfaces with $S_z({\bf r})=-0.9$ for the sinusoidal 4$Q$ state in Eq.~(\ref{eq:4qnc_ansatz}) for different topological phases in Fig.~\ref{fig:4qnc_pd}: 
(a) $m=0$ and $\tvp=\pi$ ($N_{\rm m}=48$), (b) $m=0.2$ and $\tvp=\frac{\pi}{3}$ ($N_{\rm m}=32$), (c) $m=0.6$ and $\tvp=\frac{\pi}{3}$ ($N_{\rm m}=24$), (d) $m=0.6$ and $\tvp=\pi$ ($N_{\rm m}=16$), and (e) $m=0.8$ and $\tvp=\frac{5\pi}{6}$ ($N_{\rm m}=8$).
The notations are common to those in Fig.~\ref{fig:4qch_spin}.
}
\end{figure*}

Figure~\ref{fig:4qnc_spin} showcases typical spin configurations of the sinusoidal $4Q$ states for all the topological phases in Fig.~\ref{fig:4qnc_pd}, in a similar manner to Fig.~\ref{fig:4qch_spin}.  
Figure~\ref{fig:4qnc_spin}(a) is for the $N_{\rm m}=48$ state at $m=0$ and $\tvp=\pi$. 
In this state, all the isosurfaces have the same shape and volume. 
Figure~\ref{fig:4qnc_spin}(b) is for the $N_{\rm m}=32$ state at $m=0.2$ and $\tvp=\frac{\pi}{3}$. 
The isosurfaces including the pairs of the hedgehog and antihedgehog connected by the Dirac strings with $\zeta=-1$ are small (hardly seen in the figure) compared to those including the twisted two-barred crosses, 
and they disappear by pair annihilation while increasing $m$, as shown in Fig.~\ref{fig:4qnc_spin}(c) for the $N_{\rm m}=24$ state at $m=0.6$ and $\tvp=\frac{\pi}{3}$. 
In these states, any horizontal $xy$ plane intersecting the Dirac strings with $\zeta=-1$ gives a SkL with antiskyrmions, as exemplified on the bottom planes of the cubes. 
Figure~\ref{fig:4qnc_spin}(d) is for the $N_{\rm m}=16$ state at $m=0.6$ and $\tvp=\pi$. 
In this state, the horizontal Dirac strings disappear, and all the isosurfaces become small, while they retain knoblike features as the remnant of the horizontal Dirac strings. 
In this case also, the horizontal $xy$ plane is a SkL with antiskyrmions, and the antiskyrmions appear on the $xy$ planes with almost all $z$ coordinates. 
Figure~\ref{fig:4qnc_spin}(e) is for the $N_{\rm m}=8$ state at $m=0.8$ and $\tvp=\frac{5\pi}{6}$.
In this state, the number of the Dirac strings is halved and both Dirac strings and surrounding isosurfaces are shrunk. 
Hence, the antiskyrmions are found on the horizontal planes with limited $z$ coordinates, in contrast to the case in Fig.~\ref{fig:4qnc_spin}(d). 
In all the above cases, the spin configurations have fourfold improper rotational symmetry about each vertical Dirac string, as shown in Fig.~\ref{fig:4qnc_spin}.
Moreover, the spin configurations with $\tvp=\pi$ are symmetric for the screw operation $\{C_{4z}|-\tilde{\bf a}_{\eta}\}$ and the spatial-inversion operation, as exemplified in Figs.~\ref{fig:4qnc_spin}(a) and \ref{fig:4qnc_spin}(d).
These results are consistent with the symmetry arguments in Table~\ref{tab:4q_sin_sym}.

\section{Numerical analysis of phase shift \label{sec:5}}

Thus far, we elucidated the topological properties of 2D $3Q$-SkLs and 3D $4Q$-HLs by systematically changing the phases of their constituent waves as well as the magnetization.
In this section, we study how the actual phases of the superposed waves evolve while increasing an external magnetic field based on specific model Hamiltonians.
For the 2D SkLs, we analyze the numerical data of the real-space spin configurations obtained for the Kondo lattice model in Ref.~\cite{Ozawa2017}, and extract the phases for two types of SkLs with $|N_{\rm sk}|=1$ and $2$ in Sec.~\ref{sec:5.1}. 
In Sec.~\ref{sec:5.2}, we apply similar analysis to the 3D HLs obtained for an effective spin model 
in Ref.~\cite{Okumura2020} to extract the phases for two types of HLs with $N_{\rm m}=8$ and $16$.

\subsection{Phase shift in the sinusoidal 3$Q$ state \label{sec:5.1}}

First, using the numerical data obtained for the Kondo lattice model on a 2D triangular lattice in the previous study~\cite{Ozawa2017}, we extract the phases for the $3Q$ states, and discuss their magnetic field dependences in comparison with our result in Sec.~\ref{sec:3.3}. 
The Kondo lattice model is a fundamental model for the systems in which itinerant electrons are coupled with localized spins, whose 
Hamiltonian is given by
\begin{eqnarray}
\mathcal{H}&=&-\sum_{{\bf r}_l,{\bf r}_{l'},\sigma}t_{{\bf r}_l {\bf r}_{l'}}
\hat{c}^{\dag}_{{\bf r}_{l}\sigma}\hat{c}_{{\bf r}_{l'}\sigma}
-J\sum_{{\bf r}_{l},\sigma,\sigma'} {\bf S}_{{\bf r}_l}\cdot \hat{c}^{\dag}_{{\bf r}_l \sigma} 
\boldsymbol{\sigma}_{\sigma\sigma'} \hat{c}_{{\bf r}_{l} \sigma'} \notag \\
&&
-h\sum_{{\bf r}_{l}} S_{{\bf r}_{l}}^z, 
\label{eq:KLmodel}
\end{eqnarray}
where the operator $\hat{c}^{\dag}_{{\bf r}_l \sigma}$ $(\hat{c}_{{\bf r}_l \sigma})$ creates (annihilates) an electron with spin index 
$\sigma=\pm$ at site ${\bf r}_l$. 
The first term represents the kinetic energy of the itinerant electrons and $t_{{\bf r}_l{\bf r}_{l'}}$ denotes the hopping integral 
between the sites ${\bf r}_l$ and ${\bf r}_{l'}$. 
The second term represents the spin-charge coupling with the coefficient $J$, where $\boldsymbol{\sigma}$ is the vector 
of Pauli matrices. 
The localized spins ${\bf S}_{{\bf r}_l}$ are treated as classical vectors with $|{\bf S}_{{\bf r}_l}|=1$. 
The last term denotes the Zeeman coupling to the external magnetic field $h$, which is taken into account only for the localized spins for simplicity. 
In the previous study, the ground state of the model in Eq.~(\ref{eq:KLmodel}) with the nearest-neighbor hopping $t_1=1$, the third-neighbor hopping $t_3=-0.85$, and $J=0.5$ was studied by using the numerical method based on the kernel polynomial method (KPM) and the Langevin dynamics (LD), which is called 
the modified KPM-LD method~\cite{Barros2013,Ozawa2017,Ozawa2017shape}. 
The ground state was obtained by minimizing the grand potential $\Omega$ given by $\Omega=\braket{\mathcal{H}}/N-\mu n_{\rm e}$, where $N$ is the number of sites ($N=96^2$), $\mu=-3.5$ is the chemical potential, and $n_{\rm e}$ is the electron density defined by $n_{\rm e}=\sum_{{\bf r}_l,\sigma} \braket{\hat{c}^{\dag}_{{\bf r}_l \sigma}\hat{c}_{{\bf r}_l \sigma}}/N$. 
The model was found to stabilize $3Q$ states whose wave vectors are dictated by the Fermi surfaces and given by Eq.~(\ref{eq:3Q_q_eta}) with $q=\frac{\pi}{3}$. 
When $0 \leq h \lesssim 0.00325$, the ground state obtained by the numerical simulation is a $3Q$-SkL with $|N_{\rm sk}|=2$, which is well described by a superposition of three sinusoidal waves like in Eq.~(\ref{eq:nonchiral_3Q_ansatz}).
On the other hand, it turns into a different $3Q$-SkL with $|N_{\rm sk}|=1$ for $h \gtrsim 0.00325$, and finally,  to yet another $3Q$ state with $N_{\rm sk}=0$ for $h \gtrsim 0.0065$. 
In the following, we focus on the region for $0 \leq h \leq 0.006$ since the $3Q$ state with $N_{\rm sk}=0$ for $h \gtrsim 0.0065$ cannot be well represented by Eq.~(\ref{eq:nonchiral_3Q_ansatz}). 

From the spin configurations obtained in the previous study, we extract the phases of the constituent waves in the two SkLs by assuming Eq.~(\ref{eq:nonchiral_3Q_ansatz}).  
For this purpose, we estimate the cost function defined by 
\begin{eqnarray}
&&U(\{\vp_\eta\}, m, \theta, \Gamma; \hat{\bf n}, \xi)=\notag \\
&&\quad\frac{1}{N}\sum_{{\bf r}_l} 
\left(1-{\bf S}^{\rm KPM-LD}_{{\bf r}_l} \cdot \tilde{{\bf S}}^{{\rm sin}3Q}_{{\bf r}_l}(\{\vp_\eta\}, m, \theta, \Gamma; \hat{\bf n}, \xi) \right), 
\label{eq:3q_cost} 
\end{eqnarray}
where ${\bf S}^{\rm KPM-LD}_{{\bf r}_l}$ is the spin configuration obtained by the modified KPM-LD simulation and $\tilde{{\bf S}}^{{\rm sin}3Q}_{{\bf r}_l}$ is that generated from Eq.~(\ref{eq:nonchiral_3Q_ansatz}) as 
\begin{align}
\tilde{{\bf S}}^{{\rm sin3Q}}_{{\bf r}_l}(\{\vp_\eta\}, m, \theta, \Gamma; \hat{\bf n}, \xi)= 
R\left(\hat{\bf n}, \xi\right){\bf S}^{{\rm sin}3Q}_{{\bf r}_l}(\{\vp_\eta\}, m, \theta, \Gamma).  
\label{eq:3q_fit_ansatz}
\end{align}
Here, ${\bf S}^{{\rm sin}3Q}_{{\bf r}_l}(\{\vp_\eta\}, m, \theta, \Gamma)$ is given by Eq.~(\ref{eq:nonchiral_3Q_ansatz}) 
with $q=\frac{\pi}{3}$, and $R\left(\hat{\bf n}, \xi\right)$ denotes the 3D rotation matrix about the unit vector 
$\hat{\bf n}$ with the angle $\xi$. 
We note that the model in Eq.~(\ref{eq:KLmodel}) does not change the energy by flipping all the $y$ components of 
the localized spins, which corresponds to the change of $\Gamma$ between $0$ and $1$ in Eq.~(\ref{eq:nonchiral_3Q_ansatz}); 
hence, the ground states with $\Gamma=0$ and $1$ are energetically degenerate. 
By minimizing $U$ in Eq.~(\ref{eq:3q_cost}) for the numerical data $\{{\bf S}^{\rm KPM-LD}_{{\bf r}_l}\}$ at each value of $h$, we obtain the optimal values of the phases, 
$\vp_{\eta}^{*}$, as well as the other parameters. 
For comparison with $\tvp$ in Sec.~\ref{sec:3.3}, 
we define the sum of $\vp_{\eta}^{*}$ in the form of 
\begin{eqnarray}
\tvp^{*} = \pi - \left|\pi - {\rm Mod}\left[\sum_{\eta=1}^{3} \vp^{*}_{\eta}, 2\pi\right]\right|, 
\label{eq:phase_sum_3q}
\end{eqnarray}
paying attention to the sixfold rotational symmetry of the model and the symmetry by the transformation from $\vp_{\eta}$ to $-\vp_{\eta}$, in addition to the $2\pi$ periodicity. 

\begin{figure}[tb]
\centering
\includegraphics[width=1.0\columnwidth]{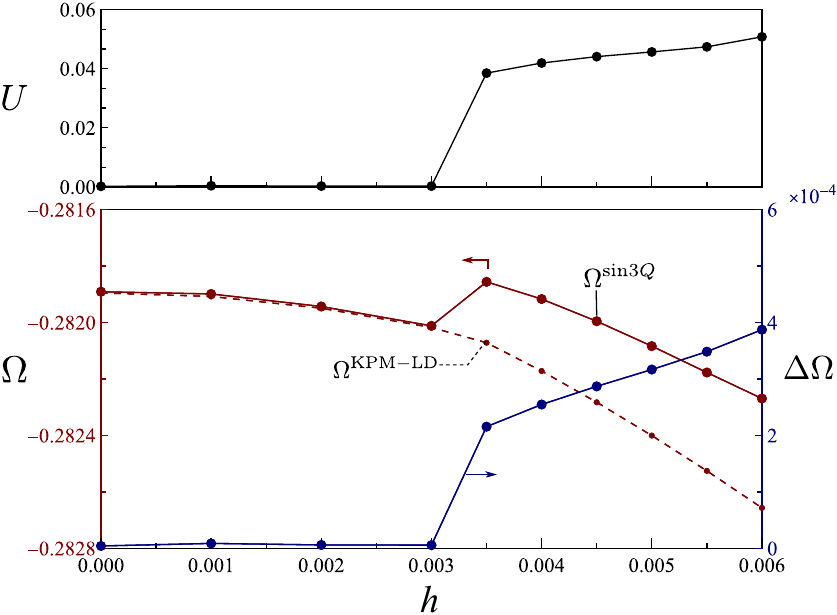}
\caption{
\label{fig:3q_fit_cost}
Magnetic field dependences of (top) the cost function $U$ in Eq.~(\ref{eq:3q_cost}) and (bottom) the grand potential calculated from $\{{\bf S}_{{\bf r}_l}^{\rm KPM-LD}\}$ and $\{\tilde{\bf S}_{{\bf r}_l}^{{\rm sin}3Q}\}^{*}$, and the difference of the grand potential, $\Delta\Omega$.
}
\end{figure}

We plot the $h$ dependence of the optimal values of $U$ in the upper panel of Fig.~\ref{fig:3q_fit_cost}. 
We obtain sufficiently small values less than $4\times 10^{-4}$ for $h < 0.00325$. 
On the other hand, $U$ suddenly increases to $\simeq 4\times 10^{-2}$ at $h = 0.0035$ and shows a gradual increase while increasing $h$. 
The sudden increase of $U$ is related with the topological phase transition at $h \simeq 0.00325$ as discussed below. 
The values for $h> 0.00325$ are, however, still very small, a few percent. 
The results indicate that the optimal spin state $\{\tilde{\bf S}_{{\bf r}_l}^{{\rm sin}3Q}\}^{*}$ reproduces well $\{{\bf S}_{{\bf r}_l}^{\rm KPM-LD}\}$ for all $h$. 

In the lower panel of Fig.~\ref{fig:3q_fit_cost}, we show the $h$ dependences of the grand potential calculated from the numerically obtained ground state $\{{\bf S}_{{\bf r}_l}^{\rm KPM-LD}\}$ and the optimal state $\{\tilde{\bf S}_{{\bf r}_l}^{{\rm sin}3Q}\}^{*}$. 
We also plot the difference of the grand potential $\Delta \Omega=\Omega^{{\rm sin}3Q} - \Omega^{{\rm KPM-LD}}$. 
The results indicate that the optimal states $\{\tilde{\bf S}_{{\bf r}_l}^{{\rm sin}3Q}\}^{*}$ almost perfectly reproduce $\{{\bf S}_{{\bf r}_l}^{\rm KPM-LD}\}$ for $h < 0.00325$ ($\Delta\Omega < 9.0 \times 10^{-6}$, whose relative error is less than $3.2\times10^{-3}$~\%), while slight modifications could improve the results for $h > 0.00325$ ($2.0 \times 10^{-4} < \Delta\Omega < 4.0 \times 10^{-4}$, whose relative error is less than $0.14$~\%). 
We return to this point in the end of this subsection. 

\begin{figure}[tb]
\centering
\includegraphics[width=1.0\columnwidth]{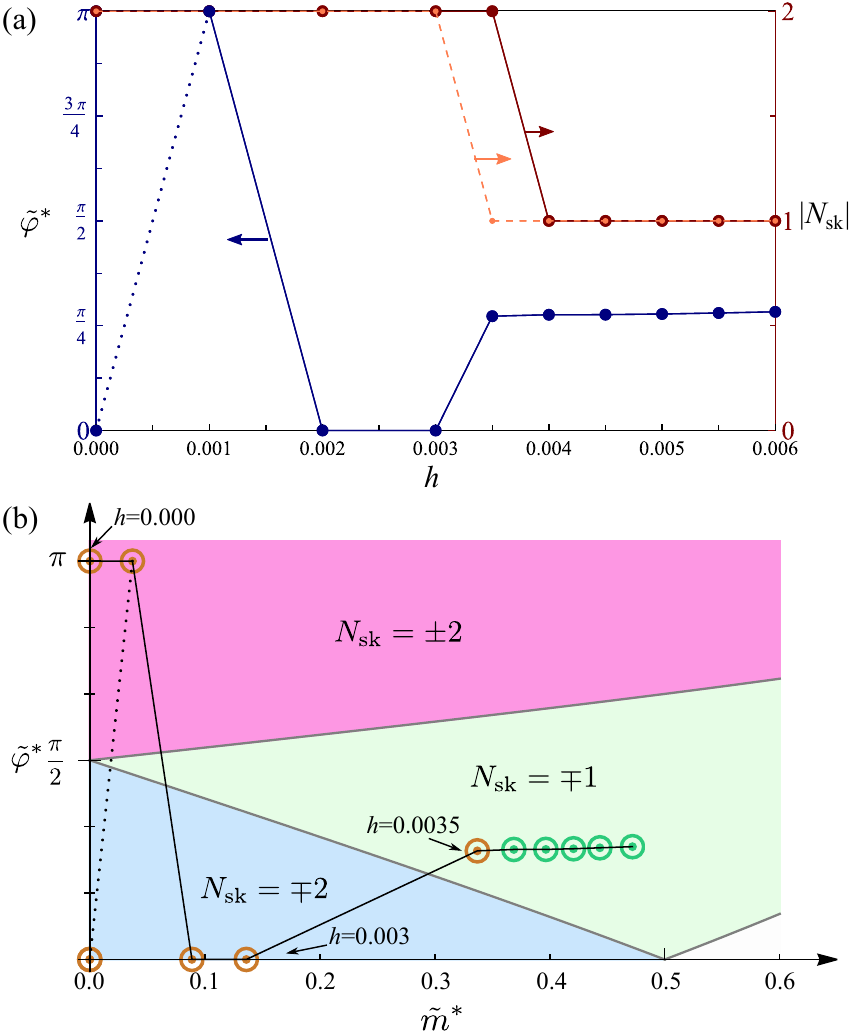}
\caption{
\label{fig:3q_fit_phase}
(a) Magnetic field dependences of $\tvp^{*}$ and $|N_{\rm sk}|$ calculated from the optimal spin textures 
$\{\tilde{{\bf S}}_{{\bf r}_l}^{{\rm sin}3Q}\}^*$. 
For comparison, $|N_{\rm sk}|$ obtained by the modified KPM-LD simulations is plotted by the orange points connected by the dashed line. 
At $h=0$, we plot the results of $\tvp^*$ at both $\tvp^*\simeq 0$ and $\pi$ because of the time-reversal symmetry; the solid and dotted lines are the guides for the eye (see the text). 
(b) Evolution of $\tilde{m}^*$ and $\tvp^{*}$ obtained by the optimization. 
The background is a part of the phase diagram in Fig.~\ref{fig:3qsin_Nsk}. 
The orange and green circles represent that the data are for $|N_{\rm sk}|=2$ and $|N_{\rm sk}|=1$, respectively. 
}
\end{figure}

Figure~\ref{fig:3q_fit_phase}(a) shows the $h$ dependences of $\tvp^{*}$ and $|N_{\rm sk}|$ calculated from $\{\tilde{{\bf S}}_{{\bf r}_l}^{{\rm sin}3Q} \}^*$. 
For comparison, we plot $|N_{\rm sk}|$ obtained by the modified KPM-LD simulations. 
For the $|N_{\rm sk}|=2$ state at $h=0$, we obtain $\tvp^{*}\simeq\pi$, but $\tvp^{*}$ is equivalent to $\tvp^{*} +\pi$ at $h=0$ 
because of time-reversal symmetry; to show this explicitly, we plot the data points at both $\tvp^{*}\simeq0$ and $\pi$. 
By increasing $h$, we find that $\tvp^{*}\simeq0$ tends to be favored in the region with $|N_{\rm sk}|=2$, 
whereas $\tvp^{*}$ becomes $\simeq \pi$ only at $h=0.001$.
This is presumably due to the failure of the modified KPM-LD simulation in the small $h$ region where the states with $\tvp^{*}=0$ and $\pi$ have a small energy difference. 
When $h$ is increased above $0.003$, $\tvp^{*}$ is shifted to $\sim \frac{\pi}{4}$, 
and becomes almost constant in the region with $|N_{\rm sk}|=1$. 
This clearly shows that the topological transition with the change of $|N_{\rm sk}|$ caused by the magnetic field is accompanied by the phase shift. 
We note that $|N_{\rm sk}|$ for the optimal spin configuration remains $2$ at $h=0.0035$ where the phase is shifted already, 
while $|N_{\rm sk}|$ is reduced to $1$ for the modified KPM-LD result; see below. 

Figure~\ref{fig:3q_fit_phase}(b) shows the evolution  
of $\tilde{m}^*$ and $\tvp^{*}$ obtained for 
$\{\tilde{\bf S}_{{\bf r}_l}^{{\rm sin}3Q} \}^*$, plotted on the phase diagram in Fig.~\ref{fig:3qsin_Nsk}. 
Note that $\tilde{m}^*$ obtained by the fitting does not correspond to the magnetization in the simulation data; see also Eq. (\ref{eq:mtilde}). 
The orange and green circles represent $|N_{\rm sk}|=2$ and 1, respectively. 
The result indicates that $|N_{\rm sk}|$ calculated from $\{\tilde{\bf S}_{{\bf r}_l}^{{\rm sin}3Q} \}^*$ is consistent with the phase diagram calculated in Sec.~\ref{sec:3.3.2} and the topological transition from $|N_{\rm sk}|=2$ to 1 is associated with the phase shift, except for $h = 0.0035$. 
When $h = 0.0035$, $\{\tilde{\bf S}_{{\bf r}_l}^{{\rm sin}3Q} \}^*$ gives $|N_{\rm sk}|=2$, although the values of $\tilde{m}^*$ and $\tvp^*$ are in the $|N_{\rm sk}|=1$ region, 
as shown in Fig.~\ref{fig:3q_fit_phase}(b). 

The discrepancy at $h = 0.0035$ is presumably due to the insufficient approximation by Eq.~(\ref{eq:3q_fit_ansatz}). 
Throughout the above analysis, we use Eq.~(\ref{eq:3q_fit_ansatz}) for two different topological phases with different $|N_{\rm sk}|$. 
However, it was recently pointed out that the spin state with $|N_{\rm sk}|=1$ is better described by a different form of a superposition of sinusoidal and cosinsoidal waves~\cite{Hayami2021locking}. 
An extension including such a superposition reconciles the discrepancy of $N_{\rm sk}$ at $h=0.0035$ in Fig.~\ref{fig:3q_fit_phase}, and at the same time, suppresses the increases of $U$ and $\Delta\Omega$ in the $|N_{\rm sk}|=1$ region~\cite{Shimizu2021fit}.
We here, however, restrict ourselves to Eq.~(\ref{eq:3q_fit_ansatz}) to discuss the phase shift within the same form of the constituent waves as in Sec.~\ref{sec:3.3}.

\subsection{Phase shift in the screw 4$Q$ state \label{sec:5.2}}

Next, we perform similar analysis for the 3D $4Q$ states on a 3D simple cubic lattice obtained in the previous study~\cite{Okumura2020} 
for an effective spin model derived from the Kondo lattice model with an antisymmetric spin-orbit coupling~\cite{Hayami2018, Okumura2020}.
The Hamiltonian reads 
\begin{eqnarray}
\mathcal{H}&=&2\sum_{\eta}\left[
-J{\bf S}_{{\bf q}_{\eta}}\cdot{\bf S}_{-{\bf q}_{\eta}}
+\frac{K}{N}({\bf S}_{{\bf q}_{\eta}}\cdot{\bf S}_{-{\bf q}_{\eta}})^2 \right. \notag \\
&& \qquad \  \left.
-iD\frac{{\bf q}_{\eta}}{|{\bf q}_{\eta}|}\cdot({\bf S}_{{\bf q}_{\eta}}\times{\bf S}_{-{\bf q}_{\eta}})
\right]
-h\sum_{l}S_{{\bf r}_l}^z, 
\label{eq:4q_ham}
\end{eqnarray}
where ${\bf S}_{{\bf q}_{\eta}}=\frac{1}{\sqrt{N}}\sum_l {\bf S}_{{\bf r}_l}e^{-i{\bf q}_{\eta}\cdot{\bf r}_l}$; 
the first, second, and third terms represent the bilinear, biquadratic, and DM-type interactions in momentum space, while the last term is the Zeeman coupling. 
The parameters are set as $J=1$, $K=0.6$, $D=0.3$, and $N=16^3$. 
The first sum in Eq.~(\ref{eq:4q_ham}) is taken for a particular set of the wave vectors ${\bf q}_{\eta}$ shown in Fig.~\ref{fig:4q_setup}(a) 
and $q=|{\bf q}_{\eta}|=\frac{\sqrt{3}\pi}{4}$, i.e., $L=8$.
The model was found to stabilize $4Q$ states 
for $0 \leq h \lesssim 1.395$, by using the variational calculation for $h=0$ and the simulated annealing for $h\geq 0$~\cite{Okumura2020}. 
When $h=0$, the ground state is given by a superposition of four proper screws, similar to the one discussed in Sec.~\ref{sec:4.2}. 
When $0 \leq h \lesssim 0.575$, the ground state obtained by the simulated annealing is a $4Q$-HL with $N_{\rm m}=16$. 
By increasing $h$, half of the hedgehog-antihedgehog pairs cause pair annihilation at $h \simeq 0.575$, which changes the spin state to a different $4Q$-HL with $N_{\rm m}=8$. 
The remaining hedgehogs and antihedgehogs pair annihilate at $h \simeq 1.395$ and the system turns into a topologically trivial $4Q$ state with $N_{\rm m}=0$, and finally, into the forced ferromagnetic state for $h \gtrsim 1.395$. 
In the following, we focus on the region for $0\leq h\leq 1.2$ where the spin configurations can be well described by Eq.~(\ref{eq:4qchiral_ansatz}).

Following the procedure in Sec.~\ref{sec:5.1}, we extract the phases of the constituent waves in the two $4Q$-HLs by assuming that the ground state is well approximated by 
a superposition of proper screws like in Eq.~(\ref{eq:4qchiral_ansatz}).
In this case, we estimate the cost function defined by 
\begin{eqnarray}
&U&(\vp_1, \vp_2, \vp_3, \vp_4, m)=\notag \\
&&\frac{1}{N}\sum_{{\bf r}_l} 
\left(1-{\bf S}^{\rm SA}_{{\bf r}_l} \cdot {\bf S}^{{\rm scr}4Q}_{{\bf r}_l}(\vp_1, \vp_2, \vp_3, \vp_4, m) \right), 
\label{eq:4q_cost}
\end{eqnarray}
where ${\bf S}^{\rm SA}_{{\bf r}_l}$ is the spin configuration obtained by the simulated annealing 
and ${\bf S}^{{\rm scr}4Q}_{{\bf r}_l}$ is that generated from Eq.~(\ref{eq:4qchiral_ansatz}) with ${\bf e}_{\eta}^2 \rightarrow -{\bf e}_{\eta}^2$ 
[in the previous study~\cite{Okumura2020}, $D$ was taken to be positive in Eq.~(\ref{eq:4q_ham}), which prefers left-handed screws].
We define the sum of $\vp_{\eta}^*$ in the same form as in Eq.~(\ref{eq:phase_sum_3q}), with the summation of $\eta$ from 1 to 4. 

\begin{figure}[tb]
\centering
\includegraphics[width=1.0\columnwidth]{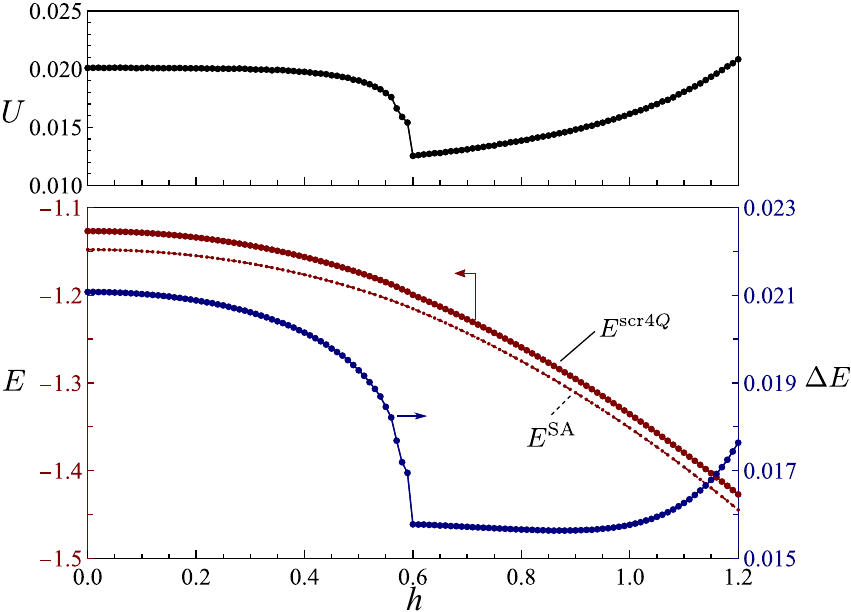}
\caption{
\label{fig:4q_fit_cost}
Magnetic field dependences of (top) the cost function $U$ in Eq.~(\ref{eq:4q_cost}) 
and (bottom) the energy per site calculated from $\{{\bf S}_{{\bf r}_l}^{\rm SA}\}$ and $\{{\bf S}_{{\bf r}_l}^{{\rm scr}4Q}\}^*$, 
and the difference of the energy, $\Delta E$.
}
\end{figure}

The upper panel of Fig.~\ref{fig:4q_fit_cost} shows the $h$ dependence of the optimal values of $U$. 
In the whole range of $h$, the optimal $U$ is less than $0.021$, indicating that the optimal $\{{\bf S}_{{\bf r}_l}^{{\rm scr}4Q}\}^*$ reproduces well the numerically obtained ground state $\{{\bf S}_{{\bf r}_l}^{\rm SA}\}$.
The lower panel of Fig.~\ref{fig:4q_fit_cost} shows the $h$ dependences of the energy per site calculated from $\{{\bf S}_{{\bf r}_l}^{{\rm scr}4Q}\}^*$ and $\{{\bf S}_{{\bf r}_l}^{\rm SA}\}$, $E^{{\rm scr}4Q}$ and $E^{{\rm SA}}$, respectively, and their difference, $\Delta E=E^{{\rm scr}4Q}-E^{{\rm SA}}$. 
We find that the energy is well reproduced for all $h$ (the relative error is less than $1.9$~\%.) 
Both $U$ and $\Delta E$ show rapid changes when approaching $h \sim 0.6$, which 
is related to the topological transition discussed below. 
We note that $U$ as well as $\Delta E$ shows a hump at $h \simeq 0.58$; we return to this point in the end of this subsection. 
 
\begin{figure}[tb]
	\centering
	\includegraphics[width=1.0\columnwidth]{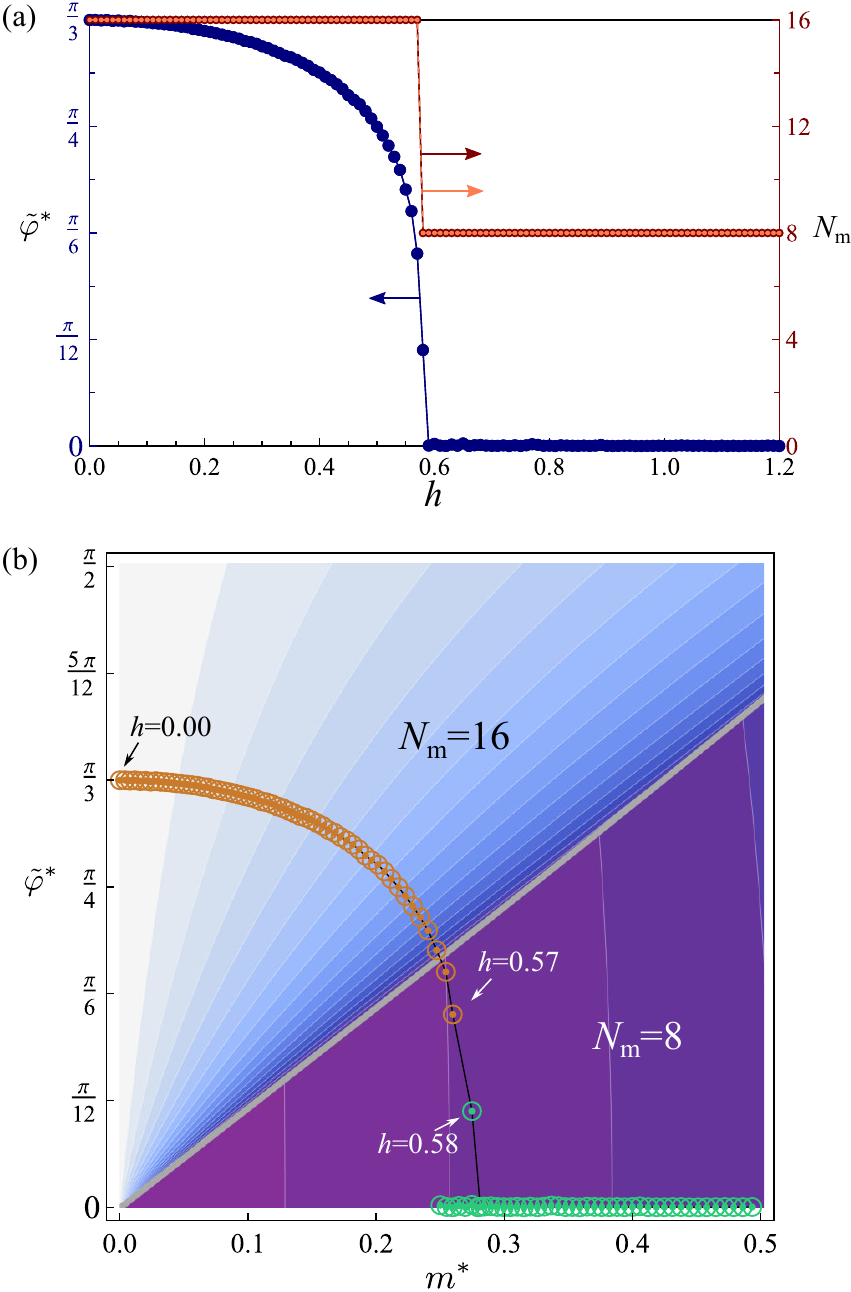}
	\caption{
	\label{fig:4q_fit_phase}
	(a) Magnetic field dependences of $\tvp^{*}$ and $N_{\rm m}$ calculated from the optimal spin textures $\{{\bf S}_{{\bf r}_l}^{{\rm scr}4Q}\}^*$, which are represented by blue and red points, respectively. 
For comparison, $N_{\rm m}$ obtained by the simulated annealing is plotted by the orange points. 
	(b) Evolution of $m^*$ and $\tvp^{*}$ on the phase diagram in Fig.~\ref{fig:4qch_pd}. 
	The background is an enlarged figure of the phase diagram in Fig.~\ref{fig:4qch_pd}. 
	The orange and green circles for each data represent $N_{\rm m}=16$ and $N_{\rm m}=8$, respectively. 
	}
\end{figure}

Figure~\ref{fig:4q_fit_phase}(a) shows the $h$ dependences of $\tvp^{*}$ and $N_{\rm m}$ calculated from the optimal $\{{\bf S}_{{\bf r}_l}^{{\rm scr}4Q}\}^*$.  
The results of $N_{\rm m}$ well reproduce those for $\{{\bf S}_{{\bf r}_l}^{\rm SA}\}$ plotted by the orange points in the figure. 
For the $N_{\rm m}=16$ state at $h=0$, we obtain $\tvp^{*} \simeq \frac{\pi}{3}$, while it is equivalent to $\tvp^{*}\simeq\pi$ and $\frac{4\pi}{3}$ 
because of the threefold rotational symmetry about the [111] axis. 
By increasing $h$, $\tvp^{*}$ gradually decreases from $ \simeq \frac{\pi}{3}$, but rapidly reduces to $\simeq 0$ when approaching $h \simeq 0.585$; $\tvp^{*} \simeq 0$ for $h \gtrsim 0.585$. 
The rapid change of $\tvp^{*}$ appears to occur as a precursor of the topological change from $N_{\rm m}=16$ to $8$.

Figure~\ref{fig:4q_fit_phase}(b) shows the evolution of $m^{*}$ and $\tvp^{*}$ obtained from the optimal $\{{\bf S}_{{\bf r}_l}^{{\rm scr}4Q}\}^*$, plotted on the phase diagram in Fig.~\ref{fig:4qch_pd}. 
The orange and green circles represent the phases with $N_{\rm m}=16$ and 8 in Fig.~\ref{fig:4q_fit_phase}(a), respectively. 
The result indicates that $N_{\rm m}$ calculated from $\{{\bf S}_{{\bf r}_l}^{{\rm scr}4Q}\}^*$ is almost consistent with the phase diagram 
calculated in Sec.~\ref{sec:4.2.2} and the topological transition from $N_{\rm m}=16$ to 8 is correlated with the rapid phase shift to $\simeq 0$. 
We note that two orange points with $N_{\rm m}=16$ are obtained in the $N_{\rm m}=8$ region of the phase diagram. 
As in the $3Q$ case in Sec.~\ref{sec:5.1}, we speculate that the discrepancy is presumably due to the insufficiency of Eq.~(\ref{eq:4qchiral_ansatz}). 
Another possible reason is that the phase diagram in Fig.~\ref{fig:4qch_pd} is calculated in continuous space, while the analysis in this section is done for the discrete lattice system; the phase boundary between $N_{\rm m}=16$ and $8$ may be shifted by the discretization. 

Near the topological transition, we note that $m^*$ increases monotonically up to $h=0.58$, but slightly decreases at $h \simeq 0.59$ after $\tvp^*$ reduces to $\simeq 0$, and increases again for larger $h$, as shown in Fig.~\ref{fig:4q_fit_phase}(b). 
In the previous study, it was pointed out that the system shows the first-order phase transition at $h \simeq 0.595$ accompanied by a jump of the net magnetization~\cite{Okumura2020}. 
Hence, the nonmonotonic behavior of $m^*$ implies that $\{{\bf S}_{{\bf r}_l}^{{\rm scr}4Q}\}^*$ is insufficient to approximate the sudden change of the spin texture through the first-order transition. We speculate that the hump in $U$ in the top panel of Fig.~\ref{fig:4q_fit_cost} appears to be related with this issue.

\section{Discussion \label{sec:6}}

Through this study, we clarified the effect of the phase shift on the spin textures, the symmetry, the topological properties, and the emergent magnetic field of the 2D $3Q$-SkLs and the 3D $4Q$-HLs, by developing the systematic way to deal with the phase degree of freedom, the hyperspace representation. 
Our complete phase diagrams in terms of the sum of phases of the constituent waves, $\tvp$, and the magnetization $m$ provide a ``guiding map'' for searching topologically nontrivial phases and 
novel topological phase transitions, which would shed light on the engineering of the topological properties of the multiple-$Q$ spin states. 

In the case of the 2D $3Q$-SkLs, we unveiled the parameter regions for the SkLs with high skrmion number $|N_{\rm sk}|=2$, in addition to those for the conventional ones with $|N_{\rm sk}|=1$. 
Different values of $N_{\rm sk}$ bring about different emergent electromagnetic phenomena, e.g., in the topological Hall effect~\cite{Loss1992, Ye1999, Bruno2004, Onoda2004, Binz2008, Nakazawa2019} 
and the anomalous Nernst effect~\cite{Mizuta2016,Hirschberger2020TNE}, since $N_{\rm sk}$ is related to the scalar spin chirality by Eq.~(\ref{eq:Nsk}). 
Moreover, the dynamics of the skyrmions also shows different aspects depending on $N_{\rm sk}$, as discussed in Refs.~\cite{Thiele1973, Everschor2011, Schulz2012, Seki2016skyrmions, Zhang2017}. 
While it is usually difficult to directly measure the phase degree of freedom in experiments, our results indicate that the information of the phases 
can be obtained by such transport and optical responses. 

In the 3D $4Q$-HLs, our results also revealed that the phase shift gives rise to a variety of topological phases with different number of the hedgehogs and antihedgehogs, $N_{\rm m}$, ranging from $8$ to $48$ per unit cube. 
In this case also, different values of $N_{\rm m}$ and different distributions of the hedgehogs and antihedgehogs affect the emergent electromagnetic phenomena, such as the topological Hall effect~\cite{Kanazawa2012, Kanazawa2016} and the thermoelectric effect
~\cite{Shiomi2013, Fujishiro2018}.  
This means that such responses can be good probes of the phase shifts in the $4Q$-HLs. 
Most interestingly, we discovered the appearance of the horizontal Dirac strings, which give rise to unconventional pair creation and fusion of the hedgehogs and antihedgehogs. 
Our results indicate that both pair creation and fusion do not affect the emergent magnetic field. 
It would be interesting to explore the emergent electromagnetic phenomena specific to the hidden topological objects. 

A crucial question is how to control the phase degree of freedom. 
Once one can establish a systematic way to cause the phase shift, it is possible to generate intriguing emergent electromagnetic phenomena associated with the topological changes, which would lead to new functionalities of the multiple-$Q$ spin textures. 
In the present study, by analyzing the previous numerical data, we demonstrated that the external magnetic field causes characteristic changes in the phase degree of freedom associated with the topological transitions. 
The phase shifts are, however, limited in the narrow regions of the topological phase diagrams, and there remain wide interesting parameter regions, e.g., the topological changes in the 2D $3Q$-SkLs caused by the pair annihilation of the hedgehogs and antihedgehogs in the 3D hyperspace (black dots in Figs.~\ref{fig:3qscr_Nsk} and \ref{fig:3qsin_Nsk}), and the maxima of the emergent magnetic field by the fusion of the hedgehogs and antihedgehogs in the 3D $4Q$-HLs (crossing points of the orange 
lines in Figs.~\ref{fig:4qch_pd} and \ref{fig:4qnc_pd}). 
It was recently pointed out that the sinusoidal $3Q$ state changes the phase from 
$\tvp=0$ (or $\pi$) to $\tvp=\frac{\pi}{2}$ 
by effective six-spin interactions arising from the entropic contribution or the spin-charge coupling~\cite{Hayami2021phase}
. 
In general, the interactions which can modulate the sum of phases in an $N_Q Q$ spin texture are given by $N_Q$ ($2N_Q$) multiple-spin interactions when $N_Q$ is even (odd). 
Such higher-order multiple spin interactions have been discussed as an origin of noncoplaner spin textures~\cite{Akagi2012, Muhlbauer2009, Binz2006-1, Binz2006-2, Park2011, Hayami2017, Grytsiuk2020, Okumura2020}. 
Further studies on the higher-order interactions are desired as the key ingredients to control the phase degree of freedom. 

While we have considered the 2D 3$Q$ and 3D 4$Q$ states as typical examples of the multiple-$Q$ spin textures with the phase degree of freedom in this study, one can extend the current analysis to other spin textures, such as the 2D sextuple-$Q$ state~{\cite{Okada2018} and the 3D sextuple-$Q$ states~\cite{Binz2006-1, Binz2006-2, Ritz2013}.
In such general cases, the number of the phase degree of freedom can be more than one; for instance, for the 3D sextuple-$Q$ states, 
the additional degree of freedom is $N_Q-d=6-3=3$. 
Such extensions would bring further intriguing topological phenomena beyond the present cases with only one phase degree of freedom. 

\section{Summary \label{sec:7}}

To summarize, we have theoretically studied the effect of phase shifts on the magnetic and topological properties of the multiple-$Q$ spin textures. 
We established the generic framework to systematically deal with the phase shifts by introducing a hyperspace with the additional dimension representing the phase degree of freedom in Sec.~\ref{sec:2}. 
In this framework, we can regard a multiple-$Q$ spin texture with phase variables as the one on an intersection of the corresponding spin texture in the hyperspace. 
Topological objects in the hyperspace characterize the topological defects in the original spin textures: For instance, the Dirac strings in the 3D hyperspace define the cores of 2D skyrmions and antiskyrmions, and the closed loops composed of the singularities and the membranes of the downward spins (Dirac planes) in the 4D hyperspace define the hedgehogs and antihedgehogs, and the Dirac strings connecting them, respectively, in the 3D HLs. 
Thus, we can discuss not only the magnetic textures but also the topological properties from the configuration of topological objects in the hyperspace. 

In Sec.~\ref{sec:3}, we have elucidated the effect of phase shifts on the 2D $3Q$ states composed of three proper screws or three sinusoidal waves, by analyzing the 3D $3Q$ states obtained by the hyperspace representation. 
The 3D $3Q$ states involve the hedgehogs and antihedgehogs, and the Dirac strings connecting them, whose configurations in the hyperspace change with the magnetization $m$. 
We elucidate that the topological defects evolve in a different way between the screw and sinusoidal cases, which leads to distinct topological phase diagrams for the 2D $3Q$ states while changing the sum of phases, $\tvp$. 
For the screw case, we clarified that the major portions of the phase diagram are occupied by the SkLs with $|N_{\rm sk}|=1$, whose structures are ubiquitously found in the chiral magnets under the magnetic field, and the remaining small regions with nonzero magnetization realize the SkLs with high topological numbers, $|N_{\rm sk}|=2$. 
In contrast, we discovered that the regions of the SkLs with $|N_{\rm sk}|=2$ are extended in the sinusoidal case, including all the states with zero magnetization. 
The results indicate that the types of the superposed waves crucially affect the topology of the multiple-$Q$ spin textures through the phase degree of freedom. 

In Sec.~\ref{sec:4}, we have studied the 3D $4Q$ states. 
In this case, the hyperspace representation is given by the 4D loop lattices, whose patterns evolve with $m$.
In the case of the $4Q$ states composed of the proper screws, we found the topological phases with the number of hedgehogs and antihedgehogs per unit cube $N_{\rm m}=8$, 16, 32, and 48 while changing $\tvp$ and $m$. 
On the other hand, for the $4Q$ states composed of the sinusoidal waves, we obtained the topological phases with $N_{\rm m}=8$, 16, 24, 32, and 48. 
In the former, the large portions of the phase diagram as a function of $\tvp$ and $m$ are occupied by the phases with $N_{\rm m}=8$ and $16$, while for the latter, the major portions are occupied by the phases with larger $N_{\rm m}$. 
The emergent magnetic field $\bar{b}_z$ is always negative for the former, but it changes the sign depending on $\tvp$ and $m$ for the latter. 
Interestingly, we discovered that unusual Dirac strings appear on the planes perpendicular to the magnetization direction, and their evolution leads to unconventional topological phenomena: 
pair creation of hedgehogs and antihedgehoges, which increases $N_{\rm m}$ with $m$, in the screw case, 
and fusion of three hedgehogs and antihedgehogs, which maximizes the amplitude of $\bar{b}_z$, in both screw and sinusoidal cases.
These topological phenomena caused by the horizontal Dirac strings have not been found in the 3D HLs~\cite{Kanazawa2016, Zhang2016, Shimizu2021moire}. 
Our finding indicates the importance of the phases for such unexplored topological transitions. 

Finally, in Sec.~\ref{sec:5}, 
we have studied how the phases evolve with an external magnetic field, 
by fitting the spin configurations obtained by the numerical simulations for the microscopic models. 
For the sinusoidal $3Q$ states, by analyzing the results for the Kondo lattice model on the triangular lattice, 
we found that $\tvp^*$ is shifted from $\sim 0$ to $\sim \frac{\pi}{4}$ while increasing the magnetic field, accompanied by the topological transition with reduction of $|N_{\rm sk}|$ from 2 to 1. 
Meanwhile, for the screw $4Q$ states, from the results obtained for the effective spin model on the 3D cubic lattice, 
we elucidated that $\tvp^{*}$ is shifted from $\sim\frac{\pi}{3}$ to $\sim0$ accompanied by the topological transition with reduction of $N_{\rm m}$ from $16$ to $8$. 
These results not only demonstrate that the phase shifts can be caused by the external magnetic field but also suggest further variety of phase shifts depending on the situations. 

Our results have unveiled the unconventional topological phases and topological transitions by the comprehensive study of the phase degree of freedom in the multiple-$Q$ spin textures. 
In order to access such interesting physics, it is crucial to establish the way of controlling the phase variables. 
Once one can control the phase degree of freedom, it is possible to flexibly change the symmetry, the topological properties, and the emergent magnetic field. 
Such changes by the phase shifts are important for not only the magnetic properties including the spin excitation spectra~\cite{Kato2021} 
but also the electronic quantum transport phenomena, as the electronic band structure of conduction electrons is modulated by the spin texture. 
Furthermore, dynamics related with the phase degree of freedom is also an interesting issue since the dynamical change of the spin texture gives rise to not only the emergent magnetic field but also the emergent electric field. 
Such dynamical control would produce electromagnetic phenomena beyond the conventional electromagnetism. 
Our findings provide a guiding map for the future studies.

\begin{acknowledgments}
The authors thank R. Ozawa for providing the numerical data, and Y. Fujishiro, M. Hirschberger, S. Hayami, N. Kanazawa, K. Nakazawa, and R. Yambe for fruitful discussions.
This research was supported by Grant-in-Aid for Scientific Research Grants (Nos. JP18K03447, JP19H05822, JP19H05825, and JP21J20812), JST CREST (Nos. JP-MJCR18T2 and JP-MJCR19T3), and the Chirality Research Center in Hiroshima University and JSPS Core-to-Core Program, Advanced Research Networks. K.S. was supported by the Program for Leading Graduate Schools (MERIT-WINGS). Parts of the numerical calculations were performed in the supercomputing systems in ISSP, the University of Tokyo.
\end{acknowledgments}

\bibliography{ref}

\begin{thebibliography}{101}
\bibitem [{2-1()}]{2-1-1_note}%
  \BibitemOpen
  \href@noop {} {}\bibinfo {note} {In discrete lattice systems, the phase
  degree of freedom is relevant to the symmetry even when $N_Q \leq
  d$.}\BibitemShut {Stop}%
\bibitem [{22_()}]{22_note}%
  \BibitemOpen
  \href@noop {} {}\bibinfo {note} {While one can freely choose ${\bf Q}_{\eta}$
  as long as they are linearly independent, we assume that their projections
  onto the original space are ${\bf q}_{\eta}$ in the following
  analyses.}\BibitemShut {Stop}%
\bibitem [{1_n()}]{1_note}%
  \BibitemOpen
  \href@noop {} {}\bibinfo {note} {One can obtain the hyperspace position ${\bf
  R}$ for any real-space position ${\bf r}$ and phases $\vp_{\eta}$ from
  Eq.~(\ref{eq:r2R}), but the inverse mapping is not unique, as different sets
  of ${\bf r}$ and $\vp_{\eta}$ can give the same ${\bf R}$.}\BibitemShut
  {Stop}%
\bibitem [{3_n()}]{3_note}%
  \BibitemOpen
  \href@noop {} {}\bibinfo {note} {The skymion number is not well defined when
  the hedgehogs and/or antihedgehogs in the hyperspace are involved in the 2D
  intersection.}\BibitemShut {Stop}%
\end{thebibliography}

\end{document}